\documentclass[a4paper,11pt]{article}
\usepackage[centertags]{amsmath}
\usepackage{amsfonts} \usepackage{amssymb} \usepackage{amsthm}
\allowdisplaybreaks[1]
\usepackage{graphicx}
\makeatletter
\def\section{\@startsection {section}{1}{\z@}{-3.5ex plus -1ex minus
 -.2ex}{2.3ex plus .2ex}{\large\bf}}
\def\subsection{\@startsection{subsection}{2}{\z@}{-3.25ex plus -1ex 
minus -.2ex}{1.5ex plus .2ex}{\normalsize\bf}}
\makeatother
\makeatletter

\@addtoreset{equation}{section}

\makeatother

\newcommand{\eq}[1]{Eq.~(\ref{#1})}

\newcommand{\ii}{\mathrm{i}}
\newcommand{\rme}{\mathrm{e}}

\newcommand{\vev}[1]{\langle #1 \rangle}
\newcommand{\Tr}{\mathrm{Tr}\,}

\newcommand{\cF}{{\mathcal{F}}}

\newcommand{\cN}{{\mathcal{N}}}

\newcommand{\cP}{{\mathcal{P}}}

\newcommand{\cS}{{\mathcal{S}}}

\newcommand{\re}{\mbox{Re}\,}

\newcommand{\one}{{\rm 1\kern -.9mm l}}

\newcommand{\ft}[2]{{\textstyle\frac{#1}{#2}}}

\newcommand{\ch}[2]{{\textstyle{\Big[\begin{array}{c} #1\\ #2\end{array}\Big]}}}

\newcommand{\be}{\begin{equation}}
\newcommand{\ee}{\end{equation}}
\newcommand{\p}{\partial}

\newcommand{\wF}{{\widetilde F}}
\newdimen\tableauside\tableauside=1.0ex
\newdimen\tableaurule\tableaurule=0.4pt
\newdimen\tableaustep
\def\phantomhrule#1{\hbox{\vbox to0pt{\hrule height\tableaurule
width#1\vss}}}
\def\phantomvrule#1{\vbox{\hbox to0pt{\vrule width\tableaurule
height#1\hss}}}
\def\sqr{\vbox{%
  \phantomhrule\tableaustep
\hbox{\phantomvrule\tableaustep\kern\tableaustep\phantomvrule\tableaustep}%
  \hbox{\vbox{\phantomhrule\tableauside}\kern-\tableaurule}}}
\def\squares#1{\hbox{\count0=#1\noindent\loop\sqr
  \advance\count0 by-1 \ifnum\count0>0\repeat}}
\def\tableau#1{\vcenter{\offinterlineskip
  \tableaustep=\tableauside\advance\tableaustep by-\tableaurule
  \kern\normallineskip\hbox
    {\kern\normallineskip\vbox
      {\gettableau#1 0 }%
     \kern\normallineskip\kern\tableaurule}%
  \kern\normallineskip\kern\tableaurule}}
\def\gettableau#1 {\ifnum#1=0\let\next=\null\else
  \squares{#1}\let\next=\gettableau\fi\next}
\newcommand{\Yfund}{\tableau{1}}

\newcommand{\eu}{\epsilon_1}
\newcommand{\ed}{\epsilon_2}

\def\XXint#1#2#3{{\setbox0=\hbox{$#1{#2#3}{\int}$}
     \vcenter{\hbox{$#2#3$}}\kern-.5\wd0}}

\begin{document}

\begin{titlepage}
\centerline{\Large \bf Non-perturbative studies of $\mathcal{N}=2$ conformal }
\vskip 0.3cm
\centerline{\Large \bf quiver gauge theories}
\vskip 1.2cm
\begin{center}
{\bf S. K. Ashok$^1$, M. Bill\'o$^2$, E. Dell'Aquila$^1$, 
M. Frau$^{2}$, R. R. John$^1$, A. Lerda$^{2}$}\\
\vskip 0.7cm
$^1${\emph{Institute of Mathematical Sciences, \\
C.I.T. Campus, Taramani\\
Chennai, India 600113}}
\\
\vskip 0.2cm
$^2$ {\emph{Universit\`a di Torino, Dipartimento di Fisica
\\ and I.N.F.N. - sezione di Torino\\
Via P. Giuria 1, I-10125 Torino, Italy}}
\vskip 0.4cm
{\tt sashok,edellaquila,renjan@imsc.res.in,\\ 
billo,frau,lerda@to.infn.it} 
\end{center}
\vskip 0.8cm
\begin{abstract}
We study $\cN=2$ super-conformal field theories in four dimensions that correspond to mass-deformed linear quivers with $n$ gauge groups 
and (bi-)fundamental matter. We describe them using Seiberg-Witten curves
obtained from an M-theory construction and via the AGT correspondence. We take particular care in obtaining the detailed 
relation between the parameters appearing in these descriptions and the physical quantities of the quiver gauge theories. 
This precise map allows us to efficiently reconstruct the non-perturbative prepotential that encodes
the effective IR properties of these theories. We give explicit expressions in the cases $n=1,2$, also in the presence of an $\Omega$-background in the Nekrasov-Shatashvili limit. 
All our results are successfully checked against those of the direct microscopic evaluation of the prepotential \`a la 
Nekrasov using localization methods.
\end{abstract}
\vskip 1cm
\begin{flushright}
\emph{Dedicated to the memory of Tullio Regge}
\end{flushright}
\end{titlepage}

\renewcommand{\thefootnote}{\arabic{footnote}}
\setcounter{footnote}{0} \setcounter{page}{1} 

\tableofcontents
\vspace{1cm}
\section{Introduction and summary}
\label{secn:intro}
Superconformal field theories (SCFT) with $\cN=2$ supersymmetry in four dimensions have attracted a lot of attention, and tremendous progress has been made in describing them and in uncovering their duality structure \cite{Gaiotto:2009we}. Various approaches have been pursued: the geometric description of the low-energy effective action \`a la Seiberg-Witten (SW)\ \cite{Seiberg:1994rs, Seiberg:1994aj}, the exact computation of instanton corrections by means of localization techniques \cite{Nekrasov:2002qd, Nekrasov:2003rj},  the relations to integrable models \cite{Nekrasov:2009rc}, the 2d/4d correspondence also known as the AGT correspondence \cite{Alday:2009aq,Alday:2009fs},
the use of $\beta$-ensembles and matrix model techniques \cite{Dijkgraaf:2009pc, Cheng:2010yw}. Moreover, the string 
embedding of such theories via geometric engineering has led to the possibility of expressing some relevant observables 
via topological string amplitudes \cite{Antoniadis:2010iq}-\nocite{Huang:2011qx}\cite{Antoniadis:2013bja}%
\footnote{We refer to the series of recent review articles \cite{Teschner:2014oja}-\nocite{Gaiotto:2014bja, Tachikawa:2014dja, Maruyoshi:2014eja}\cite{Gukov:2014gja} for an extensive discussion of these topics.}, and further insights have been obtained
by considering several aspects of the gauge/gravity relation and holography in this context
\cite{Buchel:2000cn}-\nocite{Pilch:2000ue,Bertolini:2000dk,Polchinski:2000mx,Billo:2001vg,Benini:2008ir,Cremonesi:2009hq,Billo:2011uc,Billo:2012st,Buchel:2013fpa,Bigazzi:2013xia}\cite{Conde:2013wpa} .
The profound interplay among these various approaches is one of the most fruitful lessons to be learned from studying $\cN=2$ SCFT's. 
Let us emphasize that this interplay relies crucially on the precise relation between the parameters used in the various approaches, uncovered and verified through the analysis of examples of increasing complexity. This is the main rationale behind the work we present here. 

{From} a purely gauge-theoretic point of view, mass-deformed conformal quiver theories have been studied in \cite{Fucito:2012xc}-\nocite{Nekrasov:2012xe}\cite{Nekrasov:2013xda} through limit-shape equations obtained from the saddle-point 
analysis \cite{Fucito:2011pn} of Nekrasov partition functions. This has led to a deeper understanding of the SW geometry of conformal gauge theories and it has clarified the relation between gauge theories, integrable systems and the quantization of various moduli spaces. Our goal in this work is more pragmatic: we discuss and compare various approaches available to study the conformal quiver theories, find the detailed map between the parameters that appear in these approaches, and propose an efficient way to calculate the prepotential of the gauge theory.  

A particularly simple class of $\cN=2$ SCFT's  are those of the so-called class $\cS$ \cite{Gaiotto:2009we}, which arise as compactifications of a $(2,0)$ 6-dimensional theory and admit various weakly-coupled descriptions related by S-dualities. Each of these descriptions contains products of 
$\mathrm{SU}(N_i)$ gauge groups plus matter 
arranged in representations such that all $\beta$-functions vanish. 
Here we will focus on class $\cS$ theories that have a weak-coupling realization in terms of linear quivers with 
$n$ SU(2) gauge groups and matter in fundamental or bi-fundamental representations. For these theories 
one can apply localization techniques \cite{Nekrasov:2002qd,Nekrasov:2003rj}%
\footnote{See also \cite{Bruzzo:2002xf}-\nocite{Losev:2003py, Nekrasov:2004vw}\cite{Fucito:2004gi}.}
to compute microscopically the prepotential $F$ as an expansion in powers 
of the instanton weights $q_i$, with coefficients depending on the masses and on the eigenvalues $a_i$ of the 
vacuum expectation value $\vev{\Phi_{i}}$ of the adjoint scalar of the $i$-th gauge group.
In Appendix~\ref{Nekrasov} we briefly describe this computation and give the expression of the non-perturbative prepotential for the first few instanton numbers. These explicit results provide a very concrete testing ground for any description of the IR regime of these theories.

Localization computations require the introduction of the $\Omega$-deformation parameters $\epsilon_1$ and  $\epsilon_2$, which encode an explicit breaking of the SO(4) Euclidean space-time symmetry. The logarithm of the resulting 
partition function describes the prepotential $F$ in the limit $\epsilon_{1,2}\to 0$, plus a series of $\epsilon$-corrections which correspond to deformations of the gauge theory in the presence of constant backgrounds for bulk
fields, like for example the graviphoton \cite{Antoniadis:1993ze,Ooguri:2004zv}. 
The study of such $\epsilon$-deformations represents an important line of research, and various methods have been used to tackle it 
\cite{Billo:2006jm}-\nocite{Ito:2008qe}\cite{Hellerman:2011mv}. 

In this paper we will consider two distinct approaches to the study of linear quivers: first, we study the IR description of the SU(2)$^n$ theories using the SW curve obtained via an M-theory construction \cite{Witten:1997sc}; next, we use the AGT correspondence \cite{Alday:2009aq,Alday:2009fs} and analyze chiral conformal blocks of Liouville theory in two dimensions. We then show that these two approaches are equivalent to the microscopic multi-instanton calculations  of Nekrasov. To this end, we need all observables, whether arising from M-theory or from the Liouville theory, to be expressed in terms of the physical masses and bare coupling constants of the gauge theory. Therefore we work out the precise and explicit map between the geometric parameters of M-theory, the parameters of the Liouville conformal field theory and the  physical parameters of the gauge theory.

Let us now briefly describe the content of this paper. We begin with the SW curve description of the quiver theories. In general, for 
class $\cS$ theories the SW curves cover a base $C$ which is a Riemann surface with marked punctures whose positions parametrize 
the moduli space of the marginal UV gauge couplings. For the quiver theories we consider, $C$ is a sphere with $(n+3)$ 
punctures and the expression of the SW curve and of the corresponding SW differential $\lambda$ can be derived starting 
from a NS5-D4 system uplifted to M-theory, as originally shown in \cite{Witten:1997sc} and studied in 
great detail in \cite{Gaiotto:2009we,Bao:2011rc}. 
In Section~\ref{secn:Mtheory} we revisit this procedure for a generic quiver with $n$ nodes and derive 
explicitly the curves for generic $n$ in the massless case and for $n=1,2$ in the presence of masses. 
These curves are of the form \cite{Gaiotto:2009we}:
\be
x^2(t) = \frac{\cP_{2n+2}(t)}{t^2(t-t_1)^2\cdots(t-t_n)^2(t-1)^2}~, 
\label{cintr}
\ee
where the $t_i$'s are the positions of the punctures which are related to the 
gauge theory couplings as $q_i = t_i/t_{i+1}$, while $\cP_{2n+2}(t)$ is a polynomial of degree $(2n+2)$ which
depends on the $q_i$'s, on the masses and on the Coulomb branch parameters $u_i$. 
If we are to check the curve (\ref{cintr}) against the microscopic prepotential $F$, we have to take into account the fact that the prepotential
depends on the eigenvalues $a_i$. In the SW approach, the variables $a_i$ and their duals $a_i^D = \partial F/\partial a_i$ 
correspond to periods of the differential $\lambda$ over a symplectic basis of cycles on the SW curve, and are thus functions of the 
parameters $u_i$ appearing in (\ref{cintr}). By inverting the functions $a_i(u)$ to express $u_i$ in terms of $a_i$, we can recast 
the dual periods $a_i^D(u)$ as functions of $a_i$ and hence compute the IR couplings $\tau_{ij} = \partial a_i^D/\partial a_j = \partial^2 F/(\partial a_i\partial a_j)$. 
Integrating this formula twice we obtain $F$ as a function of the $a_i$'s, and we can then compare it with the Nekrasov prepotential.

This procedure is in practice rather cumbersome, just because the integrals leading to the dual periods $a_i^D$ are often difficult to compute. Various strategies have been developed to reconstruct the prepotential from the SW curve avoiding the direct computation of the dual periods. A central r\^ole in these strategies is played by relations of the Matone type \cite{Matone:1995rx} which, in the class of theories we study, take the form 
\begin{equation}
\label{defUi}
U_i = q_i\frac{\partial F}{\partial q_i}~,
\end{equation}
where $U_i= \vev{\Tr\Phi_{i}^2}$ is the gauge-invariant modulus of the $i$-th gauge group.
If we know the relation between the parameters $u_i$'s appearing in the curve and the physical moduli $U_i$'s, after inverting the periods $a_i$ as discussed above, we can directly obtain the $U_i$'s as functions of the $a_i$'s and obtain the prepotential $F$ by
integrating once the Matone-like relations with respect to (the logarithm of) the $q_i$'s. 

In recent works \cite{Marshakov:2013lga, Gavrylenko:2013dba}, it has been proposed that the $U_i$'s should be identified with 
the residues of the quadratic differential $x^2(t)$ at the various punctures of the SW curve; 
this identification yields an explicit map from the $u_i$'s (appearing in $x^2(t)$) to the $U_i$'s, thereby 
allowing for an efficient computation of the prepotential.  We show that in the mass-deformed theory, global symmetries of the quiver theory play a crucial role in deriving the precise relation between the residues of $x^2(t)$ and the prepotential of the gauge theory. Having done this, we explicitly compute 
in Sections \ref{secn:Nf4} and \ref{secn:quiver} the periods $a_i$ in the cases $n=1$ and $n=2$, 
and then reconstruct $F$. 
The prepotential we obtain in this way perfectly agrees with the microscopic results, presented in Appendix \ref{Nekrasov}.  

We also perform another consistency check on the SW description of the linear quiver, 
which provides interesting relations between the UV and the IR parameters of the gauge theory.
If we consider the hyperelliptic form of the SW curve, $y^2 = \cP_{2n+2}(t)$,
the classical Thom\ae~formul\ae~\cite{thomae,enolskii-richter} express the cross-ratios of the roots of the polynomial $\cP_{2n+2}$ 
in terms of Riemann $\Theta$-constants at genus $n$. These are constructed in terms of the period matrix $\tau_{ij}$ which
represents the matrix of low-energy effective couplings of the gauge theory and can be computed from the prepotential.
We show that the Thom\ae~formul\ae~do indeed yield the cross-ratios of the roots of $\cP_{2n+2}$, provided we relate 
the parameters $u_i$ appearing in $\cP_{2n+2}$ to the moduli $U_i$ exactly as required by the residue prescription discussed above. Thus, even if we did not assume this prescription, we would be led to it by this analysis.    
We also note that, in the massless case, the cross-ratios of the roots of $\cP_{2n+2}$ are just the UV couplings 
$q_i$, so the Thom\ae~formul\ae~express the UV couplings in terms of the IR couplings 
$\tau_{ij}$ as rational functions of Riemann $\Theta$-constants; in the SU(2) theory with $N_f=4$, these formul\ae~reduce 
to the well-known relation $q = \theta_4(\tau)^4/\theta_2(\tau)^4$ \cite{Grimm:2007tm}.

In Sections \ref{secn:AGT} and \ref{prepotAGT} we then turn to the corrections to the prepotential induced by 
the $\Omega$-deformation. 
For this purpose we use the well-established AGT correspondence \cite{Alday:2009aq} for the conformal $\mathrm{SU(2)}^{n}$ quivers. In particular, we work in the Nekrasov-Shatashvili (NS) limit, where one sets $\epsilon_2=0$, and show that in this limit 
the SW curve and the $\epsilon_1$-deformed SW differential appear naturally in the analysis of a null-vector decoupling equation satisfied by a conformal block with the insertion of a degenerate operator \cite{Alday:2009fs}. 
The deformed SW differential is then used to evaluate the periods in the $\epsilon_1$-deformed theory.  By inverting the expansion, we reconstruct the prepotential order by order in $\epsilon_1$. 
These results precisely match the prepotential calculated via Nekrasov's equivariant localization in Appendix \ref{Nekrasov}. 

Such methods have already been used in deriving the deformed prepotential of the conformal $\mathrm{SU}(2)$ theory with $N_f=4$ flavours and the ${\cal N}=2^*$ theory \cite{Mironov:2009uv}-\nocite{Mironov:2009dv, He:2010xa}\cite{He:2010if}\footnote{For these theories, the instanton contributions have been resummed into almost modular forms in \cite{Marshakov:2010fx}-\nocite{Bonelli:2011na,KashaniPoor:2012wb, Kashani-Poor:2013oza, He:2014yka}\cite{Kashani-Poor:2014mua} by writing the equations in elliptic variables and using recursion relations.}.
Our work extends these computations to the linear quiver case in the presence of masses. As for the undeformed theory, it proves sufficient to evaluate only the $a$-periods of the deformed SW differential in order to obtain the prepotential; thus the problem reduces to the calculation of a new set of integrals over an algebraic curve.

Summarizing, in this paper we investigate how the prepotential of linear SU(2)$^n$ superconformal quivers can be efficiently 
computed using their IR description through a SW curve or, in the $\Omega$-deformed case, through the AGT map. 
These computations require a careful identification between the parameters appearing in these effective descriptions and the ``physical'' parameters of the gauge theory.
This precise understanding of the mapping of parameters is preliminary to further extensions and 
developments, some of which are indicated in the final Section~\ref{secn:concl}. The four appendices we include 
contain technical details and results which are used in the main body of the paper; in particular, Appendix \ref{Nekrasov} contains the first terms in the expansion of the $\Omega$-deformed quiver prepotential obtained by localization techniques.

\section{Seiberg-Witten curves from M-theory}
\label{secn:Mtheory}
In this section we review the M-theory construction \cite{Witten:1997sc} of the Seiberg-Witten (SW) curves for $\mathcal{N}=2$
quiver gauge theories in four dimensions. This construction has been recently discussed in \cite{Bao:2011rc} and we closely follow 
this presentation, adapting it to our purposes. Our main reason for reviewing this material is to fix our conventions and set the stage
for the explicit calculations of the following sections.
  
We begin with a collection of NS5 branes and D4 branes in Type IIA string theory, arranged as shown in Tab.~1.

\begin{table}[ht]
\begin{center}
\begin{tabular}{|c|c|c|c|c|c|c|c|c|c|c||c|}
\hline
\phantom{\Big|}  & $x^0$ & $x^1$ & $x^2$ & $x^3$ & $x^4$ & $x^5$ & $x^6$ & $x^7$ & $x^8$ & $x^9$ & $x^{10}$\\
\hline
NS5 branes & $-$ & $ -$ &$ -$ & $-$ &$-$ & $-$ & $\cdot$ & $\cdot$ & $\cdot$ & $\cdot$ &$\cdot$
\\
\hline
D4 branes  & $-$ &$ -$ &$ -$ & $-$ &$\cdot $ & $\cdot$ & $-$ & $\cdot$ & $\cdot$& $\cdot$  &$-$
\\
\hline
\end{tabular}
\caption{Type IIA brane configuration: $-$ and $\cdot$ denote longitudinal and transverse directions 
respectively; the last column refers to the eleventh dimension after the M-theory uplift. \label{Tab1}}
\end{center}
\end{table}

\noindent
The first four directions $\{x^0,x^1,x^2,x^3\}$ are longitudinal for both kinds of branes and 
span the space-time $\mathbb{R}^{1,3}$ where the quiver gauge theory is defined. 
After compacting the $x^5$ direction on a circle $S^1$ of radius $R_5$, we uplift the system
to M-theory by introducing a compact eleventh coordinate $x^{10}$ with radius $R_{10}$. 
We finally minimize the world-volume of the resulting M5 branes; in this way we obtain the SW curve for 
a 5-dimensional $\mathcal{N}=1$ gauge theory defined in $\mathbb{R}^{1,3}\times S^1$ which takes the form of 
a 2-dimensional surface inside the space parameterized by $\{x^4,x^5, x^6,x^{10} \}$.
To get the curve for the $\mathcal{N}=2$ theory in four dimensions, we first perform a T-duality along $x^5$ and 
then take the limit of small (dual) radius. Thus, in terms of the dual circumference
\begin{equation}
\beta= \frac{2\pi\alpha'}{R_5}~,
\label{beta}
\end{equation}
the 4-dimensional limit corresponds to $\beta\to 0$. Let us now give some details.

\subsection{Brane solution}
\begin{figure}[ht!]
\centering
\includegraphics[width=110mm]{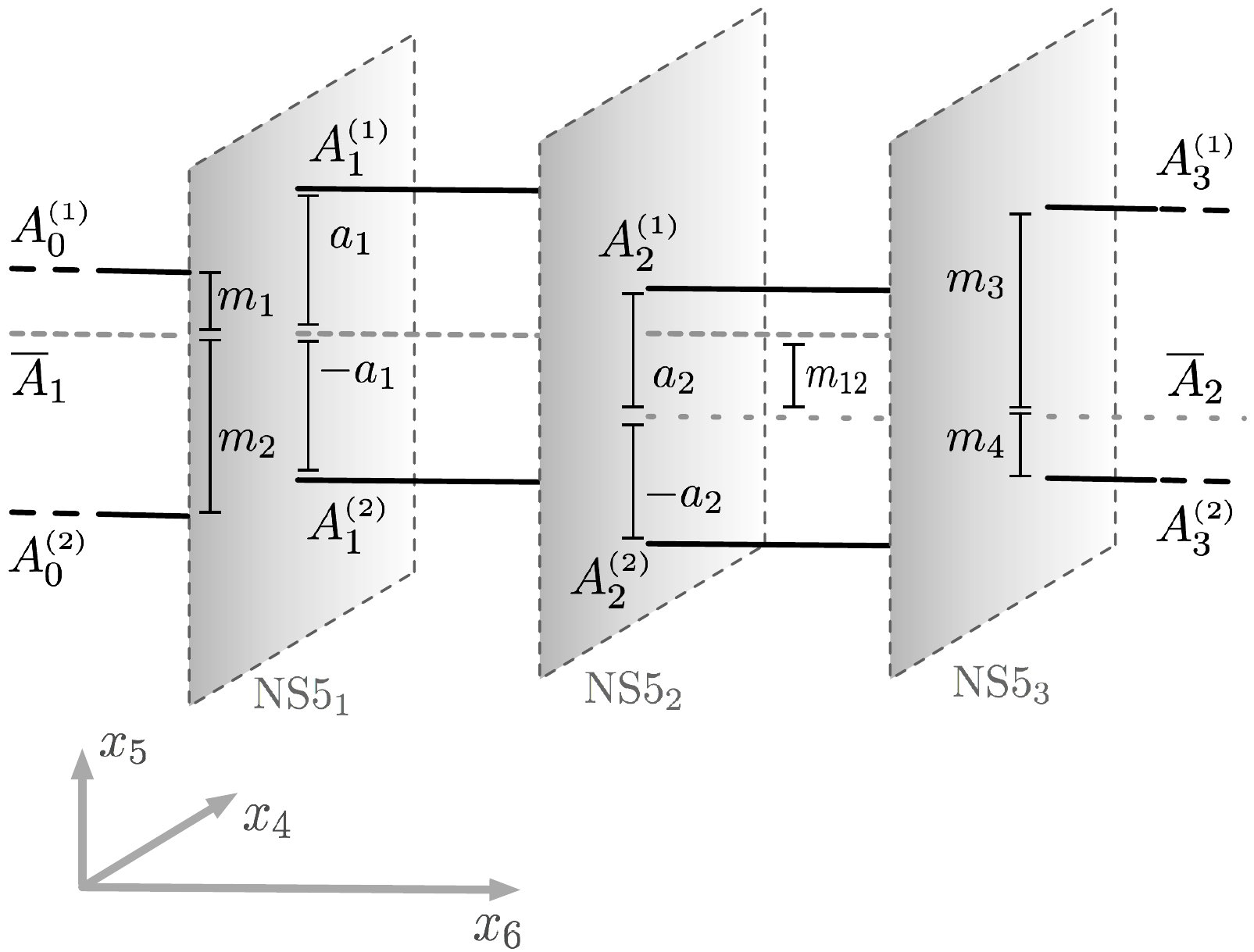}
\caption{NS5 and D4 brane set up for the conformal SU(2)$\,\times\,$SU(2) quiver theory. \label{overflow}}
\end{figure}

We want to engineer a conformal quiver with $n$ SU(2) nodes, two massive fundamental flavors attached to 
the first node, two massive fundamental flavors attached to the last node and one massive bi-fundamental 
hypermultiplet between each pair of nodes%
\footnote{With this field content, the $\beta$-function vanishes for each SU(2) factor; see (\ref{betaa}).}.
To do so we consider a brane system in Type IIA consisting of:
\begin{itemize}
\item $n+1$ NS5 branes separated by finite distances along the $x^6$ direction; we denote them as NS5$_i$ with
$i=1,\ldots,n+1$.
\item Two semi-infinite D4 branes ending on NS5$_1$ and two semi-infinite D4 branes ending on NS5$_{n+1}$; we
call them flavour branes.
\item Two finite D4 branes stretching between NS5$_i$ and NS5$_{i+1}$ for $i=1,\ldots,n$; we will refer to
them as colour branes.
\end{itemize}
In Fig.~\ref{overflow} we have represented, as an example, the set-up for the 2-node quiver theory ($n=2$).

The brane configuration is best described in terms of the complex combinations
\begin{equation}
x^4+\ii \,x^5 \,\equiv \,2\pi\alpha' v\qquad\mbox{and}\qquad x^6+\ii\, x^{10} \,\equiv\, s~,
\label{vs}
\end{equation}
or their exponentials
\begin{equation}
w\,\equiv\,\rme^{\frac{2\pi\alpha' v}{R_5}}=\,\rme^{\beta v} \qquad\mbox{and}\qquad
t\,\equiv\,\rme^{\frac{s}{R_{10}}}
\label{wt}
\end{equation}
which are single-valued under integer shifts of $x^5$ and $x^{10}$ along the respective circumferences. 
Notice that we have introduced factors of $\alpha'$ to
assign to $v$ scaling dimensions of a mass; this choice will be particularly convenient for our later purposes. 
For each NS5$_{i}$ the variable $s_{i}$ 
satisfies the Poisson equation in the $v$-plane
\cite{Witten:1997sc}
\begin{equation}
\nabla^2 \,s_{i}= f_{i} 
\label{poisson}
\end{equation}
where the source term in the right hand side describes the pulling on the $i$-th NS5 brane due to the D4 branes terminating 
on it from each side. For our configuration this is simply a sum of four $\delta$-functions localized at the relevant D4 positions
in the $v$-plane. We denote the positions of the flavour D4 branes on the left by $\big(A^{(1)}_0, A^{(2)}_0\big)$,
those of the flavour D4 branes on the right by $\big(A^{(1)}_{n+1},A^{(2)}_{n+1}\big)$,
and those of the colour D4 branes between NS5$_i$ and NS5$_{i+1}$ by $\big(A^{(1)}_i,A^{(2)}_i\big)$. 
Since $x^5$ is compact, we have to take into account also the infinite images of these brane 
positions and hence the solution of the Poisson equation (\ref{poisson}) is
\begin{equation}
\label{sol}
\begin{aligned}
\frac{s_{i}}{R_{10}} &= \sum_{k=-\infty}^{\infty}\!\Bigg\{\!\log\Big[\beta\big(v-A^{(1)}_{i-1}\big)
-2\pi \ii k\Big]+
\log\Big[\beta\big(v-A^{(2)}_{i-1}\big)-2\pi \ii k\Big]\\
&\quad\quad - \log\Big[\beta\big(v-A^{(1)}_i\big)-2\pi \ii k\Big]-\log\Big[\beta\big(v-A^{(2)}_i\big)
-2\pi \ii k\Big]\Bigg\}+\mbox{const.}
\end{aligned}
\end{equation}
for $i=1,\ldots,n+1$. 
Using the identity
\begin{equation}
\prod_{k=1}^{\infty}\left(1+\frac{x^2}{k^2}\right) = \frac{\sinh \pi x}{\pi x}~,
\end{equation}
and exponentiating the above result, this can be rewritten as
\begin{equation}
\begin{aligned}
\rme^{\frac{s_i}{R_{10}}}
=\,t_{i}~\frac{\sinh\left(\frac{\beta}{2}\big(v-A^{(1)}_{i-1}\big)\right)
\sinh\left(\frac{\beta}{2}\big(v-A^{(2)}_{i-1}\big)\right)}
{\sinh\left(\frac{\beta}{2}\big(v-A^{(1)}_i\big)\right)\sinh\left(\frac{\beta}{2}\big(v-A^{(2)}_i\big)\right)}~,
\end{aligned}
\end{equation}
where $t_{i}$ is related to the integration constant in (\ref{sol}).
The asymptotic positions of the NS5 branes can be obtained by taking the limits $\re v\rightarrow -\infty$ 
({\it i.e.} $w\rightarrow 0$) and $\re v\rightarrow +\infty$ ({\it i.e.} $w\rightarrow \infty$) 
and are given by
\begin{equation}
\label{tasymp}
\begin{aligned}
\rme^{\frac{s_i}{R_{10}}}\,\Big|_{w\to 0} = t_{i}\,\sqrt{\frac{\widetilde{A}^{(1)}_{i-1}\,
\widetilde{A}^{(2)}_{i-1}}
{\widetilde{A}^{(1)}_i\widetilde{A}^{(2)}_i}}\,\equiv\,t^{(0)}_i~,\qquad
\rme^{\frac{s_i}{R_{10}}}\,\Big|_{w\to \infty} = t_{i}\,\sqrt{\frac{\widetilde{A}^{(1)}_i\widetilde{A}^{(2)}_i}
{\widetilde{A}^{(1)}_{i-1}\widetilde{A}^{(2)}_{i-1}}}\,\equiv\,t^{(\infty)}_i~.
\end{aligned}
\end{equation}
Here we have introduced tilded variables according to
\begin{equation}
\widetilde{A}=\rme^{\beta\, A}
\label{tilde}
\end{equation}
for any given $A$. 

As argued in \cite{Witten:1997sc}, the difference in the asymptotic positions of the NS5 branes is related 
to the complexified UV coupling constant of the gauge theory on the color D-branes; more precisely
if we define
\begin{equation}
\tau_i=\frac{\theta_i}{\pi}+\ii\,\frac{8\pi}{g_i^2}
\label{taualpha}
\end{equation}
where $\theta_i$ and $g_i$ are the $\theta$-angle and the Yang-Mills coupling for the SU(2) theory of the $i$-th
node, we have
\begin{equation}
\label{qalpha}
\pi \ii \tau_i \,\sim \,\frac{s_i-s_{i+1}}{R_{10}}~.
\end{equation}
However, since the distance between the NS5 branes is different in the two asymptotic regions $\re v\rightarrow \pm \infty$, there is some ambiguity in this definition. We fix it as in \cite{Witten:1997sc, Bao:2011rc} and use
\begin{equation}
\label{Calfa}
q_i=\rme^{\pi\ii\tau_i}=\frac{t_{i}}{t_{i+1}}
\quad\text{or, equivalently,}\quad t_{i}=t_{n+1}\,\prod_{j=i}^n
q_j~.
\end{equation}
The overall constant $t_{n+1}$ drops out from all equations and can 
be set to 1 without any loss of generality.
In subsequent sections we will confirm that the above identification of the UV coupling constants is fully consistent 
with the Nekrasov multi-instanton calculations.

\subsection{The 5-dimensional curve}
The general SW curve for the 5-dimensional theory defined on the color D4 branes takes the form of a polynomial equation 
\cite{Witten:1997sc} in the $t$ and $w$ variables introduced in (\ref{wt}):
\be
\sum_{p,q} C_{p,q} \,t^p w^q = 0 ~.
\label{curve0}
\ee
Since there are always only two D4 branes in each region and in total we have $(n+1)$ NS5 branes, the polynomial in (\ref{curve0})
must be of degree 2 in $w$ and of degree $(n+1)$ in $t$. Of course, there are two equivalent ways of writing it. One is:
\be
\mathcal{C}_1: \quad w^2  \,Q_2(t)+ w\,Q_1(t)  + Q_0(t) = 0 ~,
\ee
where the $Q$'s are polynomials in $t$ of degree $(n+1)$; the other is:
\be
\mathcal{C}_2: \quad t^{n+1}\,P_{n+1}(w)+t^n\,P_n(w)+\cdots t\,P_1(w) + P_0(w) = 0 ~,
\ee
where each of the $P$'s is a polynomial of degree $2$ in $w$. 
Using the known solutions of $t$ when $w\to 0$ or $w\to \infty$, the form $\mathcal{C}_1$ can be written as
\be
\label{curveC1}
\mathcal{C}_1: \quad w^2\,\prod_{i=1}^{n+1}\big(t-t^{(\infty)}_i\big)
 + w\,Q_2(t) + d'\, \prod_{i=1}^{n+1}\big(t-t^{(0)}_i\big)= 0 ~.
\ee
Having fixed to 1 the coefficient of the highest term $w^2t^{n+1}$, in (\ref{curveC1}) there are 
$(n+3)$ undetermined constants in this equation: $d'$ and the $(n+2)$ coefficients of $Q_2$.
On the other hand, using the fact that when $t\to 0$ and $t\to \infty$ there are two flavour branes at 
$w=(\widetilde{A}^{(1)}_0, \widetilde{A}^{(2)}_0)$ and $w=(\widetilde{A}^{(1)}_{n+1}, 
\widetilde{A}^{(2)}_{n+1})$ respectively, we can write the form $\mathcal{C}_2$ of the curve as
\be
\label{curveC2}
\mathcal{C}_2: \quad t^{n+1}\,\prod_{\alpha=1}^{2}\big(w-\widetilde{A}^{(\alpha)}_{n+1}\big)
+t^n\,P_n(w)+\cdots t\, P_1(w)  + d\,\prod_{\alpha=1}^{2}\big(w-\widetilde{A}^{(\alpha)}_0\big) = 0~.
\ee
Again we have fixed to 1 the coefficient of the highest term $w^2\,t^{n+1}$, but in this form there are $(3n+1)$ undetermined
parameters: $d$ and the three coefficients for each of the $n$ polynomials $P_k$'s.

Equating the two forms (\ref{curveC1}) and (\ref{curveC2}) allows us to find relations that determine some of the curve
parameters: for instance, by comparing the coefficients of $w^2\,t^0$ and $w^0\,t^{n+1}$ in the two expressions we get
\begin{equation}
d= (-1)^{n+1}\prod_{i=1}^{n+1}t^{(\infty)}_{i}~,\qquad d'=\widetilde{A}^{(1)}_{n+1}
\widetilde{A}^{(2)}_{n+1}~.
\label{dd'}
\end{equation}
Similarly, by comparing the coefficients of $w\,t^0$ and $w\,t^{n+1}$ we find that the undetermined 
polynomial $Q_2(t)$ in (\ref{curveC1}) takes the form
\be
\label{Q2pol}
Q_2(t) = -\big(\widetilde{A}^{(1)}_{n+1}+\widetilde{A}^{(2)}_{n+1}\big)\,t^{n+1} + \sum_{k=1}^n c_k\,
t^k  + (-1)^n \big(\widetilde{A}^{(1)}_0+\widetilde{A}^{(2)}_0\big) 
\prod_{i=1}^{n+1}t^{(\infty)}_{i} ~.
\ee
Proceeding in a similar way one can fix the coefficients of $w^2$ and $w^0$ in the $n$ quadratic polynomials 
$P_i$'s of (\ref{curveC2}).
In the end, all but $n$ parameters in the SW curve are fixed; the 
$n$ free coefficients that remain parametrize the Coulomb branch of the $\mathrm{SU}(2)^n$ quiver gauge theory. 
One subtlety is that the constant terms in (\ref{curveC1}) and (\ref{curveC2}) match only if the following identity is satisfied:
\be
\widetilde{A}^{(1)}_0\,\widetilde{A}^{(2)}_0\, \prod_{i=1}^{n+1}
t^{(\infty)}_{i}= \widetilde{A}^{(1)}_{n+1}\,\widetilde{A}^{(2)}_{n+1}\, 
\prod_{i=1}^{n+1}t^{(0)}_{i}~.
\ee
Using the explicit expressions (\ref{tasymp}) for the asymptotic positions of the NS5 branes, we can check that this 
is identically satisfied and both sides are equal to 
$(\widetilde{A}^{(1)}_0\,\widetilde{A}^{(2)}_0\widetilde{A}^{(1)}_{n+1}\,\widetilde{A}^{(2)}_{n+1})^{1/2}$. 
This shows that indeed the two forms $\mathcal{C}_1$ and $\mathcal{C}_2$ of the SW curve are fully equivalent.

\subsection{The 4-dimensional curve}

We now dimensionally reduce to four dimensions by first performing a T-duality and then taking the 
limit $\beta\rightarrow 0$. To find explicit expressions it necessary to introduce the physical parameters 
of the 4-dimensional theory and rewrite the geometric positions of the various branes in terms of these. 
In order to do this, for each pair of colour D4 branes we define the center of mass and
relative positions in the $v$-plane according to
\begin{equation}
\label{aAalpha}
A^{(1)}_i= a_i+\bar{A}_i~,\qquad
A^{(2)}_i= -a_i+\bar{A}_i
\end{equation}
for $i=1,\ldots,n$. The relative position $a_i$ is identified with the vacuum expectation value of the adjoint scalar field
$\Phi_i$ of the $i$-th SU(2) factor in the quiver theory. Furthermore we remove the global U(1) factor by requiring 
\begin{equation}
\bar{A}_1+\cdots+\bar{A}_n=0~,
\label{U1}
\end{equation}
and identify the relative positions of the centers of mass with the physical masses of the bi-fundamental hypermultiplets, {\it i.e.}
\begin{equation}
m_{i,i+1} = \bar{A}_{i}-\bar{A}_{i+1}
\label{massbif}
\end{equation}
for $i=1,\ldots,n-1$. Finally, the physical masses of the fundamental hypermultiplets attached to the first and the
last NS5 branes are related to the positions of the flavour D4 branes measured with respect to the first 
and last center of mass in the $v$-plane, namely
\begin{equation}
m_1=A^{(1)}_0-\bar{A}_1~,~~~m_2=A^{(2)}_0-\bar{A}_1~,~~~m_3=A^{(1)}_{n+1}-\bar{A}_{n}~,~~~
m_4=A^{(2)}_{n+1}-\bar{A}_{n}~.
\label{massfun}
\end{equation} 
All this is displayed in Fig.~\ref{overflow} for the case $n=2$.

Given this set-up, it is rather straightforward to obtain the 4-dimensional SW curve. 
However, in general it is not so simple to write explicit expressions in terms of the relevant physical parameters. 
Thus, we discuss in detail the following three cases:
\begin{itemize}
\item the conformal $\mathrm{SU}(2)^n$ quiver with massless hypermultiplets;
\item the SU(2) theory with $N_f=4$ massive fundamental flavours;
\item the SU(2)$\,\times\,$SU(2) quiver theory with generically massive hypermultiplets. 
\end{itemize}

\subsubsection*{$\bullet$ The conformal $\mathrm{SU}(2)^n$ quiver}
When all matter hypermultiplets are massless the curve equation drastically simplifies. Indeed, 
all stacks of colour branes have the same center of mass positions, so that (\ref{U1}) implies
that $\bar{A}_{i}=0$ for $i=1,\ldots,n$. Moreover, setting to zero the four fundamental masses 
implies that $A^{(1)}_{0}=A^{(2)}_{0}=A^{(1)}_{n+1}=A^{(2)}_{n+1}=0$.
Using this, we have
\begin{equation}
t^{(0)}_i=t^{(\infty)}_i=t_{i}
\end{equation}
where the constants $t_{i}$ are defined  in terms of the gauge couplings $q_i$ according to (\ref{Calfa}).
The 5-dimensional curve (\ref{curveC1}) then becomes
\begin{equation}
w^2\,\prod_{i=1}^{n+1}(t-t_{i})-2\,w\Big(
 t^{n+1}-\frac{1}{2} \sum_{k=1}^n c_k\,
t^k  - (-1)^{n} \prod_{i=1}^{n+1}t_{i}\Big)+
\prod_{i=1}^{n+1}(t-t_{i})= 0~.
\end{equation}
We now take the 4-dimensional limit $\beta\to 0$ after writing 
$c_{k}=c_{k0}+c_{k1}\beta+c_{k2}\beta^2+\cdots$ and $w=\rme^{\beta v}$. The $\mathcal{O}(\beta^0)$ and 
$\mathcal{O}(\beta^1)$ terms
yield algebraic constraints for $c_{k0}$ and $c_{k1}$ that can be easily solved. Instead, the $\mathcal{O}(\beta^2)$ term leads to
the 4-dimensional SW curve. Writing $v=x\, t$ and setting $t_{n+1}=1$, the curve becomes
\begin{equation}
x^2(t)=\frac{\mathcal{P}_{n-1}(t)}{t\,(t-t_{1})\cdots(t-t_{n})(t-1)}
\label{curvequiver}
\end{equation}
where $\mathcal{P}_{n-1}(t)$ is a polynomial of degree $n-1$, whose $n$ coefficients parametrize the Coulomb branch of the
$\mathrm{SU}(2)^n$ theory.
This is precisely the form of the SW curve discussed in \cite{Gaiotto:2009we}.

When the matter multiplets are massive, things become more involved. 
While it is always quite straightforward to write formal expressions,
it is not always immediate to identify the meaning of the various coefficients in terms of the physical parameters of the
gauge theory. Thus to avoid clumsy general expressions we discuss in detail the cases with $n=1$ and $n=2$.

\subsubsection*{$\bullet$ The SU(2) theory with $N_f=4$}
When $n=1$ the formul\ae~(\ref{aAalpha})-(\ref{massfun}) lead to
\begin{equation}
\label{ASU2}
A_0^{(1)}=m_1~,\quad
A_0^{(2)}=m_2~,\quad
A_1^{(1)}=a~,\quad
A_1^{(2)}=-a~,\quad
A_2^{(1)}=m_3~,\quad
A_2^{(2)}=m_4~,\quad
\end{equation}
where $a$ is the vacuum expectation of the adjoint scalar field $\Phi$.
Then the curve (\ref{curveC1}) becomes
\begin{equation}
w^2\big(t-t_1^{(\infty)}\big)\!\big(t-t_2^{(\infty)}\big)
-w\Big[\big(\widetilde{m}_3+\widetilde{m}_4\big)t^2-c\,t+\big(\widetilde{m}_1+\widetilde{m}_2\big)
t_1^{(\infty)}t_2^{(\infty)}\Big]+
\widetilde{m}_3 \widetilde{m}_4\big(t-t_1^{(0)}\big)\!\big(t-t_2^{(0)}\big)=0
\label{curveSU24}
\end{equation}
where, according to (\ref{tasymp}) and (\ref{Calfa}),
\begin{equation}
t_1^{(0)}=q\,\sqrt{\widetilde{m}_1\widetilde{m}_2}~,\quad
t_1^{(\infty)}=\frac{q}{\sqrt{\widetilde{m}_1\widetilde{m}_2}}~,\quad
t_2^{(0)}=\frac{1}{\sqrt{\widetilde{m}_3\widetilde{m}_4}}~,\quad
t_2^{(\infty)}=\sqrt{\widetilde{m}_3\widetilde{m}_4}
\label{t12}
\end{equation}
and we are using the tilded variables $\widetilde m_i$ according to the notation introduced in \eq{tilde}. 
To obtain the 4-dimensional curve we expand $w$, $c$ and all tilded variables
in powers of $\beta$. The $\mathcal{O}(\beta^0)$  and $\mathcal{O}(\beta^1)$ terms can be set to zero by suitably choosing the 
first two coefficients in the expansion of $c$, while the $\mathcal{O}(\beta^2)$ term yields the SW curve for
the SU(2) $N_f=4$ theory. The result is
\cite{Eguchi:2009gf,Eguchi:2010rf,Bao:2011rc} 
\begin{equation}
\label{curveSU24v}
v^2(t-q)(t-1)-v\Big[(m_3+m_4)t^2-q\sum_{f=1}^4 m_f\,t+q(m_1+m_2)\Big]
+m_3m_4 \,t^2+u \,t+ q\,m_1m_2=0~.
\end{equation}
Here we have absorbed all terms linear in $t$ and independent of $v$ by redefining $c$ into a new parameter $u$.
A simple dimensional analysis reveals that $u$ has dimensions of $(\mathrm{mass})^2$.
As pointed out in \cite{Gaiotto:2009we} it is a bit arbitrary to define the origin for this $u$ parameter when masses are present.
Here we fix such arbitrariness by requiring 
\begin{equation}
u\,\big|_{q\to 0}=\,a^2~.
\label{uto0}
\end{equation}
Shifting away the linear term in $v$ in (\ref{curveSU24v}) and writing $v=x\,t$, we get
\cite{Eguchi:2009gf,Eguchi:2010rf,Bao:2011rc}
\begin{equation}
x^2(t)=\frac{\mathcal{P}_4(t)}{t^2(t-q)^2(t-1)^2}
\label{su2curve}
\end{equation}
where $\mathcal{P}_4(t)$ is a fourth-order polynomial in $t$ of the form
\begin{equation}
\mathcal{P}_4(t)=-u\,t\,(t-q)(t-1)+\mathcal{M}_4(t)
\end{equation}
where we have collected in $\mathcal{M}_4(t)$ all terms that depend on the masses. The explicit expression of this polynomial
is given in (\ref{P4nf4}). 
Using it and choosing a specific determination for the square-root, one easily finds
\begin{equation}
\begin{aligned}
\mathrm{Res}_{\,t=0}\left(x(t)\right)&=\frac{m_1-m_2}{2}~,\qquad
\mathrm{Res}_{\,t=q}\left(x(t)\right)=\frac{m_1+m_2}{2}~,\\
\mathrm{Res}_{\,t=1}\left(x(t)\right)&=\frac{m_3+m_4}{2}~,\qquad
\mathrm{Res}_{\,t=\infty}\left(x(t)\right)=\frac{m_4-m_3}{2}~.
\end{aligned}
\label{residues}
\end{equation}

\subsubsection*{$\bullet$ The SU(2)$\,\times\,$SU(2) quiver theory}
For a 2-node quiver (see Fig.~\ref{overflow}), the formul\ae~(\ref{aAalpha})-(\ref{massfun}) read
\begin{equation}
\begin{aligned}
A_1^{(0)}&=m_1+\frac{m_{12}}{2}~,~\, A_2^{(0)}=m_2+\frac{m_{12}}{2}~,~\, 
A_1^{(1)}=a_1+\frac{m_{12}}{2}~,~\, A_2^{(1)}=-a_1+\frac{m_{12}}{2}~,\\
A_1^{(2)}&=a_2-\frac{m_{12}}{2}~,~\, A_2^{(2)}=-a_2-\frac{m_{12}}{2}~,~\, 
A_1^{(3)}=m_3-\frac{m_{12}}{2}~,~\, A_2^{(3)}=m_4-\frac{m_{12}}{2}
\end{aligned}
\end{equation}
where $a_1$ and $a_2$ are the vacuum expectation values of the adjoint scalars $\Phi_1$ and $\Phi_2$ of the two SU(2) factors. 
With this configuration the 5-dimensional curve (\ref{curveC1}) becomes
\begin{equation}
\begin{aligned}
w^2\prod_{i=1}^{3}\big(t-t^{(\infty)}_i\big)&
- w\,\Big(\frac{\widetilde{m}_3+\widetilde{m}_4}{\sqrt{\widetilde{m}_{12}}}\,t^3-c_2\,t^2-c_1\,t
\\&~~
-\sqrt{\widetilde{m}_{12}}\big(\widetilde{m}_1+\widetilde{m}_2\big)\prod_{i=1}^{3}t^{(\infty)}_i\Big)
+ \frac{\widetilde{m}_3\widetilde{m}_4}{\widetilde{m}_{12}} 
 \,\prod_{i=1}^{3}\big(t-t^{(0)}_i\big)= 0 
 \end{aligned}
\end{equation}
where the asymptotic values are
\begin{equation}
\begin{aligned}
t_1^{(0)}&=t_1\sqrt{\widetilde{m}_1\widetilde{m}_2}~,\quad~
t_2^{(0)}=t_2\,\widetilde{m}_{12}~,\quad~
t_3^{(0)}=\frac{1}{\sqrt{\widetilde{m}_3\widetilde{m}_4}}~,\\
t_1^{(\infty)}&=\frac{t_1}{\sqrt{\widetilde{m}_1\widetilde{m}_2}}~,\qquad~
t_2^{(\infty)}=\frac{t_2}{\widetilde{m}_{12}}~,\quad~
t_3^{(\infty)}=\sqrt{\widetilde{m}_3\widetilde{m}_4}
\end{aligned}
\end{equation}
with
\begin{equation}
t_1=q_1q_2~,\qquad t_2=q_2~.
\end{equation}
We now take the 4-dimensional limit $\beta\to 0$, proceeding as in the previous examples. The resulting SW curve is
\begin{eqnarray}
&&v^2(t-t_1)(t-t_2)(t-1)\nonumber\\
&&- v\,\Bigg[(m_3+m_4-m_{12})\,t^3-\bigg(\Big(\sum_{f=1}^4m_f-m_{12}\Big)\,t_1
+(m_3+m_4+m_{12})\,t_2-m_{12}\bigg) t^2\nonumber\\
&&~+\bigg((m_1+m_2-m_{12})\,t_1+m_{12}\,t_2+
\Big(\sum_{f=1}^4m_f+m_{12}\Big)\,t_1\,t_2\bigg)\,t
-(m_1+m_2+m_{12})\,t_1\,t_2\Bigg]
\nonumber\\
&&+\Bigg[\Big(m_3-\frac{m_{12}}{2}\Big)\Big(m_4-\frac{m_{12}}{2}\Big)\,t^3-\Big(\frac{m_{12}^2}{4}-u_2\Big)\,t^2
+\Big(\frac{m_{12}^2}{4}-u_1\Big)t_2\,t\nonumber\\
&&~-\Big(m_1+\frac{m_{12}}{2}\Big)\Big(m_2+\frac{m_{12}}{2}\Big)\,t_1\,t_2
 \Bigg]= 0~.
 \label{cmass}
 \end{eqnarray}
Here we have exploited the freedom to redefine the arbitrary coefficients $c_1$ and $c_2$ 
into the parameters $u_1$ and $u_2$ for which we require the following classical limit
\begin{equation}
u_1\,\big|_{q_1,q_2\to 0}=\,a_1^2\quad\quad\mbox{and}\qquad u_2\,\big|_{q_1,q_2\to 0}=\,a_2^2~.
\label{u12to0}
\end{equation}
 In Section~\ref{secn:quiver} we will confirm the validity of this requirement.
 
In order to put the curve in a more convenient form, we shift away the linear term in $v$ in (\ref{cmass}) and
then write $v=x\,t$, obtaining
\begin{equation}
\label{cmassg}
x^2(t) =  
\frac{\mathcal{P}_6(t)}{t^2\,(t-t_{1})^2(t-t_{2})^2(t-1)^2}~,
\end{equation}
where $\mathcal{P}_6(t)$ is a polynomial of degree six in $t$ of the form
\begin{equation}
\label{P60}
\mathcal{P}_6(t)=-t\,(u_2\,t-t_2\,u_1)(t-t_{1})(t-t_{2})(t-1)+\mathcal{M}_6(t)
\end{equation}
with $\mathcal{M}_6(t)$ containing all mass-dependent terms. The explicit expression of this polynomial is given in
(\ref{P6}). Using it we find
\begin{eqnarray}
\mathrm{Res}_{\,t=0}\left(x(t)\right)&=&\frac{m_1-m_2}{2}~,\quad
\mathrm{Res}_{\,t=t_1}\left(x(t)\right)=\frac{m_1+m_2}{2}~,\quad
\mathrm{Res}_{\,t=t_2}\left(x(t)\right)=m_{12}~,\nonumber\\
&&\!\!\!\!\!\!\!\!\!\!\!\mathrm{Res}_{\,t=1}\left(x(t)\right)=\frac{m_3+m_4}{2}~,\quad
\mathrm{Res}_{\,t=\infty}\left(x(t)\right)=\frac{m_4-m_3}{2}~.\label{residues2}
\end{eqnarray}

\subsection{From the 4-dimensional curve to the prepotential}

The spectral curve  (\ref{cmassg}) encodes all relevant information about the effective quiver gauge theory 
through the SW differential 
\begin{equation}
\label{lambdadef}
\lambda = x(t) dt~.
\end{equation}
If we differentiate $\lambda$ with respect to $u_1$ and $u_2$, we get 
(up to normalizations which are irrelevant for our present purposes)
\begin{equation}
\label{omeamass}
\frac{\partial\lambda}{\partial u_1} \simeq\frac{dt}{y}~,\qquad
\frac{\partial\lambda}{\partial u_2} \simeq \frac{t\,dt}{y}~,
\end{equation}
where
\begin{equation}
\label{g2eqmass}
y^2 = \mathcal{P}_6(t)~.
\end{equation}
This is the standard equation defining a genus-2 Riemann surface. Such a surface admits a canonical symplectic basis with two pairs of 1-cycles $(\alpha_1, \alpha_2)$ and $(\beta_1, \beta_2)$ whose intersection matrix 
is $\alpha_i\cap \alpha_j =\beta_i\cap \beta_j =\delta_{ij}$. The periods of the SW differential $\lambda$
along these cycles represent the quantities $a_i$ and $a_i^D$ in the effective gauge theory, namely
\begin{equation}
\label{aad}
a_i = \frac{1}{2\pi\ii}\oint_{\alpha_i} \lambda ~,\qquad
a_i^D = \frac{1}{2\pi\ii}\oint_{\beta_i} \lambda ~.
\end{equation}
Through these relations, $a_i$ and $a_i^D$ are determined as functions of the $u_i$'s 
(and, of course, of the UV couplings $q_i$ and of the mass parameters). Inverting these relations, one 
can express the $u_i$'s in terms of the $a_i$'s and, substituting them into the dual periods, obtain $a_i^D(a)$. 
Since 
\begin{equation}
\label{aDF}
a_i^D(a) = \frac{\partial F}{\partial a_i}~,
\end{equation}
one can reconstruct in this way the prepotential $F$ (up to $a$-independent terms). 
By comparing this prepotential with the one obtained from the multi-instanton calculus via localization one can therefore test the validity of the proposed form of the SW curve.

However, an alternative and more efficient approach has been presented in
\cite{Marshakov:2013lga, Gavrylenko:2013dba} in which the difficult computations of the dual periods $a_i^D$ 
are avoided and the effective prepotential is directly put in relation with the residues of the quadratic differential $x^2(t) dt^2$ in the following way
\be
\label{marsrel}
\mathrm{Res}_{\,t=t_i}\left(x^2(t)\right)
= \frac{\p {\widetilde F}}{\p t_i}~.
\ee
As we will show in more detail below, assuming this relation and just computing the $\alpha$-periods of the SW differential we 
can readily reconstruct $\widetilde F$ from the spectral curve and check that it coincides with the 
effective prepotential $F$ computed via localization up to mass-dependent but $a$-independent
shifts (so that $\widetilde F$ and $F$ encode the same effective gauge couplings); 
the expression of these shifts is however rather interesting, and we will comment on this in the next sections.

\section{The SU(2) theory with $N_f=4$}
\label{secn:Nf4}
We show how to derive the effective prepotential for the SU(2) $N_f=4$ theory starting
from the curve (\ref{su2curve}) and the residue formula (\ref{marsrel})
which in this case reads
\begin{equation}
\mathrm{Res}_{\,t=q}\left(x^2(t)\right)=\frac{\partial\widetilde{F}}{\partial q}~.
\label{marshnf4}
\end{equation}
In doing this we do not only provide a generalization of the results presented in \cite{Gavrylenko:2013dba},
but also set the stage for the discussion of the quiver theory in the next section.
 
Using the curve (\ref{su2curve}) and the explicit expression of the polynomial $\mathcal{P}_4$
reported in (\ref{P4nf4}), the above residue formula leads to
\begin{equation}
q(1-q)\frac{\partial\widetilde{F}}{\partial q}=u-\frac{1-q}{2}\left(m_1^2+m_2^2\right)+\frac{q}{2}	\left(m_1+m_2\right)\left(m_3+m_4\right)+q\left(m_1m_2+m_3m_4\right)~.
\label{ftilde1}
\end{equation}
Combining this with the residues (\ref{residues}) amounts to rewrite the SW curve (\ref{su2curve}) as
\begin{equation}
\begin{aligned}
x^2(t)=&\,
\frac{(m_1-m_2)^2}{4t^2}+\frac{(m_1+m_2)^2}{4(t-q)^2}+\frac{(m_3+m_4)^2}{4(t-1)^2}
-\frac{m_1^2+m_2^2+2m_3m_4}{2t(t-1)}\\
&
+\frac{q(q-1)}{t(t-q)(t-1)}\,\frac{\p\widetilde{F}}{\p q}~.
\end{aligned}
\label{curveSU2f}
\end{equation} 
We now clarify the meaning of $\widetilde{F}$.
Imposing in (\ref{ftilde1}) the boundary value (\ref{uto0}) for $u$, we easily find
\begin{equation}
q\frac{\partial\widetilde{F}}{\partial q}\,\Big|_{q\to 0}=\,a^2-\frac{1}{2}\left(m_1^2+m_2^2\right)~,
\end{equation}
from which we deduce that $\widetilde{F}$ cannot be directly identified with the effective gauge theory 
prepotential, whose classical term is in fact
$F_{\mathrm{cl}}= a^2 \log q$. Therefore, to ensure the proper classical limit we shift $\widetilde F$ according to
\begin{equation}
\widetilde{F}= \widehat{F}-\frac{1}{2}\left(m_1^2+m_2^2\right)\log q~,
\label{Fhat}
\end{equation}
and rewrite (\ref{ftilde1}) as
\begin{equation}
q(1-q)\frac{\partial\widehat{F}}{\partial q}=u+\frac{q}{2}	\left(m_1+m_2\right)\left(m_3+m_4\right)
+q\left(m_1m_2+m_3m_4\right)~.
\label{fhat1}
\end{equation}
The function $\widehat{F}$ has the correct classical limit, but it is not yet the gauge theory prepotential since it is determined by an 
equation in which the four masses do not appear on equal footing. 
There are two independent ways to remedy this and restore complete symmetry among the
flavors, namely by redefining $\widehat{F}$ as%
\footnote{All other possibilities can be seen as linear combinations of these two. It is interesting to observe that the shifts in
(\ref{fhatI}) and (\ref{fhatII}) are directly related to the so-called U(1) dressing factors used
in the AGT correspondence \cite{Alday:2009aq}.} 
\begin{eqnarray}
\mathrm{I)}&:&\quad\widehat{F}=F_{\mathrm{I}}+\frac{1}{2}\log(1-q)\left(m_1+m_2\right)\left(m_3+m_4\right)~,
\label{Fhat1}\\
\mathrm{II)}&:&\quad\widehat{F}=F_{\mathrm{II}}-\frac{1}{2}\log(1-q)\left(m_1m_2+m_3m_4\right)~.
\label{Fhat2}
\end{eqnarray}
 In this way, from (\ref{fhat1}) we get
 \begin{eqnarray}
\mathrm{I)}&:&\quad q(1-q)\frac{\partial{F_{\mathrm{I}}}}{\partial q}\equiv (1-q)\,U_{\mathrm{I}}
=u+ q\sum_{f<f'}m_fm_f'~,\label{fhatI}\\
\mathrm{II)}&:&\quad q(1-q)\frac{\partial{F_{\mathrm{II}}}}{\partial q}\equiv (1-q)\,U_{\mathrm{II}}
=u+ \frac{q}{2}\sum_{f<f'}m_fm_f'~.\label{fhatII}
\end{eqnarray}
The minor difference in the numerical coefficient in front of the mass terms in these two equations is, actually, quite significant.
In fact, as we will see, $F_{\mathrm{I}}$ is the Nekrasov prepotential for the SU(2) $N_f=4$ theory, while $F_{\mathrm{II}}$ is 
the SO(8) invariant prepotential that can be derived from the SW curve of \cite{Seiberg:1994aj} expressed 
in terms of the UV coupling $q$.

\begin{figure}
\begin{center}
\def\svgwidth{14cm}
\begingroup%
  \makeatletter%
  \providecommand\rotatebox[2]{#2}%
  \ifx\svgwidth\undefined%
    \setlength{\unitlength}{656.01045375bp}%
    \ifx\svgscale\undefined%
      \relax%
    \else%
      \setlength{\unitlength}{\unitlength * \real{\svgscale}}%
    \fi%
  \else%
    \setlength{\unitlength}{\svgwidth}%
  \fi%
  \global\let\svgwidth\undefined%
  \global\let\svgscale\undefined%
  \makeatother%
  \begin{picture}(1,0.40244627)%
    \put(0,0){\includegraphics[width=\unitlength]{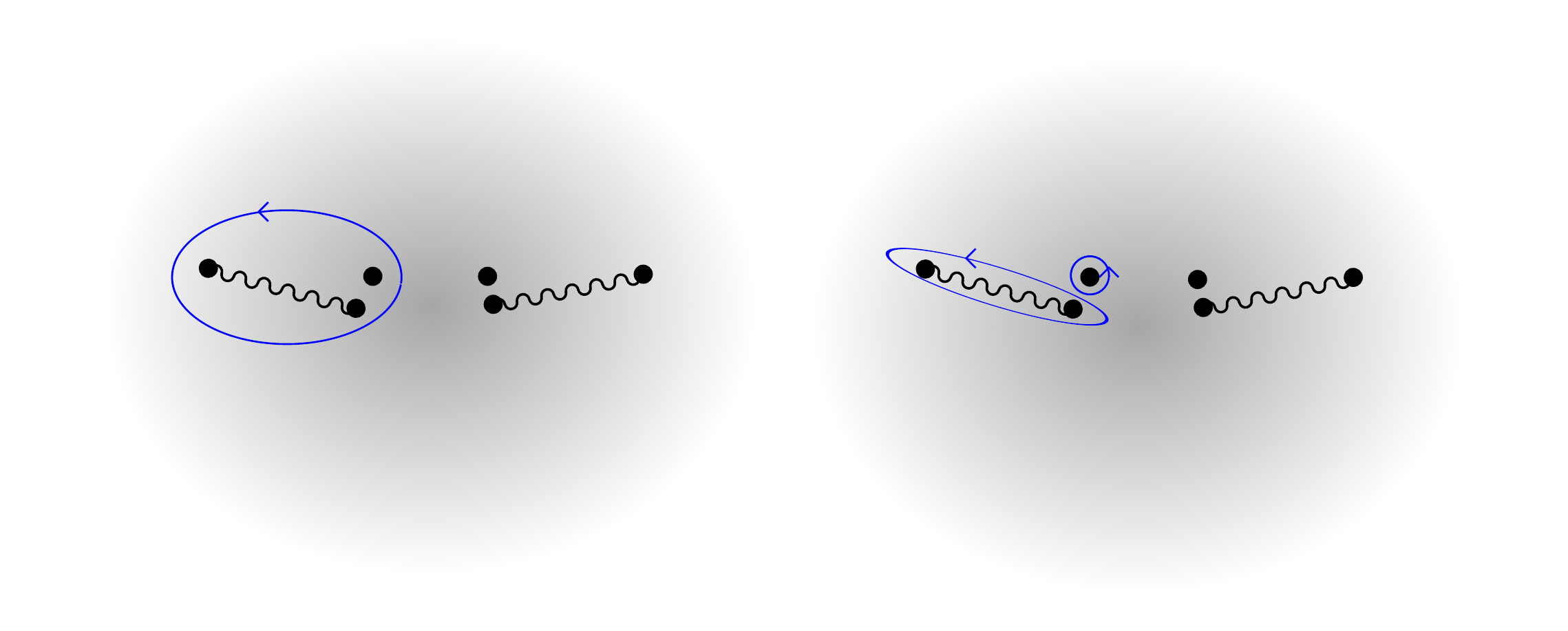}}%
    \put(0.12195561,0.20122314){\makebox(0,0)[lb]{\smash{$0$}}}%
    \put(0.2177183,0.18523397){\makebox(0,0)[lb]{\smash{$e_2$}}}%
    \put(0.13415053,0.28049017){\makebox(0,0)[lb]{\smash{$\alpha$}}}%
    \put(0.22539656,0.24163919){\makebox(0,0)[lb]{\smash{$q$}}}%
    \put(0.57928876,0.20073775){\makebox(0,0)[lb]{\smash{$0$}}}%
    \put(0.67505147,0.18474858){\makebox(0,0)[lb]{\smash{$e_2$}}}%
    \put(0.59148369,0.28000478){\makebox(0,0)[lb]{\smash{$\alpha$}}}%
    \put(0.68272972,0.2411538){\makebox(0,0)[lb]{\smash{$q$}}}%
    \put(0.47519787,0.22596318){\makebox(0,0)[lb]{\smash{$\equiv$}}}%
    \put(0.31097698,0.18293075){\makebox(0,0)[lb]{\smash{$e_3$}}}%
    \put(0.40243894,0.23780792){\makebox(0,0)[lb]{\smash{$\infty$}}}%
    \put(0.29878206,0.23780792){\makebox(0,0)[lb]{\smash{$1$}}}%
    \put(0.76385132,0.18091617){\makebox(0,0)[lb]{\smash{$e_3$}}}%
    \put(0.85531327,0.23579334){\makebox(0,0)[lb]{\smash{$\infty$}}}%
    \put(0.75165639,0.23579334){\makebox(0,0)[lb]{\smash{$1$}}}%
  \end{picture}%
\endgroup%
\end{center}
\caption{Branch-cuts and singularities of the $\alpha$-period of SW differential $\lambda$ of the SU(2) $N_f=4$ theory.}
\label{fig:lambda-a-cycles-nf4}
\end{figure}
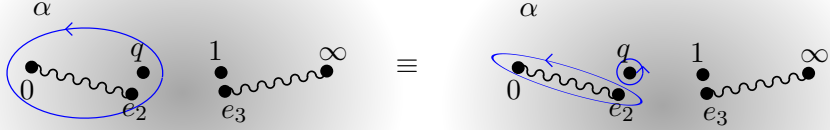

To verify this statement in an explicit way, we take
\begin{equation}
m_1=m_2=m~,\quad m_3=m_4=M~.
\label{masses}
\end{equation}
This is a simple choice of masses that allows us to exhibit all non-trivial features of the calculation.
With these masses the curve (\ref{su2curve}) becomes
\begin{equation}
x^2(t)=\frac{\mathcal{P}_2(t)}{t(t-1)^2(t-q)^2}
\label{curvemM}
\end{equation}
where
\begin{eqnarray}
\mathcal{P}_2(t)\!\!&=&\!\!-C t^2+\Big(u(1+q)-q(m-M)^2+q^2(m+M)^2\Big)t
-q\Big(u-\left(1-q\right)m^2+2qmM\Big)\nonumber\\
&=&C(e_2-t)(t-e_3)\phantom{\Big |}
\end{eqnarray}
with
\begin{equation}
C=u+2qmM-M^2(1-q)~.
\label{Ccoeff}
\end{equation}
The expressions of the two roots $e_2$ and $e_3$ can be easily obtained by solving the quadratic equation $\mathcal{P}_2(t)=0$; 
in the 1-instanton approximation we find%
\footnote{Here and in the following, for brevity we explicitly exhibit the results only up to one or two instantons, but we have checked 
that everything works also for higher instanton numbers.}
\begin{equation}
\begin{aligned}
e_2&=q\Big(1-\frac{m^2}{u}+q\frac{m^2\left(u^2+M^2 u+2mMu-m^2M^2\right)}{u^3}+\ldots\Big)~,\\
e_3&=1+\frac{M^2}{u-M^2}+q\frac{M^2\left(m^2M^2-m^2u-2mMu-u^2\right)}{u(u-M^2)}+\ldots
\end{aligned}
\label{e2e3}
\end{equation}
The SW differential associated to the spectral curve (\ref{curvemM}) is
\begin{equation}
\lambda= x(t) dt = \sqrt{\frac{(e_2-t)(t-e_3)}{t}}\,\frac{\sqrt{C}\, dt}{(1-t)(t-q)}~;
\label{lambdapp}
\end{equation}
it possess four branch points at $t=0,\,e_2,\,e_3$ and $\infty$ and two simple poles at $t=q$ and $1$. This 
singularity structure is shown in Fig.~\ref{fig:lambda-a-cycles-nf4}. The cross-ratio of the four
branch points is
\begin{eqnarray}
\zeta &=&\frac{e_2}{e_3}= q\left(1-\frac{(m^2+M^2)u-m^2M^2}{u^2}\right)
\label{zeta1}\\
&&~+
q^2\left(\frac{(m^2+M^2)u^3+2mM(m^2+M^2)u^2-2m^2M^2(m+M)^2u+2m^4M^4}{u^4}\right)+\ldots
\nonumber
\end{eqnarray}
In the massless limit, note that the cross ratio reduces to the Nekrasov counting parameter $q$, as expected. 
As always, we identify the $\alpha$-period of the SW differential with the vacuum expectation value $a$, namely
\begin{equation}
a=\frac{1}{2\pi\ii}\oint_\alpha\lambda = ~
\mathrm{Res}_{\,t=q}\left(\lambda\right)+
\frac{\sqrt{C}}{\pi}\int_0^{e_2}\sqrt{\frac{e_2-t}{t}}\,\frac{\sqrt{e_3-t}}{(1-t)(q-t)}\,dt~.
\label{anf4}
\end{equation}

It is important to stress that the $\alpha$-cycle corresponds to a closed contour encircling both the branch cut from $0$ to $e_2$ 
and the simple pole of $\lambda$ at $t=q$, see Fig.~\ref{fig:lambda-a-cycles-nf4}. With this prescription, the $\alpha$-cycle has a smooth limit when the masses are set to zero.   
This explains the two terms on the right hand side of (\ref{anf4}): the residue over the pole in $t=q$, which in view 
of (\ref{residues}) is simply $m$, and the integral over the branch cut. 
This integral is explicitly evaluated in Appendix~\ref{secn:integral}
(see in particular (\ref{afinapp})); in the final result
the mass term coming from the residue is canceled and we are left with
\begin{equation}
\begin{aligned}
a &= \frac{\sqrt{C(e_3-q)}}{1-q}+\frac{\sqrt{C}}{1-q}\sum_{n,\ell=0}^\infty
(-1)^\ell\binom{1/2}{n+1}\binom{1/2}{n+\ell+1}\frac{e_2^{n+1}\,q^\ell}{e_3^{n+\ell+1/2}}\\
&~~~~-\frac{\sqrt{C}}{1-q}\sum_{n=0}^\infty\sum_{\ell=0}^n
(-1)^{(n+\ell)}\binom{1/2}{n+1}\binom{1/2}{\ell}\frac{e_2^{n+1}}{e_3^{\ell-1/2}}~.
\end{aligned}
\label{afin}
\end{equation} 
Exploiting the expressions of the roots $e_2$ and $e_3$, it is not difficult to realize that the right hand side of (\ref{afin})
has an expansion in positive powers of $q$ and that only a finite number of terms contribute to a given order, {\it{i.e.}} to a given
instanton number. 
For example, using (\ref{Ccoeff}) and (\ref{e2e3}), up to one instanton we find
\begin{equation}
a=\sqrt{u}\left(1+q\,\frac{u^2+m^2u+4mMu+M^2 u-m^2M^2}{4u^2}+\ldots\right)~,
\label{a1inst}
\end{equation}
which can be inverted leading to
\begin{equation}
u=a^2\left(1-q\,\frac{a^4+m^2a^2+4mMa^2+M^2a^2-m^2M^2}{2a^2}+\ldots\right)~.
\label{u1inst}
\end{equation}

This result allows us to finally obtain the prepotential. Inserting it into (\ref{fhatI}) we get
\begin{equation}
F_{\mathrm{I}}-a^2\log q=q\left(\frac{a^2}{2}+\frac{m^2+4mM+M^2}{2}+\frac{m^2 M^2}{2a^2}\right)+\ldots
\label{Fnek}
\end{equation}
On the right hand side we recognize the 1-instanton prepotential for the SU(2) $N_f=4$ theory obtained in Nekrasov's
approach described in Appendix \ref{Nekrasov}%
\footnote{For the explicit expression see for example Section 7 and Appendix D of \cite{Billo:2010mg}, keeping in mind that
$m_i^{\mathrm{there}}=\sqrt{2}\,m_i^{\mathrm{here}}$.}. This instanton prepotential follows from that of the U(2) theory
after decoupling the U(1) contribution and, as is well known, does not possess the SO(8) 
flavor symmetry of the effective theory; however the terms which spoil this symmetry are all 
$a$-independent (like for example the pure mass terms in (\ref{Fnek})) and therefore are not physical. 
On the other hand, if we insert (\ref{u1inst}) into (\ref{fhatII}) we get
\begin{equation}
F_{\mathrm{II}}-a^2\log q=q\left(\frac{a^2}{2}+\frac{m^2 M^2}{2a^2}\right)+\ldots
\label{Fsw}
\end{equation}
which is the 1-instanton term of the SO(8) invariant prepotential following from the SW curve of \cite{Seiberg:1994aj}.
In this respect it is worth recalling that this curve, differently from (\ref{su2curve}), is parametrized in terms of 
the IR coupling of the massless theory $Q^{(0)}$ which is related to the UV coupling $q$ by \cite{Grimm:2007tm}
\begin{equation}
q=\frac{\theta_2^4}{\theta_3^4}\big(Q^{(0)}\big)~.
\label{qq0}
\end{equation}
As shown for example in \cite{Billo:2010mg,Billo':2011pr}, if one rewrites the prepotential derived from the SW curve
in terms of $q$ using (\ref{qq0}) one can precisely recover the above SO(8) invariant result.

The last ingredient is the perturbative 1-loop contribution which is given by%
\footnote{See also (\ref{Fpertapp}), with obvious modifications, in the limit $\eu,\ed\to 0$.} 
\begin{equation}
F_{\mathrm{pert}}= -2a^2\log\frac{4a^2}{\Lambda^2}+\frac{1}{4}\sum_{i=1}^4\left[(a+m_i)^2\log\frac{(a+m_i)^2}{\Lambda^2}
+(a-m_i)^2\log\frac{(a-m_i)^2}{\Lambda^2}\right]~.
\label{F1loopapp}
\end{equation}
{From} the complete prepotential $\mathcal{F}=F+F_{\mathrm{pert}}$ 
one obtains the IR effective coupling $Q$ of the massive theory by means of
\begin{equation}
Q=\rme^{\pi\ii\tau}\qquad\mbox{with}\qquad \pi\ii\tau=\frac{1}{2}\frac{\partial^2\cF}{\partial a^2}~.
\label{Qapp}
\end{equation}
Notice that both $F_{\mathrm{I}}$ and $F_{\mathrm{II}}$ lead to the same $Q$ since they only differ by $a$-independent terms.
For our specific mass choice (\ref{masses}), up to 1 instanton we find
\begin{equation}
Q=\frac{q}{16}\left(1-\frac{m^2+M^2}{a^2}+\frac{m^2M^2}{a^4}\right)\left(1+q\,\frac{a^4+3m^2M^4}{2a^4}+\ldots\right)~.
\label{QmM}
\end{equation}
As is well-known, given $Q$ one can obtain the cross-ratio $\zeta$ of the four roots $e_i$ of the associated 
SW torus by means of the uniformization formula
\begin{equation}
\label{UVIRnf4}
\zeta=\frac{(e_1-e_2)(e_3-e_4)}{(e_1-e_3)(e_2-e_4)}
= \frac{\theta_2^4}{\theta_3^4}\big(Q\big)
\end{equation}
which is the massive analogue of the massless relation (\ref{qq0}).
Using (\ref{QmM}) and expanding the Jacobi $\theta$-functions we find
\begin{equation}
\begin{aligned}
\zeta&=q\left(1-\frac{m^2+M^2}{a^2}+\frac{m^2M^2}{a^4}\right)\\
&~~~+
q^2\left(\frac{m^2+M^2}{2a^2}-\frac{m^4+M^4}{2a^6}-\frac{m^2M^2(m^2+M^2)}{2a^6}+\frac{m^4M^4}{a^8}\right)+\ldots
\end{aligned}
\label{zeta2}
\end{equation}
It is not difficult to check that this expression exactly agrees with the cross-ratio (\ref{zeta1}) upon using the relations between $a$
and $u$ given in (\ref{a1inst}) and (\ref{u1inst}), thus confirming in full detail the consistency of the calculations.

\section{The SU(2)$\,\times\,$SU(2) quiver theory}
\label{secn:quiver}
\begin{figure}[ht]
\centering
\includegraphics[width=90mm]{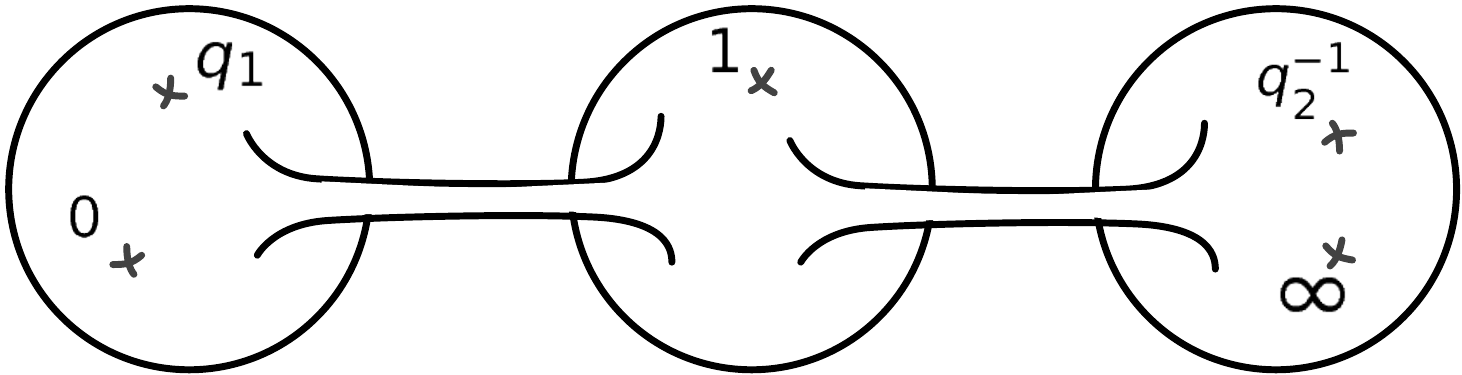}
\caption{Punctured sphere in the weak-coupling limit \label{overflow1}}
\end{figure}

We now consider the 2-node quiver theory whose SW curve takes the form (see (\ref{cmassg}))
\begin{equation}
\label{cmassg1}
x^2(t)=  
\frac{\mathcal{P}_6(t)}{t^2\,(t-q_1 q_2)^2(t-q_2)^2(t-1)^2}~,
\end{equation}
where the sixth-order polynomial $\mathcal{P}_6(t)$ is given in (\ref{P6}).
In the following it will be useful to use yet another form of the curve that can be obtained from (\ref{cmassg1}) by
performing the rescaling $(x,t) \rightarrow (x\,{q_2}^{-1}, t\,q_2)$. This yields
\be
\label{cmassg2}
x^2(t) =  \frac{p_6(t)}{t^2\,(t-q_1)^2(t-1)^2(q_2 t-1)^2}
\ee
where 
\begin{equation}
\label{p6rescale}
p_6(t)= \mathcal{P}_6(q_2t)\,q_2^{-4} = (u_1-u_2 t) \,t(t-1)(t-q_1)(q_2 t-1) +  \mathcal{M}_6(q_2 t)\,q_2^{-4}~.
\end{equation}
In this form the two SU(2) factors appear on the same footing and their weak coupling limit is simply
described by $q_1$ and $q_2$ approaching zero.
In this limit the punctured sphere which corresponds to the denominator of (\ref{cmassg2}) looks as depicted 
in Fig.~\ref{overflow1}.  

\begin{figure}[ht]
\begin{center}
\def\svgwidth{9cm}
\begingroup%
  \makeatletter%
  \providecommand\rotatebox[2]{#2}%
  \ifx\svgwidth\undefined%
    \setlength{\unitlength}{379.0994644bp}%
    \ifx\svgscale\undefined%
      \relax%
    \else%
      \setlength{\unitlength}{\unitlength * \real{\svgscale}}%
    \fi%
  \else%
    \setlength{\unitlength}{\svgwidth}%
  \fi%
  \global\let\svgwidth\undefined%
  \global\let\svgscale\undefined%
  \makeatother%
  \begin{picture}(1,0.67530816)%
    \put(0,0){\includegraphics[width=\unitlength]{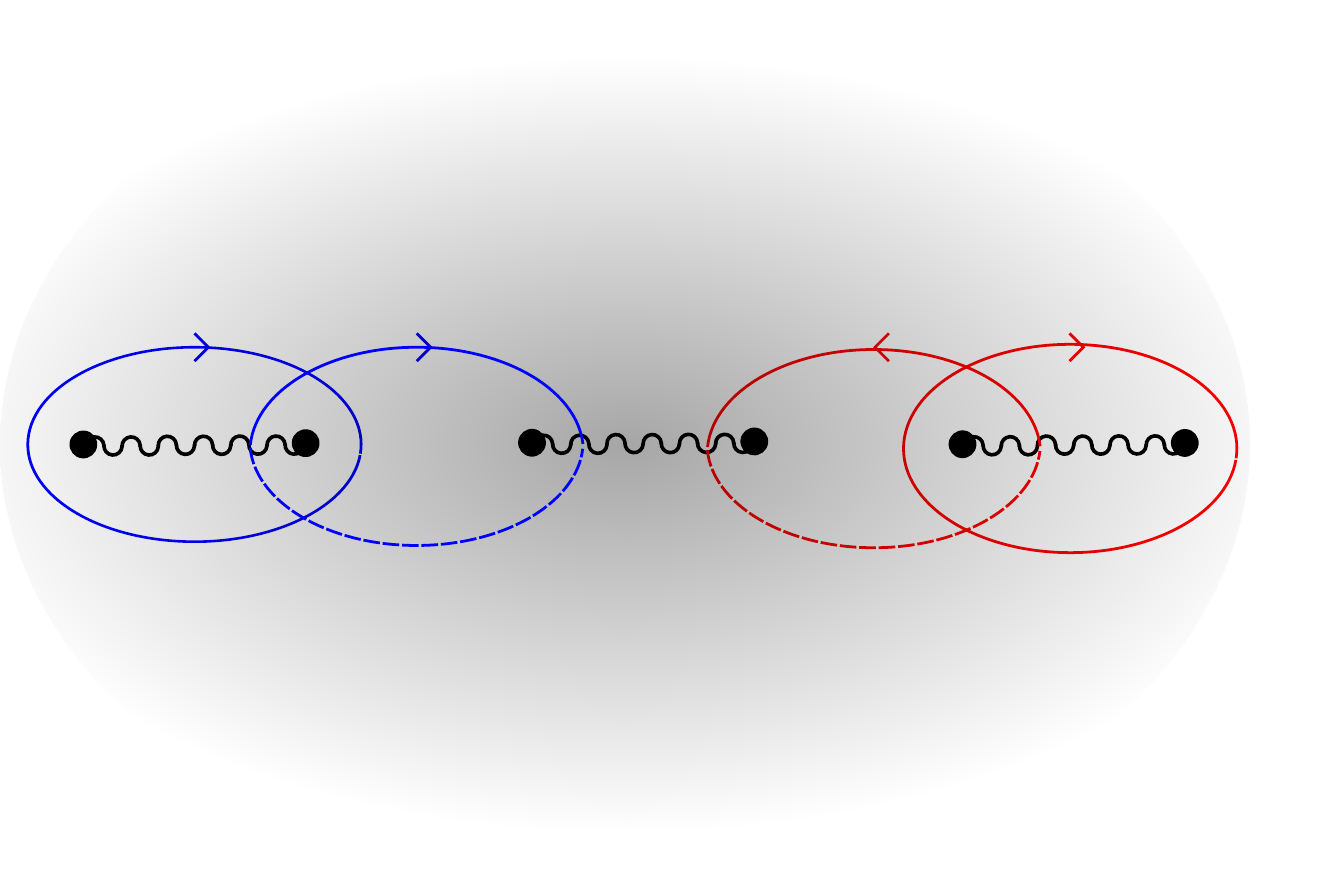}}%
    \put(0.04261623,0.2894488){\makebox(0,0)[lb]{\smash{$0$}}}%
    \put(0.22143736,0.2894488){\makebox(0,0)[lb]{\smash{$\zeta_1$}}}%
    \put(0.37985849,0.2894488){\makebox(0,0)[lb]{\smash{$1$}}}%
    \put(0.59423095,0.2894488){\makebox(0,0)[lb]{\smash{$\widehat\zeta$}}}%
    \put(0.71750075,0.2894488){\makebox(0,0)[lb]{\smash{$\zeta_2$}}}%
    \put(0.86521924,0.2894488){\makebox(0,0)[lb]{\smash{$\infty$}}}%
    \put(0.0531888,0.42405333){\makebox(0,0)[lb]{\smash{$\alpha_1$}}}%
    \put(0.87577056,0.42205936){\makebox(0,0)[lb]{\smash{$\alpha_2$}}}%
    \put(0.58033359,0.42206465){\makebox(0,0)[lb]{\smash{$\beta^2$}}}%
    \put(0.33765321,0.42206465){\makebox(0,0)[lb]{\smash{$\beta^1$}}}%
  \end{picture}%
\endgroup%
\end{center}
\caption{The structure of branch-cuts and a basis of cycles for the Riemann surface described by \eq{hyperred}.}
\label{fig:cycles}
\end{figure}
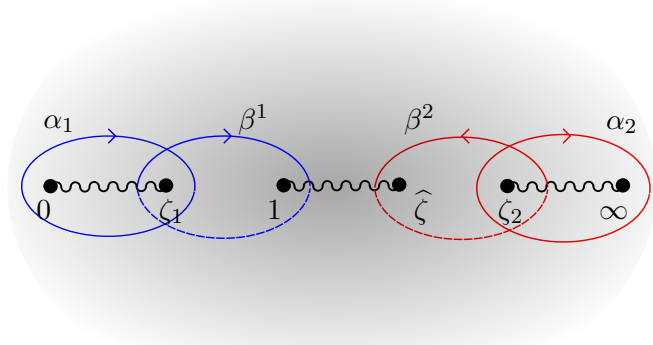

In general the polynomial $p_6(t)$ defined in (\ref{p6rescale}) is of order 6, and thus
the hyperelliptic equation (see (\ref{g2eqmass})) identifying the genus-2 SW curve can be written as
\begin{equation}
\label{sw2roots}
y^2(t) = p_6(t) = c\, \prod_{i=1}^6 (t - e_i)
\end{equation}  
where $e_i$'s are the six roots of the polynomial, which clearly are branch points for the function $y(t)$.   
With a projective transformation we can always fix three of them, say $e_1$, $e_3$ and $e_6$, in $0$, $1$ and $\infty$ and lower 
by one the degree of the polynomial in the right hand side; 
if we call $\zeta_1,\zeta_2$ and $\widehat{\zeta}$ the remaining three parameters, 
corresponding to three independent an harmonic ratios of the $e_i$'s, the equation (\ref{sw2roots}) reduces to
\begin{equation}
\label{hyperred}
y^2(t) = c\, t\big(t-1\big)\big(t-\zeta_1\big)\big(t-\zeta_2\big)\big(t-\widehat{\zeta}\,\big)~.
\end{equation}
When the curve is put in this form, we can choose a symplectic basis of cycles $\{\alpha_i,\beta^i\}$
in the Riemann sphere parametrized by the $t$ variable as shown in Fig.~\ref{fig:cycles}, and then proceed to compute the periods of the 
SW differential and finally derive the effective prepotential. 

However, for generic values of the masses of the matter hypermultiplets this method is not practical since one is not able 
to find the  roots of $p_6(t)$ in closed form and only a perturbative approach in the masses is viable to 
derive the effective prepotential. On the other hand we can exploit the residue conditions (\ref{marsrel}), which 
after the rescalings we have performed, take the form
\begin{equation}
\label{resx2massive}
\begin{aligned}
\mathrm{Res}_{\,t=q_1}\left(x^2(t)\right) =
\frac{\partial \widetilde F}{\partial q_1}~,\qquad
\mathrm{Res}_{\,t=1/q_2}\left(x^2(t)\right) =
- q_2^2\, \frac{\partial \widetilde F}{\partial q_2}~,
\end{aligned}
\end{equation} 
and through them obtain some information on the prepotential directly from the quadratic differential.
Evaluating the residues using the curve equation (\ref{cmassg2}), we find
\begin{align}
q_1(1-q_1)(1-q_1q_2)\,\frac{\partial \wF}{\partial q_1}&=
u_1-q_1u_2-\frac{1}{2}\big(m_1^2+m_2^2\big)\cr
&+\frac{q_1}{4}\Big(\!(m_1+m_2)^2
+(m_1+m_2-m_{12})^2\Big)\cr
&+\frac{q_1q_2}{2}\big(m_1+m_2\big)\Big(m_{12}+\sum_{f=1}^4m_f\Big)\label{res1}\\
&-\frac{q_1^2q_2}{4}\Big(m_{12}^2+2\big(m_1+m_2-m_{12}\big)\sum_{f=1}^4m_f+4m_3m_4\Big)~,\cr
q_2(1-q_2)(1-q_1q_2)\,\frac{\partial \wF}{\partial q_2}&=
u_2-q_2u_1+m_3m_4+\frac{q_2}{4}m_{12}\big(m_{12}+2 m_3+2m_4\big)\cr
&+\frac{q_1q_2}{2}\big(m_3+m_4\big)\big(m_1+m_2-m_{12}\big)\label{res2}\\
&-\frac{q_1q_2^2}{4}\Big(\!m_{12}^2+2m_{12}\sum_{f=1}^4m_f+2\big(m_1+m_2\big)\big(m_3+m_4\big)+
4m_1m_2\Big)\,.\nonumber
\end{align}
Combining these with the residues (\ref{residues2}) suitably rescaled for the new poles, we can rewrite the curve as
\begin{equation}
\begin{aligned}
x^2(t)=
&~
\frac{(m_1-m_2)^2}{4t^2}+\frac{(m_1+m_2)^2}{4(t-q_1)^2}+\frac{m_{12}^2}{(t-1)^2}
+\frac{(m_3+m_4)^2}{4(t-\frac{1}{q_2})^2}\\
& -\frac{m_1^2+m_2^2+2m_3m_4+2m_{12}^2}{2t(t-1)}
+\frac{q_1(q_1-1)}{t(t-q_1)(t-1)}\,\frac{\p\widetilde{F}}{\p q_1}+
\frac{q_2(1-\frac{1}{q_2})}{t(t-1)(t-\frac{1}{q_2})}\,\frac{\p\widetilde{F}}{\p q_2}
\end{aligned}
\label{curvequiverf}
\end{equation}
which is a simple generalization of (\ref{curveSU2f}). We now investigate the meaning of the function
$\widetilde{F}$ appearing in the last two terms of (\ref{curvequiverf}).
If we impose the boundary conditions (\ref{u12to0}) on the $u_i$'s, from (\ref{res1}) and (\ref{res2}) we obtain
\begin{equation}
\begin{aligned}
q_1\frac{\partial\widetilde{F}}{\partial q_1}\,\Big|_{q_1,q_2\to 0}&=\,a_1^2-\frac{1}{2}\big(m_1^2+m_2^2\big)~,\\
q_2\frac{\partial\widetilde{F}}{\partial q_2}\,\Big|_{q_1,q_2\to 0}&=\,a_2^2+m_3m_4~.
\end{aligned}
\end{equation}
Thus, in order to match with the classical prepotential $F_{\mathrm{cl}}=a_1^2\log q_1+a_2^2\log q_2$,
we are led to the following redefinition
\begin{equation}
\label{Fhatquiver}
\wF=\widehat{F}-\frac{1}{2}\big(m_1^2+m_2^2\big)\log q_1+m_3\,m_4\log q_2~.
\end{equation}
Just as we did for the SU(2) $N_f=4$ theory discussed in Section \ref{secn:Nf4}, here too we have to make sure that all symmetries
of the quiver model are correctly implemented. If we just focus on the first group factor, 
we obtain an SU(2) theory with coupling $q_1$ and four effective flavors with masses $\{m_1,m_2,a_2+m_{12},-a_2+m_{12}\}$.
Therefore, according to (\ref{Fhat1}) we have to redefine $\widehat{F}$ by the term
\begin{equation}
\frac{1}{2}\left(m_1+m_2\right)\left(a_2+m_{12}-a_2+m_{12}\right)\log(1-q_1)=
\left(m_1+m_2\right)m_{12}\log(1-q_1)~.
\end{equation}
Likewise, if we focus on the second group factor, we find an SU(2) theory with coupling $q_2$ and 
four effective flavors with masses $\{a_1-m_{12},-a_1-m_{12},m_3,m_4\}$; finally if we consider the quiver as whole,
we have a "diagonal" SU(2) theory with coupling $q_1q_2$ and four masses given by $\{m_1,m_2,m_3,m_4\}$. All in all, in order to
implement all symmetries of the quiver diagram and its subdiagrams, we must redefine $\widehat{F}$ according to
\begin{equation}
\begin{aligned}
\widehat{F}=F&+\left(m_1+m_2\right)m_{12}\log(1-q_1)-m_{12}\left(m_3+m_4\right)\log(1-q_2)\\
&+\frac{1}{2}\left(m_1+m_2\right)\left(m_3+m_4\right)\log(1-q_1q_2)~.
\end{aligned}
\label{Fquiver}
\end{equation}
It is interesting to observe that these logarithmic terms are like the U(1) dressing factors commonly
used in the context of the AGT correspondence
\cite{Alday:2009aq}.
Quite remarkably, if we combine (\ref{Fhatquiver}) and (\ref{Fquiver}), the two very asymmetric equations 
(\ref{res1}) and (\ref{res2}) acquire a symmetric structure. 
Indeed, if we set
\begin{equation}
\label{defFi}
U_i =q_i\frac{\p F}{\p q_i}\qquad\quad\mbox{for}~i=1,2~,
\end{equation}
then equation (\ref{res1}) becomes
\begin{align}
(1-q_1)(1-q_1q_2)\,U_1=&\,
u_1-q_1u_2 +\frac{q_1}{4}\Big(m_{12}\big(m_{12}+2m_1+2m_2\big)+4m_1m_2\Big)\cr
&\,+\frac{q_1q_2}{2}\Big(\big(m_1+m_2\big)\big(m_{12}+2m_3+2m_4\big)+2m_1m_2\Big)\label{reshat1}\\
&\,-\frac{q_1^2q_2}{4}\Big(m_{12}\big(m_{12}+2m_1+2m_2-2m_3-2m_4\big)+4\!
\sum_{f<f'}\!m_FM_{f'}\Big)~,\notag
\end{align}
while the corresponding equation for $U_2$ following from (\ref{res2})
can be obtained from (\ref{reshat1}) with the replacements
\begin{equation}
q_1\leftrightarrow q_2~,~~u_1\leftrightarrow u_2~,~~(m_1,m_2)\leftrightarrow (m_3,m_4)~,~~
m_{12}\leftrightarrow -m_{12}~.
\label{symm}
\end{equation}
This is precisely the exchange symmetry that should hold in the 2-node quiver model under consideration.
The function $F$ therefore has all the required properties to be identified with the effective prepotential of the SU(2)$\,\times\,$SU(2)
gauge theory. To check this statement in an explicit way, we choose two mass configurations 
for which the polynomial $p_6(t)$ in (\ref{cmassg2}) can be factorized and its roots and period integrals
can be explicitly computed.
Specifically we consider the following two cases:
\begin{subequations}
\begin{align}
\mathrm{A)}:\qquad & m_1 = m_2 = m~,~~ m_3 = m_4 = m_{12}=0~,\label{masschoice}\\
\mathrm{B)}:\qquad & m_1 = m_2 = m_3 = m_4 = 0~,~~ m_{12}=M~.\label{masschoiceM}
\end{align}
\label{massAB}
\end{subequations}
As we will see, these mass configurations allow us to make the point and exhibit all relevant features while keeping the 
treatment quite simple.

\subsection{The IR prepotential from the UV curve}
\label{subsecn:prepmassless}
\paragraph{Case A):} With the masses (\ref{masschoice}) the polynomial $p_6(t)$ of the SW curve becomes
\begin{equation}
\label{p6mass}
p_6(t) = t (t-1)  (q_2 t - 1) \Big[(u_1-u_2 t)(t -q_1) 
+ m^2 q_1 \left(q_1 q_2 t + 1 - q_1 - q_1 q_2\right)\Big]~.
\end{equation} 
If we factorize the term in square brackets we immediately bring the curve to the form (\ref{hyperred}), 
with $c= -q_2 u_2$ and
\begin{equation}
\label{zetamassive}
\zeta_1  = \frac{u_1 + q_1u_2 + m^2 q_1^2 q_2 - \sqrt{D}}{2 u_2}~,~~
\widehat{\zeta}  = \frac{u_1 + q_1u_2 + m^2 q_1^2 q_2 + \sqrt{D}}{2 u_2}~,~~
\zeta_2 = \frac{1}{q_2}
\end{equation}
where
\begin{equation}
\label{disc}
D = \big(u_1 - q_1 u_2\big)^2 + 2 m^2 q_1\Big[q_1 q_2 u_1 + u_2\big(2-2q_1-2q_1q_2+q_1^2 q_2\big) \Big]
+ m^4 q_1^4 q_2^2~.
\end{equation}
Then the spectral curve (\ref{cmassg2}) reduces to%
\footnote{Note that in the massless limit we have $\zeta_1 \to q_1$ and $\widehat{\zeta}\to u_1/u_2$.}
\begin{equation}
\label{x2mass}
x^2(t) =\frac{-u_2(t-\zeta_1)(t - \widehat{\zeta})}{t (t-1) (t-q_1)^2(q_2 t - 1)}~.
\end{equation}
For later purposes it is convenient to invert the relation (\ref{reshat1}) and the corresponding one for $U_2$
in order express $u_1$ and $u_2$ in terms of $U_1$ and $U_2$. For the mass configuration (\ref{masschoice})
we get
\begin{equation}
\label{utoF}
\begin{aligned}
u_1 & = \big(1-q_1\big)\,U_1 + q_1 \big(1-q_2\big)\,U_2-m^2q_1\big(1+q_2\big)~,\\
u_2 & = \big(1-q_2\big)\,U_2 + q_2 \big(1-q_1\big)\,U_1-m^2q_1q_2~.
\end{aligned}
\end{equation}  

\begin{figure}
\begin{center}
\def\svgwidth{9cm}
\begingroup%
  \makeatletter%
  \providecommand\rotatebox[2]{#2}%
  \ifx\svgwidth\undefined%
    \setlength{\unitlength}{370.5057144bp}%
    \ifx\svgscale\undefined%
      \relax%
    \else%
      \setlength{\unitlength}{\unitlength * \real{\svgscale}}%
    \fi%
  \else%
    \setlength{\unitlength}{\svgwidth}%
  \fi%
  \global\let\svgwidth\undefined%
  \global\let\svgscale\undefined%
  \makeatother%
  \begin{picture}(1,0.6909717)%
    \put(0,0){\includegraphics[width=\unitlength]{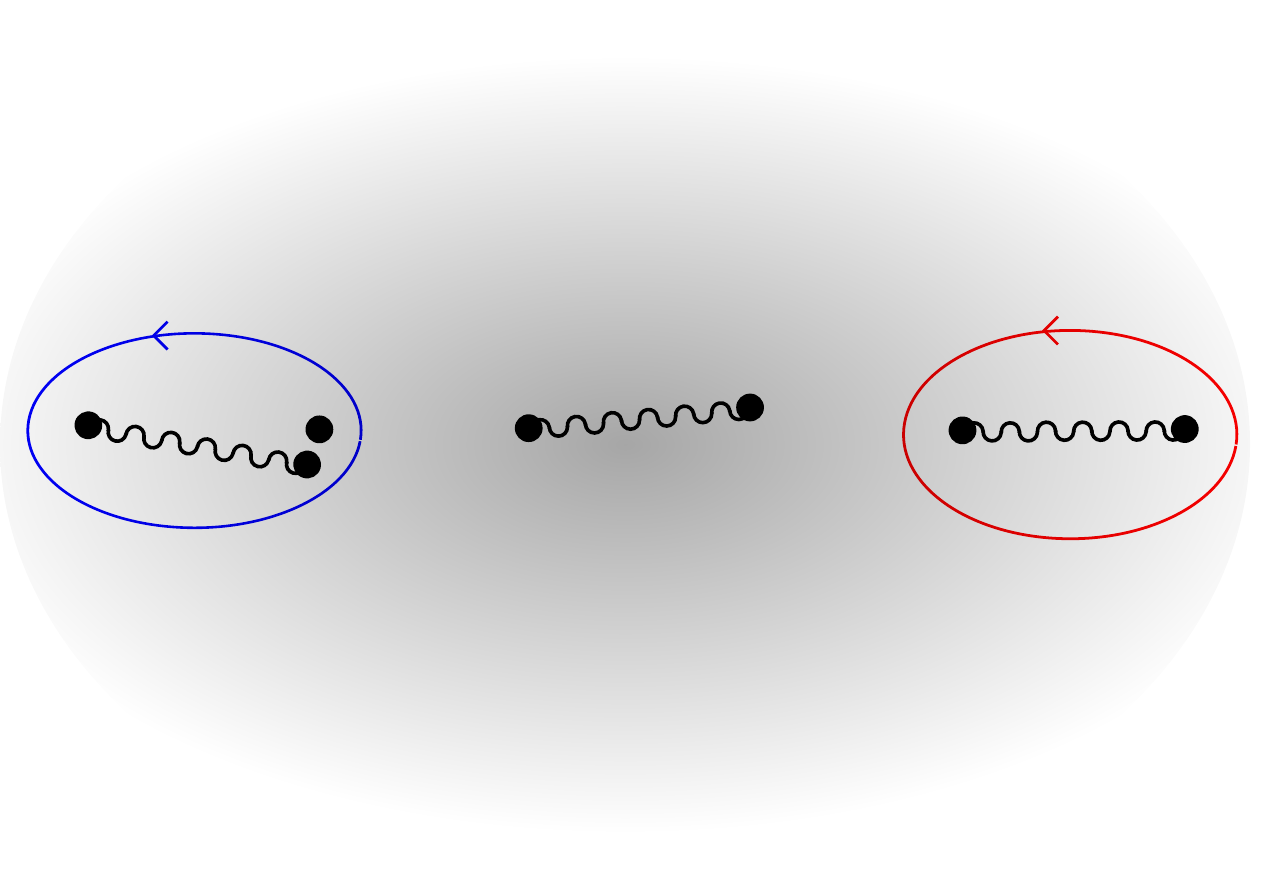}}%
    \put(0.04319543,0.31309768){\makebox(0,0)[lb]{\smash{$0$}}}%
    \put(0.2086075,0.28271579){\makebox(0,0)[lb]{\smash{$\zeta_1$}}}%
    \put(0.38866918,0.31309768){\makebox(0,0)[lb]{\smash{$1$}}}%
    \put(0.56271207,0.31656893){\makebox(0,0)[lb]{\smash{$\widehat\zeta$}}}%
    \put(0.72828306,0.30577287){\makebox(0,0)[lb]{\smash{$\zeta_2=\frac{1}{q_2}$}}}%
    \put(0.91767586,0.37787401){\makebox(0,0)[lb]{\smash{$\infty$}}}%
    \put(0.06478753,0.4534464){\makebox(0,0)[lb]{\smash{$\alpha_1$}}}%
    \put(0.83130743,0.44265034){\makebox(0,0)[lb]{\smash{$\alpha_2$}}}%
    \put(0.21169641,0.38172767){\makebox(0,0)[lb]{\smash{$q_1$}}}%
  \end{picture}%
\endgroup%
\end{center}
\caption{Branch-cuts and singularities of the $a$-periods of SW differential $\lambda$ defined in \eq{swm}.}
\label{fig:lambda-a-cycles-massive}
\end{figure}
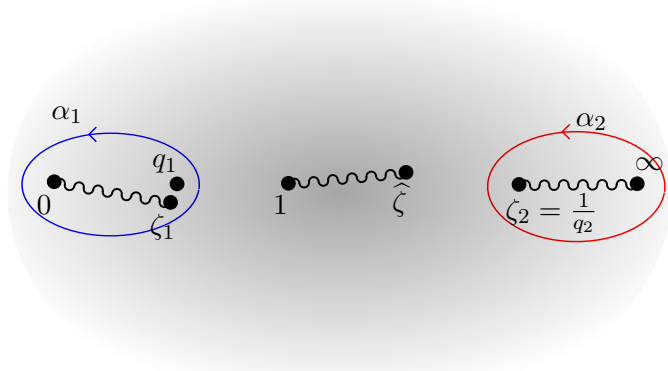

The SW differential associated to the curve (\ref{x2mass}) is
\begin{equation}
\label{swm}
\lambda = x(t) \,dt = \sqrt{ \frac{-u_2(t-\zeta_1)(t-\widehat{\zeta})}{t(t-1)(q_2 t - 1)}}\,
\frac{dt}{q_1-t}~,
\end{equation}
and its singularity structure is shown in Fig.~\ref{fig:lambda-a-cycles-massive}.

The periods of $\lambda$ along the cycles $\alpha_1$ and $\alpha_2$ are identified with the vacuum expectation values
$a_1$ and $a_2$, respectively.  Let us first consider the cycle $\alpha_1$ and note that it surrounds both the branch cut
from $0$ to $\zeta_1$ and the pole in $t=q_1$. Thus we have
\begin{equation}
\begin{aligned}
a_1&= \frac{1}{2\pi \ii}\oint_{\alpha_1} \lambda=~
\mathrm{Res}_{\,t=q_1}\left(\lambda\right)~
+\frac{1}{\pi}\int_0^{\zeta_1}
 \sqrt{\frac{\zeta_1-t}{t}}~\sqrt{\frac{u_2(\widehat{\zeta}-t)}{(1-t)(1-q_2t)}}\,\frac{dt}{q_1-t}~.
\end{aligned}
\label{a1mass}
\end{equation}
The integral over the branch cut can be evaluated as explained in Appendix~\ref{secn:integral} (see in particular Eq.~(\ref{J'})); it contains a contribution that cancels the residue
and the final result for $a_1$ is
\begin{equation}
a_1=\sqrt{\frac{u_2(\widehat{\zeta}-q_1)}{(1-q_1)(1-q_1q_2)}}
-\sum_{n,\ell=0}^\infty(-1)^n \binom{1/2}{n+1}f_{n+\ell+1}\,\zeta_1^{n+1}\,q_1^\ell
\label{a1mass1}
\end{equation}
where the $f_n$'s are the coefficients in the following Taylor expansion
\begin{equation}
\sqrt{\frac{u_2(\widehat{\zeta}-t)}{(1-t)(1-q_2t)}}=\sum_{n=0}^\infty f_n\,t^n~,
\end{equation}
namely
\begin{equation}
f_n=(-1)^n\sqrt{u_2}
\sum_{\ell,k=0}^n \binom{1/2}{\ell}\binom{-1/2}{k}\binom{-1/2}{n-\ell-k}~
\frac{q_2^k}{\widehat{\zeta}^{\,\ell-1/2}}~.
\label{taylorf}
\end{equation}
Using the expressions (\ref{zetamassive}) for the roots it is not difficult to check that $a_1$ has an expansion in positive 
powers of $q_1$ and $q_2$ and that only a finite number of terms contribute to a given instanton number.
Substituting in the result the relations (\ref{utoF}) we obtain the following weak coupling expansion%
\footnote{For brevity we display only the results up to two instantons, but we have computed also higher instanton contributions without difficulty.}
\begin{eqnarray}
a_1 &=&\sqrt{U_1}\,\Bigg(1-q_1\,\frac{\big(U_1-U_2\big)\big(U_1+m^2\big)}{4U_1^{2}} -q_1q_2\,\frac{\big(U_1+m^2\big)U_2}{4U_1^{2}}
\label{a1actual}\\ &&~~~\qquad
-q_1^2\,\frac{\big(U_1-U_2\big)\Big(U_1\big(7U_1 - 3U_2\big)
\big(U_1+2m^2\big)+3m^4(U_1-5U_2\big)\Big)}{64U_1^{4}}+ \ldots \Bigg)~.
\nonumber
\end{eqnarray}

Let us now turn to the second period $a_2$ along the cycle $\alpha_2$. Referring to Fig.~\ref{fig:lambda-a-cycles-massive} we have
\begin{equation}
\label{a2mass}
\begin{aligned}
a_2&=\frac{1}{2\pi \ii}\oint_{\alpha_2} \lambda ~=~\frac{1}{\pi}\int_{1/q_2}^\infty
\sqrt{\frac{u_2(t-\zeta_1)(t-\widehat{\zeta})}{t(t-1)(q_2t-1)}}\,
\frac{dt}{t-q_1}\phantom{\Bigg(}\\
&=\frac{1}{\pi}\int_0^1\Bigg[\,\sqrt{\frac{u_2(1-q_2\zeta_1z)(1-q_2\widehat{\zeta}z)}{(1-q_2z)}}
\,\frac{1}{(1-q_1q_2z)}\,\Bigg]\,\frac{dz}{\sqrt{z(1-z)}}
\end{aligned}
\end{equation}
where the last step simply follows from the change of integration variable: $t\to1/(q_2z)$. This integral
can be computed by expanding the factor in square brackets in powers of $z$ and then using
\begin{equation}
\label{integral01}
\int_0^{1}\frac{z^n dz}{\sqrt{z(1-z)}} = (-1)^n\,\pi\, \binom{-1/2}{n}~.
\end{equation}
Inserting the root expressions (\ref{zetamassive}) and exploiting the relations (\ref{utoF}),
we find
\begin{equation}
\label{a2actual}
\begin{aligned}
a_2=\sqrt{U_2}\,\Bigg(1-q_2\,\frac{U_2-U_1}{4U_2^{}}
-q_2^2\,\frac{7U_2^2 - 10U_1U_2+3U_1^2}
{64U_2^2}-q_1q_2\,\frac{U_1+m^2}{4U_2} + \ldots \Bigg)~.
\end{aligned}
\end{equation}

Note that the results (\ref{a1actual}) and (\ref{a2actual}) are perturbative in the instanton counting parameters $q_1$ and $q_2$,
but are exact in the mass deformation parameter $m$. We now invert these weak-coupling expansions to obtain
\begin{align}
U_1&=a_1^2+q_1\,\Big(\frac{a_1^2-a_2^2}{2}+m^2\,\frac{a_1^2-a_2^2}{2a_1^2}\Big)
+q_1q_2\,\Big(\frac{a_1^2+a_2^2}{4}
+m^2\,\frac{a_1^2+a_2^2}{4a_1^2}\Big)\notag\\
&~~~+q_1^2\,\Big(\frac{13a_1^4-14a_1^2a_2^2+a_2^4}{32a_1^2}+
m^2\,\frac{9a_1^4-6a_1^2a_2^2-3a_2^4}{16a_1^4}+m^4\,\frac{a_1^4-6a_1^2a_2^2+5a_2^4}{32a_1^6}\Big)+
\ldots~,\label{wF1exp}\\
U_2&=a_2^2+q_2\,\frac{a_2^2-a_1^2}{2}
+q_1q_2\,\Big(\frac{a_1^2+a_2^2}{4}+m^2\,\frac{a_1^2+a_2^2}{4a_1^2}\Big)
+q_2^2\,\frac{13a_2^4-14a_1^2a_2^2+a_1^4}{32a_2^2}+\ldots~.\label{wF2exp}
\end{align}
These two expressions are integrable, thus leading to the determination of $F$ (up to $q$-independent terms)%
\footnote{We note that our results differ in some numerical coefficients from those reported in \cite{Gavrylenko:2013dba} for the
massless quiver. However, we have checked that our results are consistent with the microscopic multi-instanton calculations.}: 
\begin{eqnarray}
F &=& a_1^2 \,\log q_1 + a_2^2\, \log q_2 +
q_1\,\Big(\frac{a_1^2-a_2^2}{2}+m^2\,\frac{a_1^2-a_2^2}{2a_1^2}\Big)+q_2\,\frac{a_2^2-a_1^2}{2} \nonumber\\
&&~+q_1q_2\,\Big(\frac{a_1^2+a_2^2}{4}+m^2\,\frac{a_1^2+a_2^2}{4a_1^2}\Big)
+ q_2^2\,\frac{13a_2^4-14 a_1^2a_2^2+a_1^2}{64a_2^2}\label{Fzero}\\
&&~+q_1^2\,\Big(\frac{13a_1^4-14a_1^2a_2^2+a_2^4}{64a_1^2}+
m^2\,\frac{9a_1^4-6a_1^2a_2^2-3a_2^4}{32a_1^4}+m^4\,\frac{a_1^4-6a_1^2a_2^2+5a_2^4}{64a_1^6}\Big)+
\ldots~.\nonumber
\end{eqnarray}
This precisely matches the $q$-dependent part of the prepotential derived using Nekrasov's localization techniques in the quiver theory
when we choose the masses as in (\ref{masschoice}) and set the $\Omega$-deformation parameters $\epsilon_i$ to zero 
(see Appendix \ref{Nekrasov} for details, and in particular (\ref{Fmapp})). 

Finally, adding the $q$-independent 1-loop contribution $F_{\mathrm{pert}}$ (see Eq.~(\ref{Fpertapp1})), we may obtain the
complete prepotential of the effective theory
\begin{equation}
\label{FiswF}
\begin{aligned}
\cF=&\, F+F_{\mathrm{pert}} \\
=&\,F - 2a_1^2 \log\frac{4a_1^2}{\Lambda^2} -2a_2^2 \log\frac{4a_2^2}{\Lambda^2}
+\frac{1}{2}(a_1+m)^2\log\frac{(a_1+m)^2}{\Lambda^2}\\
&~~+\frac{1}{2}(a_1-m)^2\log\frac{(a_1-m)^2}{\Lambda^2}+ a_2^2\log\frac{a_1^2}{\Lambda^2} \\
&~~+ \frac 12 (a_1+a_2)^2\log\frac{(a_1+a_2)^2}{\Lambda^2 }
+ \frac 12 (a_1-a_2)^2\log\frac{(a_1-a_2)^2}{\Lambda^2 } ~.
\end{aligned}
\end{equation}
This result represents a nice check of the spectral curve (\ref{x2mass}) 
and of the relations (\ref{resx2massive}).

Using all our findings so far, we can easily derive the weak-coupling expansions of the roots (\ref{zetamassive}) which are
\begin{align}
\zeta_1&=q_1\Big(1-\frac{m^2}{a_1^2}\Big)\Bigg(1+q_1\,m^2\,\frac{a_1^2-a_2^2}{2a_1^4}
+q_1q_2\,m^2\,\frac{a_1^2+a_2^2}{4a_1^4}
\notag\\
&~~~\qquad+q_1^2\,m^2\frac{\big(a_1^2-a_2^2\big)\big(5a_1^4+7a_1^2a_2^2+7a_1^2m^2-19a_2^2m^2\big)}{32a_1^8}
+\ldots\Bigg)~,
\label{zeta1weak}\\
\widehat{\zeta}&= \frac{a_1^2}{a_2^2}\,\Bigg(1 - q_1\,\frac{\big(a_1^2-a_2^2\big)\big(a_1^2+m^2\big)}{2a_1^4} 
-q_2\,\frac{a_1^2-a_2^2}{2a_2^2}+q_1 q_2\,\frac{\big(a_1^2-a_2^2\big)\big(a_1^2+m^2\big)}{2a_1^2a_2^2}
\notag\\
&~~~~~~~~~~~- q_1^2 \,\frac{\big(a_1^2-a_2^2\big)
\big(a_1^2-m^2\big)\big(3a_1^4 +a_1^2a_2^2+a_1^2m^2+11a_2^2m^2\big)}{32a_1^8}\notag\\
&~~~~~~~~~~~+q_2^2\, \frac{\big(a_1^2-a_2^2\big)\big(7a_1^2-11a_2^2\big)}{32a_2^4}+ \ldots\Bigg)~,
\label{zetahatweak}
\end{align}
and
\begin{equation}
\frac{1}{\zeta_2}=q_2~.\label{zeta2weak}
\end{equation}
We remark that (\ref{zeta1weak}) and (\ref{zetahatweak}) are perturbative in the $q$'s  but are exact in the mass parameter.

\paragraph{Case B):} Let us now briefly consider the second mass choice (\ref{masschoiceM}). In this case the
spectral curve (\ref{cmassg2}) becomes
\begin{equation}
x^2(t)=\frac{C(t-\zeta_3)(t-\widehat{\zeta})}{t(t-q_1)(q_2t-1)(t-1)^2}
\label{x2massM}
\end{equation}
where
\begin{equation}
\label{zetamassiveM}
\begin{aligned}
\zeta_3  &= \frac{-4u_1 -4 u_2 + M^2(4- q_1-q_2+2q_1q_2)-4 \sqrt{D}}{8 C}~,\\
\widehat{\zeta}  &= \frac{-4u_1 -4 u_2 + M^2(4- q_1-q_2+2q_1q_2)+4 \sqrt{D}}{8 C}~,
\end{aligned}
\end{equation}
with
\begin{equation}
\label{discM}
\begin{aligned}
C&=-u_2+\frac{3M^2}{4}q_2-\frac{M^2}{4}q_1q_2~,\\
D &= \frac{1}{16}\big(4u_1 +4 u_2 - M^2(4- q_1-q_2+2q_1q_2)\big)^2+C\big(4u_1-3M^2q_1+M^2q_1q_2\big)
~.
\end{aligned}
\end{equation}
As in the previous case, it will prove useful to invert the relation (\ref{reshat1}) and the corresponding one for $U_2$;
this leads to
\begin{equation}
\label{utoFM}
\begin{aligned}
u_1 & = \big(1-q_1\big)\,U_1 + q_1 \big(1-q_2\big)\,U_2-\frac{M^2}{4} q_1\big(1+q_2\big)~,\\
u_2 & = \big(1-q_2\big)\,U_2 + q_2 \big(1-q_1\big)\,U_1-\frac{M^2}{4} q_2\big(1+q_1\big)~.
\end{aligned}
\end{equation}

We now compute the $\alpha$-periods of the SW differential $\lambda=x(t)dt$, whose singularity structure is similar
to the one shown in Fig.~\ref{fig:lambda-a-cycles-massive}. The main difference is that now $t=q_1$ is a branch-point and
not a pole, while $t=1$ is a pole and not a branch-point. Taking this into account we therefore have
\begin{equation}
a_1=\frac{1}{2\pi\ii}\oint_{\alpha_1}\lambda=\frac{1}{\pi}\int_{0}^{q_1}\sqrt{\frac{C(\zeta_3-t)(t-\widehat{\zeta})}{t(q_1-t)(1-q_2t)}}\,\frac{dt}{(1-t)}~.
\label{a1M}
\end{equation}
After rescaling $t\to q_1 t$, we can easily compute the integral as discussed in the previous case expanding in powers of $t$
and exploiting (\ref{integral01}). Making use of the relations (\ref{utoFM}) to express the result in terms of $U_i$,
we obtain
\begin{equation}
\begin{aligned}
a_1 &=\sqrt{U_1}\,\Bigg(1-q_1\frac{U_1+M^2}{4U_1^{2}}
+q_2\frac{U_2}{4U_1^{2}}
 -q_1q_2\,\frac{U_2}{4U_1}
\\ &~~~~\qquad\qquad
-q_1^2\,\frac{7U_1^{2} - 10U_1U_2+3U_2^{2}+M^2
\big(14U_1-6U_2+3M^2\big)}{64U_1^{\,2}}+ \ldots \Bigg)~.
\end{aligned}
\label{a1MM}
\end{equation}
The second period $a_2$ can be calculated along the same lines and the final result can be 
obtained from (\ref{a1MM}) by simply exchanging
$q_1\leftrightarrow q_2$ and $U_1\leftrightarrow U_2$. If we invert these formul\ae~and then 
integrate over $q_1$ and $q_2$, we get
\begin{eqnarray}
F &=& a_1^2 \,\log q_1 + a_2^2\, \log q_2 +
q_1\,\frac{a_1^2-a_2^2+M^2}{2}+q_2\,\frac{a_2^2-a_1^2+M^2}{2} \nonumber\\
&&~+q_1q_2\,\frac{a_1^2+a_2^2-M^2}{4}
+ q_1^2\,\Big(\frac{13a_1^4-14 a_1^2a_2^2+a_2^2}{64a_1^2}+\frac{9M^2}{32}+\frac{M^2(M^2-2a_2^2)}{64a_1^2}\Big)
\nonumber\\
&&~+q_2^2\,\Big(\frac{13a_2^4-14 a_1^2a_2^2+a_1^2}{64a_2^2}+\frac{9M^2}{32}+\frac{M^2(M^2-2a_1^2)}{64a_2^2}\Big)+
\ldots~.\label{FzeroM}
\end{eqnarray}
This exactly matches the instanton prepotential derived using Nekrasov's approach in the quiver theory 
for the particular mass choice (\ref{masschoiceM}) as one can see by comparing with (\ref{Fmapp}).

Our results provide an explicit check of the UV equation of the SW curve and of the way in which the IR effective prepotential 
is explicitly encoded in it; this will be confirmed in Section~\ref{secn:AGT} by exploiting the AGT 
correspondence \cite{Alday:2009aq}.

\subsection{The period matrix and the roots}
\label{subsec:periodmatrix}
We now consider another approach to the derivation of  the effective gauge theory from the SW curve, 
which is based on the computation of the period matrix in terms of the roots of its defining equation (\ref{sw2roots}). 
Taking the standard basis of holomorphic differentials as
\begin{equation}
\label{defomeg2}
\omega^i = \frac{t^{i-1}\, dt}{y(t)}\quad\mbox{for}~~i=1,2~,
\end{equation}
we denote their periods along the cycles described in Fig.~\ref{fig:cycles} as follows:
\begin{equation}
 \label{perdxy}
 \int_{\alpha_j} \omega^i = \big(\Omega_{(1)}\big)^{\!ij}~,\qquad
 \int_{\beta^j} \omega^i = \big(\Omega_{(2)}\big)^{\!i}_{\,j}~.
\end{equation}
The period matrix $\tau$ of the curve is given by
\begin{equation}
\label{taudef}
\tau = \Omega_{(1)}^{-1}\,\Omega_{(2)}^{}~.
\end{equation} 
It is a symmetric matrix and has thus three independent entries $\tau_{11}$, $\tau_{22}$ and $\tau_{12}$. In terms of these
we introduce the quantities
\begin{equation}
\label{Qs}
Q_1 = \rme^{\ii\pi\tau_{11}}~,\quad
Q_2 = \rme^{\ii\pi\tau_{22}}~,\quad
\widehat{Q} = \rme^{\ii\pi\tau_{12}}
\end{equation}
which will be conveniently used in the following. Given the period matrix $\tau$, we introduce the genus-2 $\theta$-constants defined as
\begin{equation}
 \label{deftheta}
 \theta\ch{\vec{\varepsilon}}{\vec{\varepsilon'}}\equiv
\sum_{\vec{n}\in\mathbb{Z}^2} \exp\Big\{\pi\ii \big[
(\vec{n}+ \ft{\vec{\varepsilon}}{2})^t\, \tau\, (\vec{n} + \ft{\vec{\varepsilon}}{2}) 
+ (\vec{n} + \ft{\vec{\varepsilon}}{2})^t\vec{\varepsilon'} \,\big]\Big\}~,
\end{equation}
where $\vec{\varepsilon}, \vec{\varepsilon'}$ are two 2-vectors; in what follows we will only 
need to consider the case in which these vectors have integer components.

The Thomae formul\ae~\cite{thomae} can be used to express%
\footnote{See for instance \cite{enolskii-richter} and Appendix C of \cite{Martucci:2012jk}.} the anharmonic 
ratios $\zeta_1$, $\zeta_2$ and $\widehat{\zeta}$ in terms of the $\theta$-constants. 
Specifically, one has
\begin{equation}
 \label{wzeta}
  \zeta_1  =  
    \frac{\theta^2\ch{10}{00}	\,\theta^2\ch{11}{00}}{\theta^2\ch{01}{00}\,\theta^2\ch{00}{00}}~,~~~
  \zeta_2  = 
    \frac{\theta^2\ch{10}{00}	\,\theta^2\ch{00}{11}}{\theta^2\ch{01}{00}\,\theta^2\ch{11}{11}}~,~~~
  \widehat{\zeta}  =  
    \frac{\theta^2\ch{00}{11}\,\theta^2\ch{11}{00}}{\theta^2\ch{11}{11}\,\theta^2\ch{00}{00}}~.
\end{equation}
Using (\ref{Qs}) and(\ref{deftheta}), we find that $\zeta_1$, $1/\zeta_2$ and $\widehat{\zeta}$ can be expressed as infinite sums containing positive integer powers of $Q_1$ and $Q_2$, and both positive and negative powers of $\widehat{Q}$. 
Up to second order in $Q_1$ and $Q_2$, we have
\begin{align}
\zeta_1 &=  Q_1\,\frac{4(\widehat{Q} +1)^2}{\widehat{Q}}\Bigg[1- Q_1\,\frac{2(\widehat{Q} + 1)^2}{\widehat{Q}} 
+Q_2\, \frac{2(\widehat{Q} - 1)^2}{\widehat{Q}} -Q_1 Q_2\, \frac{8(\widehat{Q}^2 -1)^2}{\widehat{Q}^2}
\label{zeta2exp}\\
&~~~+Q_1^2\,\frac{3 \widehat{Q}^4+ 10 \widehat{Q}^3+ 18 \widehat{Q}^2 + 
10 \widehat{Q} + 3}{\widehat{Q}^2}+Q_2^2\,\frac{(\widehat{Q} - 1)^2
(\widehat{Q}^2-4\widehat{Q}+1)}{\widehat{Q}^2} 
 +\ldots\Bigg]~,\notag
 \end{align}
 \begin{align}
 \frac{1}{\zeta_2}  &= Q_2\,\frac{4(\widehat{Q} -1)^2}{\widehat{Q}} \Bigg[1 
 + Q_1\,\frac{2(\widehat{Q} + 1)^2}{\widehat{Q}} 
 -Q_2 \,\frac{2(\widehat{Q} - 1)^2}{\widehat{Q}} -Q_1 Q_2\,\frac{8(\widehat{Q}^2 -1)^2}{\widehat{Q}^2}
\label{invzeta5exp}\\
&~~~+Q_1^2\,\frac{(\widehat{Q} + 1)^2(\widehat{Q}^2+4\widehat{Q}+1)}{\widehat{Q}^2}
+Q_2^2\,\frac{3 \widehat{Q}^4- 10 \widehat{Q}^3+ 18 \widehat{Q}^2 - 10 \widehat{Q} + 3}{\widehat{Q}^2} +\ldots\Bigg]~,
\notag
\end{align}
and
\begin{align}
\widehat{\zeta} &= \frac{(\widehat{Q} + 1)^2}{(\widehat{Q}-1)^2}\Bigg[1-8(Q_1+ Q_2-8Q_1 Q_2) +
 (Q_1^2+Q_2^2)\frac{4(\widehat{Q}^2+8\widehat{Q}+1)}{\widehat{Q}}\ldots\Bigg]~.
 \label{zeta4exp}
\end{align}

As is well-known, the period matrix of the SW curve is identified with
the matrix of the coupling constants of the low-energy effective theory, which are expressed in terms of the 
prepotential $\cF$ according to
\begin{equation}
\label{taueff}
2\pi\ii\tau_{ij} = \frac{\partial^2 \cF}{\partial a_i\partial a_j}~.
\end{equation} 
Using the prepotential (\ref{FiswF}), from (\ref{taueff}) and (\ref{Qs})  we get
\begin{align}
Q_1 & = q_1\,\frac{(a_1^2-a_2^2)(a_1^2-m^2)}{16a_1^4}\,\Bigg[1+q_1\Big(\frac{1}{2}-\frac{3m^2a_2^2}{2a_1^4}\Big)
-\frac{q_2}{2}\notag\\
&~~~~~+q_1^2\Big(\frac{21a_1^4+3a_2^4}{64a_1^4}-m^2\,\frac{21a_1^2a_2^2+15a_2^4}{16a_1^6}
+m^4\,\frac{3a_1^4-60a_1^2a_2^2+177a_2^4}{64a_1^8}\Big)\notag\\
&~~~~~+q_2^2\,\frac{3a_1^2-3a_2^2}{32a_2^2}+q_1q_2\,\frac{3m^2a_2^2}{2a_1^4}+\ldots\Bigg]~,\label{Q1}\\
Q_2 &=q_2\,\frac{a_1^2-a_2^2}{16a_2^2}\,\Bigg[1-q_1\Big(\frac{1}{2}+\frac{m^2}{2a_1^2}\Big)+\frac{q_2}{2}
+q_2^2\,\frac{21a_2^4+3a_1^4}{64a_2^4}\notag\\
&~~~~~+q_1^2\Big(\frac{3a_2^2-3a_1^2}{32a_1^2}-m^2\,\frac{9a_2^2-a_1^2}{16a_1^4}+m^4\,\frac{15a_2^2+a_1^2}{32a_1^6}\Big)+\ldots\Bigg]~,
\label{Q2}
\end{align}
and
\begin{align}
\widehat{Q} &=\frac{a_1+a_2}{a_1-a_2} \Bigg[1+q_1\,\frac{m^2a_2}{a_1^3}-q_1q_2\,\frac{m^2a_2}{2a_1^3}-q_2^2\,\frac{a_1^3}{16 a_2^3}\notag\\
&~~~~~-q_1^2\Big(\frac{a_2^3}{16 a_1^3}-m^2\,\frac{3a_1^2a_2+6a_2^3}{8a_1^5}
-m^4\,\frac{6a_1^2a_2+8a_1a_2^2-15a_2^3}{16a_1^7}\Big)+\ldots\Bigg]~.\label{qhat}
\end{align}
These formul{\ae}~represent the explicit map between the IR effective couplings and the UV data of the quiver theory.
Inserting the above expressions into  (\ref{zeta2exp})--(\ref{zeta4exp}) 
we can derive the corresponding anharmonic ratios $\zeta_1$, $\widehat{\zeta}$ and $\zeta_2$, and
find perfect agreement with the expressions in (\ref{zeta1weak}), (\ref{zetahatweak}) and (\ref{zeta2weak})!
The same agreement is found also when we use the second mass configuration (\ref{masschoiceM}) and the corresponding
prepotential (\ref{FzeroM}), thus confirming the validity of the whole picture.

Summarizing, we have verified that the SW curve 
is correct since it reproduces the correct prepotential of the low-energy effective field theory. In doing so, 
we have also found the precise relations between the UV data, namely the instanton expansion parameters $q_1$, $q_2$ 
(which encode the UV gauge couplings) and the Coulomb branch parameters $a_1$, $a_2$ on one side, 
and the IR couplings $\tau_{11},\tau_{22}, \tau_{12}$ (or equivalently $Q_1$, $Q_2$ and $\widehat{Q}$)
on the other side. Such relations are given in (\ref{Q1})--(\ref{qhat}) which in turn follow from 
\begin{equation}
\label{relUVIR}
\zeta_1
=\, \frac{\theta^2\ch{10}{00}\,\theta^2\ch{11}{00}}{\theta^2\ch{01}{00}\,\theta^2\ch{00}{00}}\big(Q\big)~,~~~
\frac{1}{\zeta_2} = \,\frac{\theta^2\ch{01}{00}\,\theta^2\ch{11}{11}}{\theta^2\ch{10}{00}\,
\theta^2\ch{00}{11}}\big(Q\big)~,~~~
\widehat\zeta=\, 
\frac{\theta^2\ch{00}{11}\,\theta^2\ch{11}{00}}{\theta^2\ch{11}{11}\,\theta^2\ch{00}{00}}\big(Q\big)~.
\end{equation}
These relations are the genus-2 analogues of the well-known relation \cite{Grimm:2007tm} that holds in the SU(2) theory with $N_f = 4$ and links the instanton counting parameter $q$ of the UV theory to the effective IR coupling $Q$ (see (\ref{UVIRnf4}) for the
massive theory or (\ref{qq0}) for the massless one).
Note that in the SU(2), $N_f=4$ case, for purely dimensional reasons, the vacuum expectation value of the adjoint scalar cannot appear in the massless UV/IR relation but, as we have just shown, this is no longer the case for quivers with more than one node.

\section{The 2d/4d correspondence}
\label{secn:AGT}

We now consider $\Omega$-deformed quiver theories with the goal of both confirming
and extending the previous results. We will also exploit the remarkable 2d/4d correspondence proposed by Alday-Gaiotto-Tachikawa (AGT) in \cite{Alday:2009aq}. 
This correspondence states that the Nekrasov partition function of a linear quiver with gauge group 
$\mathrm{SU}(2)^{n}$ is directly related to the $(n+3)$-point spherical conformal block in two dimensional Liouville CFT. 
Let us give some details%
\footnote{For a more extended and technical discussion see for example \cite{Alba:2009ya} or the recent review \cite{Teschner:2014oja}.}. 

\subsection{The AGT map}
\begin{figure}[ht!]
\centering
\includegraphics[width=110mm]{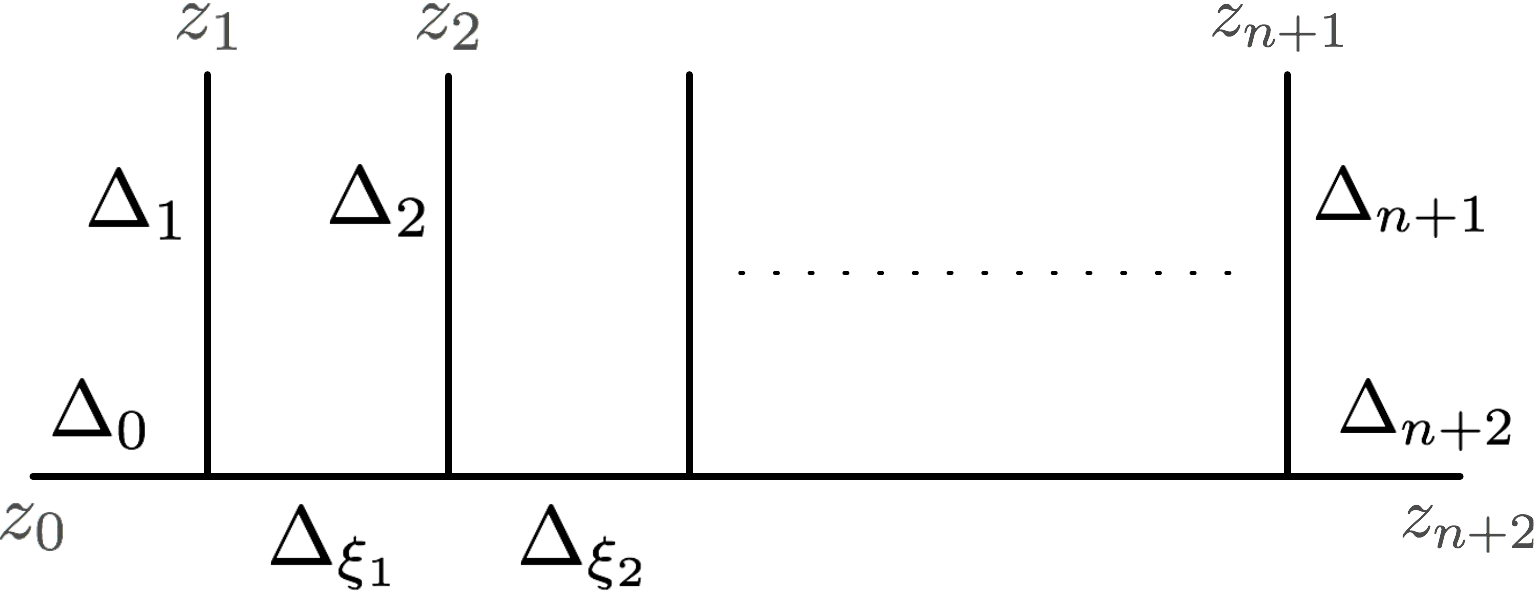}
\caption{Pair-of-pants decomposition of the spherical conformal block with $(n+3)$ punctures}
\label{figconf1}
\end{figure}

\begin{figure}[ht!]
\centering
\includegraphics[width=140mm]{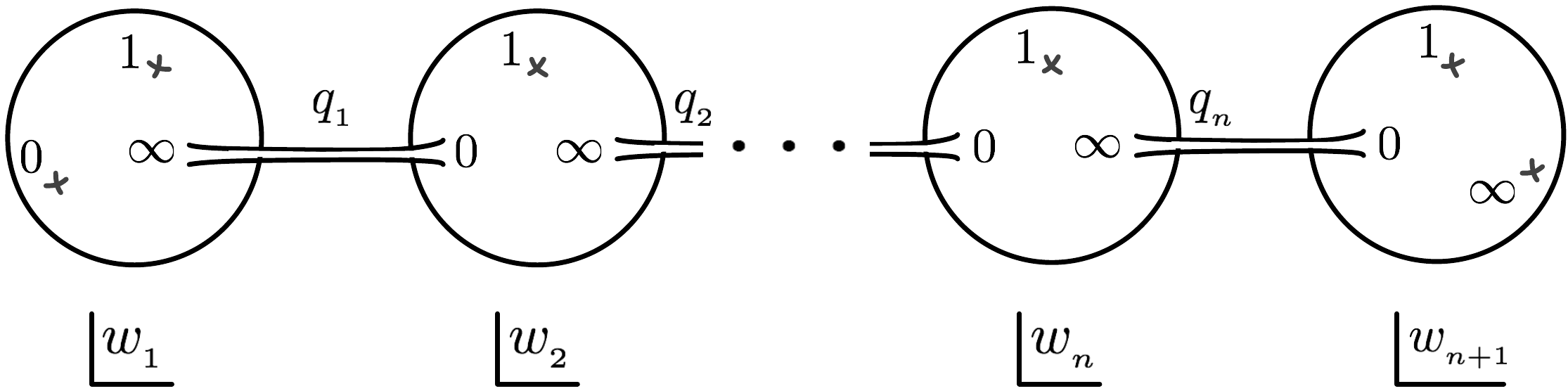}
\caption{Three-punctured spheres connected by long thin tubes, with sewing parameters $q_i$.}
\label{fissph}
\end{figure}

In 2-dimensional Liouville theory with central charge $c=1+6Q^2$, let us consider the 
conformal block
\begin{equation}
\Big\langle \prod_{i=0}^{n+2} V_{\alpha_i}(z_i) \Big\rangle_{ \{ \xi_1,\ldots,\xi_n\}}
\label{ccblock}
\end{equation}
where $V_\alpha$ denotes a primary operator with Liouville momentum $\alpha$ and conformal dimension
\begin{equation}
\Delta_\alpha=\alpha(Q-\alpha)~.
\label{deltaalpha}
\end{equation}
In (\ref{ccblock}) the subscript ${\{ \xi_1,\ldots,\xi_n\}}$ means that the correlator is computed in the specific pair-of-pants decomposition of 
the $(n+3)$-punctured sphere where only the primary field with Liouville momentum $\xi_i$
and dimension $\Delta_{\xi_i}$ plus its descendants propagate in the $i$-th internal line (see Fig. \ref{figconf1}).
Furthermore, we take the degenerate limit in which the $(n+3)$-punctured sphere reduces to a sequence of
$(n+1)$ 3-punctured spheres connected by $n$ long thin tubes with sewing parameters $q_i$, as shown in Fig.~\ref{fissph}. If we denote the local coordinates on each 3-sphere by $w_i$, then the sewing procedure requires that 
\be
\frac{w_{i+1}}{w_{i}} = q_i \quad\mbox{with}~~|q_i|<0~.
\label{sew}
\ee
In the local coordinates of each sphere, the punctures are located at $(0, 1, \infty)$; in 
particular all the unsewn external punctures are at $1$ (except for the first and the last one which are at $0$ and $\infty$ respectively). However, if we use the local coordinates of the last sphere as coordinates for the global surface, 
the sewing relations (\ref{sew}) imply that the external punctures of the first $n$ spheres are at 
\be
t_i = \prod_{j=i}^{n}q_j\qquad \text{for}\qquad i\in \{1, \ldots n\}~.
\label{t1toN}
\ee
This is precisely the same relation we found in (\ref{Calfa}).

When written in terms of the $t_i$'s, the conformal block (\ref{ccblock}) becomes \cite{Alba:2009ya}
\begin{equation}
\label{ccblockexp}
\Big\langle 
V_{\alpha_0}(0)\,
\prod_{i=1}^{n} V_{\alpha_i}(t_i)\,V_{\alpha_{n+1}}(1)\,V_{\alpha_{n+2}}(\infty) \Big\rangle_{\{\xi_1,\ldots,\xi_n\}} =
\mathcal{N}\,\mathcal{B}(t_i,\Delta_{\alpha_i},\Delta_{\xi_i})
\end{equation}
where the prefactor
\begin{equation}
\begin{aligned}
\mathcal{N}&= t_1^{-\Delta_{\alpha_0}-\Delta_{\alpha_1}+\Delta_{\xi_1}}\,\prod_{i=2}^n 
t_i^{-\Delta_{\xi_{i-1}}-\Delta_{\alpha_i}+\Delta_{\xi_{i}}}
\,=\,t_1^{-\Delta_{\alpha_0}}\,\prod_{i=i}^n 
t_i^{-\Delta_{\alpha_i}}q_i^{\Delta_{\xi_i}}
\end{aligned}
\label{N}
\end{equation}
originates from the conformal transformations that move the vertices $V_{\alpha_i}$ from 1 to $t_i$, while
$\mathcal{B}(t_i,\Delta_{\alpha_i},\Delta_{\xi_i})$ contains all other relevant information,
including the structure function coefficients and the contribution of all descendants in the internal legs.

According to \cite{Alday:2009aq}, it is possible to establish a correspondence between the conformal block (\ref{ccblockexp}) 
and the partition function of the $\epsilon$-deformed $\mathrm{SU}(2)^n$ quiver theory. To do so, one has to
identify $q_i$ with the gauge coupling of the $i$-th group factor, set
\begin{equation}
Q=\frac{\eu+\ed}{\sqrt{\eu\ed}}\,,
\label{Q}
\end{equation} 
and choose the Liouville momenta as follows:
\begin{equation}
\begin{aligned}
\alpha_0&=\frac{Q}{2}+\frac{m_1-m_2}{2\sqrt{\eu\ed}}~,\qquad
\alpha_1=\frac{Q}{2}+\frac{m_1+m_2}{2\sqrt{\eu\ed}}~,\\
\alpha_i&=\frac{Q}{2}-\frac{m_{i-1,i}}{\sqrt{\eu\ed}}\qquad\mbox{for}~i=2,\ldots,n~,\\
\xi_{i}&=\frac{Q}{2}-\frac{a_{i}}{\sqrt{\eu\ed}}\,\qquad\mbox{for}~i=1,\ldots,n~,\\
\alpha_{n+1}&=\frac{Q}{2}-\frac{m_3+m_4}{2\sqrt{\eu\ed}}~,\qquad
\alpha_{n+2}=\frac{Q}{2}-\frac{m_3-m_4}{2\sqrt{\eu\ed}}~,
\end{aligned}
\label{alphas}
\end{equation}
where the $m$'s are the fundamental or bi-fundamental masses of the matter hypermultiplets as discussed in the previous
sections, and $a_i$ is the vacuum expectation value of the adjoint scalar of the $i$-th gauge group.
{From} (\ref{deltaalpha}) and (\ref{alphas}) one can check that the 
conformal dimensions of the various operators are
\begin{equation}
\begin{aligned}
\Delta_{\alpha_0}&=\frac{(\eu+\ed)^2-(m_1-m_2)^2}{4\eu\ed}~,\qquad
\Delta_{\alpha_1}=\frac{(\eu+\ed)^2-(m_1+m_2)^2}{4\eu\ed}~,\\
\Delta_{\alpha_i}&=\frac{(\eu+\ed)^2-4m_{i-1,i}^2}{4\eu\ed}\qquad\mbox{for}~i=2,\ldots,n~,\\
\Delta_{\xi_{i}}&=\frac{(\eu+\ed)^2-4 a_{i}^2}{4\eu\ed}~~\quad\qquad\mbox{for}~i=1,\ldots,n~,\\\
\Delta_{\alpha_{n+1}}&=\frac{(\eu+\ed)^2-(m_3+m_4)^2}{4\eu\ed}~,\qquad
\Delta_{\alpha_{n+2}}=\frac{(\eu+\ed)^2-(m_3-m_4)^2}{4\eu\ed}~.
\end{aligned}
\label{Deltas}
\end{equation}
The remarkable observation of \cite{Alday:2009aq} is that%
\footnote{In our subsequent analysis we ignore the structure function coefficients in the 
conformal block $\mathcal{B}$. These are related to the 1-loop contribution to the prepotential while our focus is 
the instanton part.}
\begin{equation}
\mathcal{B}(t_i,\Delta_{\alpha_i},\Delta_{\xi_i}) 
= Z_{\mathrm{U(1)}}~\rme^{-\frac{F_{\mathrm{inst}}}{\epsilon_1 \epsilon_2}}~,
\label{AGTrel}
\end{equation}
where $F_{\mathrm{inst}}$ is the Nekrasov instanton prepotential and $Z_{\mathrm{U(1)}}$ ensures the correct decoupling of the U(1) factors. This U(1) contribution can be explicitly
computed (see for example \cite{Alba:2009ya}) and the result is
\begin{equation}
Z_{\mathrm{U(1)}}= 
\prod_{i=1}^n \prod_{j=i+1}^{n+1} \left(1 - \frac{t_i}{t_j}\right)^{-2\alpha_i(Q-\alpha_{j})}
= 
\prod_{i=1}^n \prod_{j=i+1}^{n+1} \left(1 - q_i\ldots q_{j-1}\right)^{-2\alpha_i(Q-\alpha_{j})}~.
\label{Zu1}
\end{equation}
The structure of these U(1) terms is actually quite simple: each factor in (\ref{Zu1}) can be associated to a connected 
subdiagram with four legs that is obtained by grouping together adjacent nodes of the quiver; the Liouville momenta
of the two resulting inner legs determine the exponent \cite{Alday:2009aq}.
For example, for $n=1$ we have just one diagram with one node and coupling constant
$q$; its inner legs carry momenta $\alpha_1$ and $\alpha_2$, and the corresponding U(1) factor is
\begin{equation}
(1-q)^{-2\alpha_1(Q-\alpha_2)}~.
\label{u11}
\end{equation}
For $n=2$ we have a subdiagram corresponding to the first node with coupling constant $q_1$ and inner legs
with momenta $\alpha_1$ and $\alpha_2$; a subdiagram with coupling constant $q_2$ and inner legs carrying
momenta $\alpha_2$ and $\alpha_3$, and finally a diagram with the two nodes combined, which has coupling $q_1q_2$
and inner legs with momenta $\alpha_1$ and $\alpha_3$. Thus the U(1) dressing factor is
\begin{equation}
(1-q_1)^{-2\alpha_1(Q-\alpha_2)}\,(1-q_2)^{-2\alpha_2(Q-\alpha_3)}\,(1-q_1q_2)^{-2\alpha_1(Q-\alpha_3)}~.
\label{u12}
\end{equation}
This structure, which can be easily generalized to higher values of $n$, bears a clear resemblance with that of the symmetry factors
introduced in Sections~\ref{secn:Nf4} and \ref{secn:quiver} in the redefinition of $\widehat{F}$ (see in particular (\ref{Fhat1})
and (\ref{Fquiver})). In fact, the U(1) terms (\ref{Zu1}) can be considered as the proper 
generalization in the $\epsilon$-deformed theory
of the symmetry factors discussed in the previous sections.
Finally, combining (\ref{ccblockexp}) and (\ref{AGTrel}), we can write
\begin{equation}
\label{tildeF1}
\Big\langle 
V_{\alpha_0}(0)\,
\prod_{i=1}^{n} V_{\alpha_i}(t_i)\,V_{\alpha_{n+1}}(1)\,V_{\alpha_{n+2}}(\infty) \Big\rangle_{\{\xi_1,\ldots,\xi_n\}} =
\rme^{-\frac{\widetilde{F}(\epsilon)}{\eu\ed}}
\end{equation}
where
\begin{equation}
\widetilde{F}(\epsilon)=-\eu\ed\log\mathcal{N}-\eu\ed\log Z_{\mathrm{U}(1)}+F_{\mathrm{inst}}~.
\label{tildeF2}
\end{equation}

\subsection{The UV curve}
The 2-dimensional Liouville theory also contains information about the SW curve 
of the 4-dimensional quiver gauge theory and its quantum deformation. To see this 
let us consider the normalized conformal block (\ref{ccblockexp}) with the insertion of
the energy momentum tensor, namely%
\footnote{From now on we simplify the notation by omitting the subscript ${\{\xi_1,\ldots,\xi_n\}}$ in the correlators.}
\begin{equation}
\phi^\epsilon_2(z) = 
\frac{\Big\langle V_{\alpha_{0}}(0)\prod_{i=1}^{n}\!V_{\alpha_i}(t_i)\,T(z)\,
V_{\alpha_{n+1}}(1) V_{\alpha_{n+2}}(\infty)\Big\rangle}
{\Big\langle V_{\alpha_{0}}(0)\prod_{i=1}^{n}\!V_{\alpha_i}(t_i)
V_{\alpha_{n+1}}(1) V_{\alpha_{n+2}}(\infty)\Big\rangle}
\end{equation}
with $|z|<1$.
As shown in Appendix~\ref{appT}, using the conformal Ward identities it is possible to rewrite $\phi^\epsilon_2(z)$ as
\begin{equation}
\begin{aligned}
\phi^\epsilon_2(z)=
&\,\frac{\Delta_{\alpha_0}}{z^2}+ \sum_{i=1}^{n}\frac{\Delta_{\alpha_i}}{(z-t_i)^2}+
\frac{\Delta_{\alpha_{n+1}}}{(z-1)^2}- 
\frac{\Delta_{\alpha_0}+\sum_{i=1}^{n}\Delta_{\alpha_i}+ \Delta_{\alpha_{n+1}} - \Delta_{\alpha_{n+2}}}{z(z-1)} \\
&+
\sum_{i=1}^{n}\frac{t_i(t_i-1)}{z(z-1)(z-t_i)}
\frac{\p}{\p t_i} \log\Big\langle V_{\alpha_{0}}(0)\prod_{i=1}^{n}\!V_{\alpha_i}(t_i)
V_{\alpha_{n+1}}(1) V_{\alpha_{n+2}}(\infty)\Big\rangle~.
\end{aligned}
\label{phi25}
\end{equation}
All terms on the right hand side of this equation are proportional to $1/(\eu\ed)$ since both the conformal dimensions $\Delta$'s and the logarithm of the conformal block scale in that manner. Thus the following limit 
\be
\label{classquantSW}
\lim_{\eu,\ed\rightarrow 0}\big[-\epsilon_1\epsilon_2\,\phi^\epsilon_2(z)\big] \,\equiv\, \phi_2(z)
\ee
is well-defined and non-singular.
In this limit only the mass dependent terms of the conformal weights contribute so that one finds
\begin{equation}
\begin{aligned}
\phi_2(z)=
&\,\frac{(m_1-m_2)^2}{4z^2}+ \frac{(m_1+m_2)^2}{4(z-t_1)^2}
+ \sum_{i=2}^{n} \frac{m^2_{i-1,i}}{(z-t_i)^2} + \frac{(m_3+m_4)^2}{4(z-1)^2} \\
&-\frac{m_1^2+m_2^2 +2 m_3 m_4+2 \sum_{i=2}^{n} m^2_{i-1,i}}{2z(z-1)}+
\sum_{i=1}^{n}
\frac{t_i(t_i-1)}{z(z-1)(z-t_i)}
\frac{\p {\widetilde F}}{\p t_i}
\end{aligned}
\label{ph2}
\end{equation}
where
\begin{equation}
\label{Ftilde0}
\widetilde{F}=\lim_{\eu,\ed \rightarrow 0}\widetilde{F}(\epsilon)~.
\end{equation}
$\phi_2(z)$ has the same form of $x^2(z)$ appearing in the expression of the
SW curve of the quiver theories described in the previous sections (see for example (\ref{curveSU2f}) or (\ref{curvequiverf})). 
Indeed the mass terms are exactly the ones needed to produce the correct residues of the SW differential 
and coincide with those we have written for the single node and the two-node quivers in Sections~\ref{secn:Nf4} and 
\ref{secn:quiver}. Also the other terms have the right structure, and thus what remains to be checked is whether
the function $\widetilde{F}$ in (\ref{ph2}) coincides with the analogous quantity appearing in the SW curve.
We now do this check in the three cases we have analyzed in more detail.

\subsubsection*{$\bullet$ The SU(2) theory with $N_f=4$}
For the SU(2) theory with $N_f=4$ things are particularly simple, since in this case there is only a non-trivial 
puncture at $t_1=q$ and $\widetilde{F}$ defined in (\ref{Ftilde0})
becomes
\be
\widetilde{F} = a^2 \log q - \frac{1}{2} (m_1^2 +m_2^2)\log q  +\frac{1}{2} (m_1+m_2)(m_3+m_4) \log (1-q) +
F_{\text{inst}}~.
\label{tf1}
\ee
Using (\ref{Fhat}) and (\ref{Fhat1}), one can immediately see that this agrees with the function $\widetilde{F}$
appearing in the SW curve (\ref{curveSU2f}).

\subsubsection*{$\bullet$ The SU(2)$\,\times\,$SU(2) quiver theory}
In the 2-node quiver there are two non-trivial punctures. In the above discussion 
we have located them at $t_1=q_1q_2$ and $t_2=q_2$, while in the
curve derivation of Section~\ref{secn:quiver} we have considered a different (though completely equivalent) 
configuration with punctures at $t_1=q_1$ and $t_2=1/q_2$. Thus, before comparing we have to make the
appropriate changes in the prefactor $\mathcal N$ which, being directly connected to the factorization of the conformal
block in pair-of-pants diagrams, crucially depends on where the non-trivial punctures are located. If we set the punctures at
$t_1=q_1$ and $t_2=1/q_2$, we have to use
\begin{equation}
\mathcal{N}= q_1^{-\Delta_{\alpha_0}-\Delta_{\alpha_1}+\Delta_{\xi_1}}\,
q_2^{\Delta_{\xi_2}+\Delta_{\alpha_2}-\Delta_{\alpha_3}}~.
\label{N2}
\end{equation}
The corresponding expression for $\widetilde{F}$ is then
\begin{equation}
\begin{aligned}
\widetilde{F} = &\,a_1^2 \log q_1 +a_2^2 \log q_2 - \frac{1}{2} (m_1^2 +m_2^2)\log q_1  
+m_3m_4\log q_2   \\
&+m_{12}(m_1+m_2) \log(1-q_1)-m_{12}(m_3+m_4) \log(1-q_2) \\
&+\frac{1}{2} (m_1+m_2)(m_3+m_4) \log (1-q_1q_2) +F_{\text{inst}}  
~,
\end{aligned}
\label{tf2}
\end{equation}
which exactly matches the one appearing in the M-theory derivation of the SW curve, as one can see  using (\ref{Fhatquiver}) 
and (\ref{Fquiver}). This same result can also be obtained from the general expression (\ref{ph2}) if we notice that 
under the change of variables that maps $(q_1q_2, q_2, 1)$ to $(q_1, 1, 1/q_2)$,
the term of $\phi_2(z)$ proportional to $1/(z(z-1)$ produces an extra contribution to $\widetilde{F}$ modifying its
expression and leading to (\ref{tf2}).

\subsubsection*{$\bullet$ The conformal $\mathrm{SU}(2)^n$ quiver}
When all masses are zero, ${\widetilde F}$ in (\ref{Ftilde0}) is simply
\be
\widetilde{F} = \sum_{i=1}^N a_i^2\, \log q_i + F_{\text{inst}}  \,.
\ee
Up to 1-loop $t$-independent contributions, this is precisely the prepotential $F$ of 
the conformal quiver gauge theory, and thus the corresponding SW curve can be written as
\begin{equation}
\phi_2(z)= 
\sum_{i=1}^{n}
\frac{t_i(t_i-1)}{z(z-1)(z-t_i)}
\frac{\p {F}}{\p t_i}~,
\label{ph20}
\end{equation}
confirming in this case the direct identification of the residues at 
$t_i$ with the derivatives of the gauge theory prepotential \cite{Marshakov:2013lga, Gavrylenko:2013dba}.
We can therefore say that the AGT correspondence provides the analogue of the Matone relations 
\cite{Matone:1995rx} for the quiver gauge  theory.
One can go even further and map the curve (\ref{ph20}) to that in \eqref{curvequiver} obtained using the M-theory analysis,
thus finding the explicit relation between the Coulomb parameters $u_i$ appearing there 
and the $t_i$-derivatives of the prepotential.

\section{The quiver prepotential from null-vector decoupling}
\label{prepotAGT}

We now present the derivation of the $\Omega$-deformed prepotential for the $\mathrm{SU}(2)^n$ quiver model
in the NS limit \cite{Nekrasov:2009rc} using a null-vector decoupling equation in the Liouville theory 
introduced in the previous section.
The observable we consider is the conformal block obtained by deforming (\ref{ccblockexp}) 
with the insertion of the degenerate field $\Phi_{2,1}(z)$ of the Virasoro algebra \cite{Alday:2009fs}, namely
\be
\Psi(z) = \Big\langle 
V_{\alpha_0}(0)\,\prod_{i=1}^{n} V_{\alpha_i}(t_i)\,\Phi_{2,1}(z) \,
V_{\alpha_{n+1}}(1)\,V_{\alpha_{n+2}}(\infty) \Big\rangle_{\{\xi_1,\ldots,\xi_n\}}
\label{psi}
\ee
with $|z|<1$.
The degenerate field $\Phi_{2,1}$ has conformal dimension 
\be
\Delta_{2,1}=-\frac{1}{2} - \frac{3}{4}\, \frac{\ed}{\eu}
\label{del}
\ee
and satisfies the null-vector condition
\be
\frac{\eu}{\ed}\,\frac{d^2\Phi_{2,1}(z)}{dz^2}\, +\, :\! T(z) \Phi_{2,1}(z)\!:\,= 0~.
\label{nuv}
\ee
This condition implies that $\Psi(z)$ obeys a second order differential equation that can be obtained 
from the conformal Ward identities as discussed in Appendix \ref{appT}.
If we normalize the correlator (\ref{psi}) with the unperturbed one (\ref{tildeF1}) and write
\begin{equation}
\Psi(z) = \rme^{-\frac{\widetilde{F}(\epsilon)}{\eu\ed}}\,\Phi(z)~,
\label{surfaceop}
\end{equation}
then the differential equation for $\Psi(z)$ turns into the following differential equation for $\Phi(z)$
\begin{equation}
\begin{aligned}
\Bigg[\frac{\eu}{\ed}\frac{\p^2}{\p z^2}- \frac{2z-1}{z(z-1)}\frac{\p}{\p z}+
\sum_{i=1}^{n}\Big(\frac{t_i(t_i-1)}{z(z-1)(z-t_{i})}
\frac{\p}{\p t_i}-\frac{1}{\eu\ed}\,\frac{t_i(t_i-1)}{z(z-1)(z-t_{i})}
\frac{\p {\widetilde F(\epsilon)}}{\p t_i}\Big)+\frac{ \Delta_{\alpha_0} }{z^2}\\
+\sum_{i=1}^n\frac{\Delta_{\alpha_i}}{(z-t_i)^2} + \frac{\Delta_{\alpha_{n+1}}}{(z-1)^2}
-\frac{\Delta_{\alpha_0}+\sum_{i=1}^{n}\Delta_{\alpha_i}+\Delta_{2,1} +\Delta_{\alpha_{n+1}} 
-  \Delta_{\alpha_{n+2}}} {z(z-1)} \Bigg]\,
\Phi(z)~=0~.
\end{aligned}
\label{phi1}
\end{equation}
This equation is well-suited to take the NS limit \cite{Nekrasov:2009rc} in which $\ed\to 0$ with $\eu\not=0$, provided 
we assume that
\begin{equation}
\Phi(z)= \rme^{-\frac{W(z)}{\epsilon_1}}
\label{wop}
\end{equation}
where $W(z)$ is regular in $\epsilon_1$. Multiplying (\ref{phi1}) by $(-\eu\ed)$ and sending $\ed$ to zero, the differential
equation simplifies in a few ways: the linear derivatives in $z$ and $t_i$ drop out along with the term proportional to the
conformal dimension $\Delta_{2,1}$ of the degenerate field. Furthermore, in the NS limit 
the generalized prepotential $\widetilde{F}(\epsilon)$ in (\ref{tildeF2}) becomes
\begin{equation}
\widetilde{F}(\epsilon)\,\to\,\widetilde{F}+\eu\widetilde{F}^{(1)}+\eu^2\widetilde{F}^{(2)}
\label{tildeFeps}
\end{equation}
where the $\eu$ corrections arise from the explicit $\epsilon$-dependence of the prefactors $\mathcal{N}$ and $Z_{\mathrm{U}(1)}$.
Since the terms proportional to the conformal dimensions $\Delta_{\alpha_i}$ yield contributions at most of order $\eu^2$,
in the end we obtain the Schroedinger-type differential equation:
\begin{equation}
\Big(-\epsilon_1^2\frac{d^2}{d z^2}+V(z,\epsilon_1)\Big)\Phi(z)=0~,
\label{difph}
\end{equation}
where
\be
V(z, \epsilon_1) = V^{(0)}(z)+\eu\,V^{(1)}(z)+\eu^2\,V^{(2)}(z)
\label{potential}
\ee
with
\begin{eqnarray}
V^{(0)}(z)&=&\phi_2(z)~,\nonumber \\
V^{(1)}(z)&=&\sum_{i=1}^{n}\,\frac{t_i(t_i-1)}{z(z-1)(z-t_{i})}\frac{\p \widetilde{F}^{(1)}}{\p t_i}~,\label{pot2}\\
V^{(2)}(z)&=&-\frac{1}{4z^2}-\sum_{i=1}^{n}\frac{1}{4(z-t_i)^2}-\frac{1}{4(z-1)^2}+
\frac{n+1}{4z(z-1)}+\sum_{i=1}^{n}\,\frac{t_i(t_i-1)}{z(z-1)(z-t_{i})}\frac{\p \widetilde{F}^{(2)}}{\p t_i}~.\nonumber
\end{eqnarray}
Note that $V^{(0)}$ is the SW curve of the undeformed theory.
To solve (\ref{difph}) we make a WKB-like ansatz for $\Phi(z)$ writing
\be
W(z) = \int^z \!\! P(z',\epsilon_1)\,dz'~,
\label{Phi2}
\ee
and then expand $P$ in powers of $\epsilon_1$
\be
P(z,\epsilon_1) = \sum_{n=0}^{\infty}\epsilon_1^n \, P^{(n)}(z)~.
\label{exph}
\ee
Substituting in (\ref{difph}) we find 
\be
-P(z,\epsilon_1)^2+\epsilon_1 \frac{dP(z,\epsilon_1)}{dz}+V(z,\epsilon_1) = 0 ~,	
\label{diffp}
\ee
which in turn can be solved perturbatively in $\eu$. The first few terms are
\begin{subequations}
\begin{align}
P^{(0)}(z) &= \sqrt{\phi_2(z)}~, \label{P0}\\
P^{(1)}(z) &= \frac{1}{2}\frac{d}{dz}\log P^{(0)}(z) +\frac{V^{(1)}(z)}{2P^{(0)}(z)}~, \label{P1}\\
P^{(2)}(z) &= \frac{{P^{(1)}}'(z)-{P^{(1)}}^2(z)}{2P^{(0)}(z)} +\frac{V^{(2)}(z)}{2P^{(0)}(z)}~,\label{P2}
\end{align}
\label{Pks}
\end{subequations}
and so on. Since $P^{(0)}(z)dz$ is simply the SW differential of the undeformed theory, it
is more than natural to define the deformed SW differential as
\be
\lambda(\epsilon_1) = P(z, \epsilon_1)\, dz ~.
\label{lsw}
\ee
The periods of $\lambda(\eu)$ along the $\alpha_i$-cycles can then be interpreted as the $a_i$'s in the
deformed theory, namely
\be
a_i= \frac{1}{2\pi\ii}\oint_{\alpha_i} \!\lambda(\epsilon_1) = \sum_{n=0}^\infty \eu^n\,a_i^{(n)}
\qquad\mbox{with}\qquad a_i^{(n)}=\frac{1}{2\pi\ii}\oint_{\alpha_i} \! P^{(n)}(z)\,dz~.
\label{aep}
\ee
Clearly the above integrals depend on the prepotential $F$ and its $t_i$-derivatives; therefore we can use this information to fix
the $\eu$-dependence of $F$ by demanding consistency, namely by choosing $a_i$'s as independent variables and
thus taking them to be constant. Even if it does not seem so at first sight, this procedure is fully equivalent to that used for instance in 
\cite{KashaniPoor:2012wb,Kashani-Poor:2013oza} to obtain the deformed prepotential for the $\mathcal{N}=2^*$ SU(2) theory
or the $\mathcal{N}=2$ SU(2) theory with $N_f=4$. Indeed, also in our case the periods $a_i$ which determine the
monodromy properties of the wave function $\Phi(z)$, are constant, since the $\eu$ (and $q_i$) dependence of the
prepotential is fixed precisely to achieve this goal. It is remarkable that the prepotential obtained in this way agrees
with the one computed using localization methods in the NS limit. 

\subsection{The prepotential from deformed period integrals}
We now illustrate the above procedure, focusing on the examples considered in the previous sections.

\subsubsection*{$\bullet$ The SU$(2)$ theory with $N_f=4$}

When $n=1$, the $\eu$-terms of the potential in the Schroedinger-type equation are
\begin{equation}
 \begin{aligned}
  V^{(1)}(z) &= q\,\frac{(m_1+m_2+m_3+m_4)}{2z(z-q)(z-1)}~, \\
  V^{(2)}(z) &= -\frac{1}{4z^2}-\frac{1}{4(z - q)^2} - \frac{1}{4(z - 1)^2}
   +\frac{1}{2z (z - 1)} +\frac{3q-1}{4z(z-1)(z-q)}~,
 \end{aligned}
\label{potsu2}
\end{equation}
while $V^{(0)}(z)$ is given by the SW curve $\phi_2(z)$. 

To proceed we choose the same mass configuration that we have discussed in Section~\ref{secn:Nf4}, namely
$m_1=m_2=m$, $m_3=m_4=M$, which allows us to write the curve in the factorized form
\be
\phi_2(z) =\frac{C(e_2-z)(z-e_3)}{z(z-1)^2(z-q)^2} ~.
\label{phduefac}
\ee
Here the roots $e_2$ and $e_3$ and the constant $C$ are the same as in \eqref{Ccoeff} and \eqref{e2e3}, but they are
expressed in terms of the prepotential instead of the Coulomb modulus $u$. 

At order $\eu^0$, the period has already been calculated in Section~\ref{secn:Nf4} (see \eqref{a1inst}); 
expressing it in terms of $U\equiv q\,\p F/\p q$, we have (up to 2 instantons)
\begin{eqnarray}
a^{(0)} &=& \sqrt{U} \,\bigg[
1-\frac{q}{4} \Big(1+\frac{(m^2+4mM+M^2)}{U}+\frac{m^2M^2}{U^2}\Big)\nonumber\\
&&~-\frac{q^2}{64}
\Big(7 \!+\!\frac{14m^2+48mM+14M^2}{U}\!+\!
\frac{3m^4+16m^3M+60m^2M^2 + 16mM^3 +3M^4}{U^2}\nonumber\\
&&~+\frac{6m^2M^2(m^2+8mM+M^2)}{U^3}+\frac{15m^4M^4}{U^4}
\Big)+\ldots\bigg]~.\label{a0}
\end{eqnarray}
At order $\eu$ we have instead
\be
a^{(1)}=\frac{1}{2\pi\ii}\oint_{\alpha} P^{(1)}(z) \, dz =-q\,\frac{m+M}{2\pi\sqrt{C}}\int_0^{e_2}
 \frac{dz}{\sqrt{z(e_2-z)(e_3-z)}}
\label{a111}
\ee
where in the second step we used (\ref{P1}) and discarded the total derivative term.
This integral can be evaluated as a power series and, up to two instantons, we find
\be
a^{(1)} = -q\,\frac{m+M}{2\sqrt{U}} \bigg[
 1 +q \,\frac{3U^2+U(m^2+4mM+M^2)+3m^2M^2}{4U^2}+ \ldots \bigg]~.
\ee
Using the formul\ae~in \eqref{Pks} iteratively, we can easily compute the order $\eu^2$ correction to the period and get
\begin{align}
a^{(2)} = &-\frac{q}{16U^{\frac{5}{2}}}\bigg[3U^2+m^2M^2+\frac{q}{8U^2}\Big(
17U^4+7U^3(3m^2+8mM+3M^2)\\
&+2U^2(m^4+20m^2M^2+M^4)-5Um^2M^2(m^2-8mM+M^2)+35m^4M^4\Big)+\ldots
\bigg]~.\notag
\end{align}
So far, we have calculated the period integral as an expansion of the form
\be
a= a^{(0)}(U) + \epsilon_1\, a^{(1)}(U) +  \epsilon_1^2\, a^{(2)}(U)  + \ldots 
\ee
We now invert this expression and determine how $U$ should depend on $\epsilon_1$ so that $a$ be a constant. We can do
this by writing
\begin{equation}
U= U^{(0)}+\eu\,U^{(1)}+\eu^2\,U^{(2)}+\ldots
\end{equation}
and demanding consistency order by order in $\eu$. Once $U$ is computed, we can obtain the deformed prepotential $F$
by integrating it with respect to (the logarithm of) $q$. 
The zeroth-order term that we get in this way clearly coincides with \eqref{Fnek}, 
while the first successive corrections are given by
\begin{equation}
\begin{aligned}
F^{(1)} &= q\,(m+M)+\frac{q^2}{2}\,(m+M)  +\ldots~,\\ 
F^{(2)} &= \frac{q}{8}\Big(3+\frac{m^2M^2}{a^4}\Big) + \frac{q^2}{128}\Big(23 -\frac{m^2 + M^2}{a^2} 
+ \frac{2m^4 + 16 m^2 M^2 +2 M^4}{a^4} \\
&\hspace{5cm} - \frac{15m^2 M^2(m^2 + M^2)}{a^6}  + \frac{21m^4 M^4}{a^8}\Big)+\ldots
\label{f22}
\end{aligned}
\end{equation}
These precisely match the microscopic results obtained from the Nekrasov partition function via localization methods. 

\subsubsection*{$\bullet$ The SU$(2)\times$ SU$(2)$ quiver theory}
When $n=2$ the Schroedinger problem is algebraically more complicated, but still doable. 
The $\eu$-corrections of the potential $V$ are
\begin{equation}
 \begin{aligned}
  V^{(1)}(z) =& \frac{(m_1+m_2+m_3+m_4)q_1q_2}{2z(z-1)(z-q_1q_2)} +
\frac{(m_1+m_2+2m_{12})q_1q_2}{2z(z-q_2)(z-q_1q_2)}+\frac{(m_3+m_4-2m_{12})q_2}{2z(z-1)(z-q_2)}~,\\
V^{(2)}(z) =& -\frac{1}{4z^2}- \frac{1}{4(z - q_1q_2)^2} -\frac{1}{4(z -q_2)^2} -\frac{1}{4(z - 1)^2}
   +\frac{3}{4z (z - 1)} \\
&\qquad\quad-\frac{\eta_1}{z(z-1)(z-q_2)}-\frac{\eta_2}{z(z-q_2)(z-q_1q_2)} 
\end{aligned}
\label{potentialquiver}
\end{equation}
where
\be
\eta_1=\frac{(1 - 2 (1 + q_1) q_2 + 3 q_1 q_2^2)}{2(1 -  q_1 q_2)}~,\qquad 
\eta_2=\frac{q_2 (1 + 5 q_1^2 q_2 - 3 q_1 (1 + q_2))}{4(1 -  q_1 q_2)}~.
\label{eta}
\ee
To proceed we make the simplifying mass choices discussed in Section~\ref{secn:quiver}, see (\ref{massAB}).

\paragraph{Case A):}
In our present conventions the SW curve takes the factorized form
\be
\phi_2(z)= \frac{-u_2(z-q_2\zeta_1)(z-q_2\widehat{\zeta})}{z(z-1)(z-q_1q_2)^2(z-q_2)}
\label{phi22}
\ee
where the various constants are exactly those appearing in \eqref{zetamassive}, 
with the $u_i$'s written in terms of the $U_i$'s using \eqref{utoF}. Furthermore, with this mass choice
the first-order term of the potential simplifies to
\be
V^{(1)}(z) = -\frac{mq_1q_2(1+q_2-2z)}{z(z-1)(z-q_2)(z-q_1q_2)} ~.
\label{v12a}
\ee
Using the same basis of $\alpha$-cycles discussed in Section~\ref{secn:quiver}, we
find that the first correction to the $a_1$-period takes the form
\begin{equation}
\begin{aligned}
a_1^{(1)} &= \frac{1}{2\pi\ii}\oint_{\alpha_1}\!\!P^{(1)}(z)\,dz
=-\frac{mq_1q_2}{2\sqrt{u_2}}\int_0^{q_2\zeta_1}\!\!\!\frac{dz}{\sqrt{z(q_2\zeta_1-z)}} \,
\frac{(1+q_2-2z)}{\sqrt{(1-z)(q_2-z)(q_2\widehat{\zeta}-z)}}~.
\end{aligned}
\label{a12a}
\end{equation}
Note that, unlike the case of the undeformed period (\ref{a1mass}), now there are no poles in the integrand and 
the integral can be done simply by expanding the second factor of (\ref{a12a}) in powers of $z$ and writing
the resulting integrals in terms of Euler $\beta$-functions. In this way we find%
\footnote{To keep the expressions compact we only exhibit the results up to 2 instantons. The calculations have been performed for higher instantons numbers as well.}
\begin{equation}
\begin{aligned}
a_1^{(1)} =&-
\frac{m q_1}{2\sqrt{U_1\phantom{|}}} \bigg[ 1 - q_1 \frac{U_1(U_2-3U_1)+m^2(3U_2-U_1)}{4U_1^2}+q_2+\ldots\bigg]~.
\end{aligned}
\label{a122}
\end{equation}
The first correction to the $a_2$ period can be similarly performed and we obtain
\be
a_2^{(1)}=-\frac{3 m\,q_1 q_2}{4 \sqrt{U_2\phantom{|}}} +\ldots ~.
\label{a222}
\ee
At order $\eu^2$ we find
\begin{eqnarray}
a_1^{(2)} &=&-\frac{q_1}{16\,U_1^{\frac{5}{2}}}\bigg[3U_1^2-m^2U_2 \!-q_2\Big(5U_1^2+m^2(U_1+U_2)\Big)\!
-\frac{q_1}{8}\Big(17U_1^2-7U_1U_2+2U_2^2\cr
&&\phantom{\Bigg|}~~+\frac{m^2(21U_1^2-24U_1U_2-5U_2^2)}{U_1}+\frac{m^4(2U_1^2-25U_1U_2+35U_2^2)}{U_1^2} \Big) +\ldots\bigg]~,\\
a_2^{(2)}&=&-\frac{q_2}{16 \sqrt{U_2\phantom{|}}}\bigg[3 +5 q_1+q_2\frac{2U_1^2-7U_1U_2+17U_2^2}{8U_1^2}+\ldots\bigg]~.
\end{eqnarray}
Inverting the expansion of the periods order-by-order in $\eu$, we can determine the $\eu$ dependence of $U_1$ and $U_2$.
At each order the resulting expressions turn out to be integrable and the prepotential can be recovered.
At order $\eu^0$ we get the same expression as in (\ref{Fzero}), while the corrections of order $\eu$ and $\eu^2$ are
\begin{eqnarray}
F^{(1)} &=& m\bigg(q_1+\frac{1}{2}q_1^2+q_1q_2+\ldots\bigg) \label{Fm1}~,\\
F^{(2)}&=& q_1\,\frac{3a_1^4-m^2a_2^2}{8a_1^4}+q_2\,\frac{3}{8}
+q_1q_2\,\frac{7a_1^4+m^2a_2^2}{16a_1^4}+q_2^2\,\frac{23a_2^4-a_1^2a_2^2+2a_1^4}{128a_2^4}\label{Fm2}\\
&&+q_1^2\bigg(\frac{23a_1^4-a_1^2a_2^2+2a_2^4}{128a_1^4}
-\frac{m^2(a_1^4+15a_2^4)}{128a_1^6}+\frac{m^4(2a_1^4-15a_1^2a_2^2+21a_2^4)}{128a_1^8} \bigg)+\ldots
\nonumber
\end{eqnarray}
One can check that this precisely matches the $\epsilon_1$ corrections to the prepotential obtained using Nekrasov's analysis,
thus validating the entire picture.

\paragraph{Case B):} The SW curve in this case is 
\be
\phi_2(z)=\frac{C(z-q_2\,\zeta_3)(z-q_2\,\widehat{\zeta})}{z(z-q_1q_2)(z-q_2)^2(z-1)}
\ee
where the constants are the same as in \eqref{zetamassiveM} and \eqref{discM}, provided we write
the $u_i$'s in terms of the $U_i$'s by means of \eqref{utoFM}. For this mass configuration, the first-order 
correction to the Schroedinger potential is
\be
V^{(1)}(z)=-\frac{M\,q_2\big(z(1-q_1)+q_1(1-q_2)\big)}{z(z-q_1q_2)(z-q_2)(z-1)}~,
\label{v12b}
\ee
and the $\alpha_i$-cycles are unchanged from the undeformed theory. Thus the period integrals are straightforward to perform, 
leading to the following results
\begin{eqnarray}
a_1^{(1)}&=&\frac{1}{2\pi\ii}\oint_{\alpha_1}\!\!\!P^{(1)}(z)\,dz=-\frac{Mq_2}{2\sqrt{C}}\int_0^{q_1q_2}
\!\!\frac{dz}{\sqrt{z(q_1q_2-z)}}\,
\frac{z(1-q_1)+q_1(1-q_2)}{\sqrt{(q_2\zeta_3-z)(q_2\widehat{\zeta}-z)(1-z)}}\nonumber\\
&=&-\frac{q_1M}{2\sqrt{U_1\phantom{|}}}\bigg[1+q_1\,\frac{3U_1-U_2+M^2}{4U_1}-\frac{q_2}{2}+\ldots\bigg]~.
\end{eqnarray}
At order $\eu^2$ we find
\begin{align}
a_1^{(2)}\! &= \!-\frac{q_1}{16\sqrt{U_1\phantom{|}}}\bigg[3+5q_2+q_1\,\frac{17U_1^2-7U_1U_2+2U_2^2-M^2(21U_1-4U_2)
-2M^4 }{8U_1^2}+\!\ldots\bigg]~.
\end{align}
The period integrals $a_2^{(k)}$ along the $\alpha_2$-cycle can be obtained from the above expressions by the following
symmetry operations
\be
U_1\leftrightarrow U_2~, \qquad q_1\leftrightarrow q_2~, \qquad M \leftrightarrow -M ~.
\ee
Inverting as before the map between the $a_i$'s and the $U_i$'s, and integrating with respect to the coupling constants $q_i$, 
we find that the first $\eu$-corrections to the prepotential are
\begin{equation}
\begin{aligned}
F^{(1)}&= M(q_1-q_2) +\frac{M}{2}(q_1^2-q_2^2)+ \ldots~,\\
F^{(2)}&=\frac{3(q_1+q_2)}{8}+\frac{7q_1q_2}{16} 
+ q_1^2\,\frac{23a_1^4-a_1^2a_2^2+2a_2^4-M^2(4a_1^2+a_2^2)+2M^4}{128a_1^4}\\
&\qquad+ q_2^2\frac{23a_2^4-a_1^2a_2^2+2a_1^4-M^2(4a_2^2+a_1^2)+2M^4}{128a_2^4}+\ldots~.
\end{aligned}
\end{equation}
This perfectly agrees with the Nekrasov prepotential for this mass configuration.

Combining the results for the two different mass configurations with the symmetry that exchanges the two gauge groups, 
the associated masses and coupling constants, we can therefore claim that the results following from the null-vector decoupling equation are completely consistent with the $\Omega$-deformed prepotential obtained from localization in the NS limit. 

\section{Conclusions}
\label{secn:concl}

In this paper we have considered $\mathrm{SU}(2)^n$ super-conformal linear quiver gauge theories, 
with special emphasis on the $n=1, 2$ cases, comparing two different approaches: one based on the 
analysis of the SW curves and the other based on the AGT correspondence.

Starting from the SW curves obtained from the M-theory lift of a system of NS5-D4 branes, 
we have shown how to derive efficiently the instanton expansion of the prepotential. 
We used a generalized residue prescription, along the lines suggested in \cite{Marshakov:2013lga, Gavrylenko:2013dba}, together with global symmetry considerations.
We have also shown that the cross-ratios of the branch points of the SW curve, which 
depend on the UV parameters of the theory, can be expressed in terms of $\Theta$-constants 
with period matrix $\tau_{ij}$, which encodes the IR gauge couplings, thus confirming the nice
geometric interpretation of the Nekrasov counting parameters.

We then considered the AGT correspondence, and showed that the classical SW curve encoded in this 
approach matches perfectly the one obtained via the M-theory analysis. 
Within this framework it is also possible to investigate the $\Omega$-deformed quiver 
theory, at least in the NS limit where the periods $a_i$ can be written as integrals of a deformed SW differential. 
{From} this expression we were able to extract the
expansion of  the prepotential to second order in the deformation parameter, which agrees 
completely with the microscopic evaluation of the prepotential \`a la Nekrasov. 
It is clear that our methods can be generalized in a straightforward manner to higher orders, and indeed we were able to push the calculations up to order four in a few cases. 

To compare the results obtained in the two approaches, the key point is to express all parameters in terms of gauge theory data, which are the masses and the bare coupling constants associated to each gauge group. 
In the M-theory approach, the parameters are geometric, and are related to the 
positions of the constituent branes that engineer the quiver gauge theory. 
In the Liouville theory, the natural parameters are the central charge of the CFT and the Liouville momenta of the primary operators involved in the AGT correspondence. 
After working out the detailed map between the various parameters, 
we could correctly identify the quantum mechanical system that governs the infrared dynamics of the SU$(2)^n$ quiver gauge theory in the NS limit, for the cases  $n=1, 2$. This in turn allowed us to calculate the prepotential of the gauge theory.

There are many directions that deserve to be explored. 

As mentioned in the introduction, there is a very powerful approach to the study of mass deformed conformal quiver gauge theories, which uses the limit shape equations \cite{Nekrasov:2012xe, Nekrasov:2013xda}. This method does not rely on the existence of an AGT dual. It has been shown that in the NS limit, the instanton partition function of the quiver gauge theory reduces to the wave function of some quantum mechanical system. It would be very interesting to analyze those differential equations using our simple techniques to see if they prove to be efficient in calculating the prepotential of the quiver theories. 

In all the cases discussed in this paper, we focused on mass configurations such that the SW curve can be explicitly written in a factorized form. This allowed us to compute the period integrals using relatively simple integration techniques, so that the discussion could be focused on more conceptual issues. For generic masses, we will have to use more sophisticated methods to evaluate the period integrals.  

The WKB ansatz for the wave-function which we used to obtain the deformed periods in our examples, and which is
valid only in the NS limit, would clearly work for the general linear quiver 
with SU(2) gauge group factors. Since $\epsilon_1$ appears as the Planck's constant for this quantum mechanical problem, 
it would be interesting to explore the presence of contributions that are non-perturbative in 
$\epsilon_1$, and explore their possible effects on the prepotential and their interpretation in the gauge theory (see \cite{Basar:2015xna} 
and references therein for some interesting recent work using exact WKB methods).

For conformal quiver theories with SU($N$) gauge groups, the AGT dual is the Toda CFT, which has a
$W_N$ symmetry. It would be interesting to study the null-vector decoupling equations in such theories. 
In the NS limit, the resulting differential equation will be of higher order 
and it remains to be seen if there exists a suitable WKB-type ansatz for the wave function that can be used to obtain the prepotential of such quivers. 

For conformal gauge theories with a single gauge group, such as $\mathrm{SU}(2)$ theory with $N_f=4$ and the $\mathcal{N}=2^*$ theory, there has been tremendous progress in resumming the instanton 
contributions and writing the prepotential in terms of quasi-modular functions of the coupling constant. 
This has been done both from the gauge theory perspective 
\cite{Billo:2013fi}-\nocite{Billo:2013jba}\cite{Billo:2014bja} as well as from the Liouville CFT perspective 
\cite{KashaniPoor:2012wb, Kashani-Poor:2013oza, Kashani-Poor:2014mua}. 
It would be interesting to see if similar resummations are possible for the general linear quiver.
A related question would be to understand and interpret our results in the context of topological 
string theory. Both these directions require the ability to describe  
$\Omega$-deformations beyond the NS limit $\epsilon_2= 0$, since the quantity 
$\sqrt{\epsilon_1\epsilon_2}$ plays the r\^ole of the string coupling constant for the related topological string theories. Moreover, the holomorphic/modular anomaly equation (which allows the resummation of instanton contributions in terms of suitable modular quantities) has its roots in a quantization of the moduli space for which $\epsilon_1\epsilon_2$ represents the Planck constant.
We hope to pursue some of these directions in the future.

\vskip 1.5cm
\noindent {\large {\bf Acknowledgments}}
\vskip 0.2cm
We would like to thank Dileep Jatkar, Madhusudhan Raman, Ashoke Sen and Jan Troost for useful discussions. 
The work of M.B., M.F. and A.L. is partially supported  by the Compagnia di San Paolo 
contract ``MAST: Modern Applications of String Theory'' TO-Call3-2012-0088.

\vskip 1cm

\appendix
\section{Nekrasov prepotential for quiver gauge theories}
\label{Nekrasov}

We consider $\mathcal{N}=2$ quiver theories with a gauge group of the form $\prod_i \mathrm{SU}(N_i)$, and
a matter content specified by the numbers $\{n_i\}$ of hypermultiplets in the fundamental 
representation of $\mathrm{SU}(N_i)$, and by the numbers $\{c_{ij}\}$ of 
bi-fundamental hypermultiplets which are fundamental under $\mathrm{SU}(N_i)$ and anti-fundamental under $\mathrm{SU}(N_j)$. 
The $\beta$-function coefficient for each $\mathrm{SU}(N_i)$ factor is given by
\begin{equation}
\label{betaa}
\beta_i = -2N_i +\sum_jN_j(c_{ij}+c_{ji})+ n_j ~.
\end{equation}
We restrict our attention to conformal theories such that the $\beta$-function vanishes for every node.
The basic quantity of interest is the multi-instanton partition function which, using 
localization \cite{Nekrasov:2002qd, Nekrasov:2003rj}, reduces to
\begin{equation}
\label{Zinstdef}
Z_{\mathrm{inst}}=\sum_{k_i}\,\int \prod_{i}\frac{q_i^{k_i}}{k_i!}\prod_{I_i=1}^{k_i}
\frac{d\chi_{I_i}}{2\pi i}\ z^{\mathrm{quiver}}_{\{k_i\}}~.
\end{equation}
Here we adopt the same conventions used in \cite{Billo:2012st} (see in particular Appendix A). 
For instance, in the $(k_1, k_2)$ instanton sector of a 2-node quiver theory
we have
\begin{equation}
z^{\mathrm{quiver}}_{k_1, k_2} = z_{k_1}^{\mathrm{gauge}}\, z_{k_2}^{\mathrm{gauge}}
\, z_{k_1}^{\mathrm{fund}} z_{k_2}^{\mathrm{fund}}\,
 z_{k_1,k_2}^{\mathrm{bi-fund}}~.
\end{equation}
where, in a rather obvious notation, the various factors represent the contributions of the different multiplets. 
As shown in \cite{Nekrasov:2002qd, Nekrasov:2003rj} (see also \cite{Bruzzo:2002xf,Fucito:2004gi}),
the configurations of $\chi_{I_i}$ which contribute to the integrals in (\ref{Zinstdef}) can be put
in one-to-one correspondence with a set Young tableaux $Y=\{Y_{i}\}$ containing a total 
number $k=\sum_i k_i$ of boxes, and the instanton partition function can be rewritten as
\begin{equation}
\label{Zinstfinal}
Z_{\mathrm{inst}}=
1+\sum_{Y_{i}}\prod_i q_i^{|Y_i|}Z_{\{Y_i\}} ~.
\end{equation}
Here, the $1$ represents the contribution at zero instanton number, $|Y_i|$ is the total number of boxes of the $i$-th 
Young tableau. 

There is an algorithmic way to calculate the $Z_{Y_i}$'s, using the formalism of group characters, which now we briefly describe.
For a given node $i$, we introduce the characters associated to the gauge, flavour and instanton symmetries, namely:
\begin{equation}
\label{fundchar}
W_i = \sum_{u_i=1}^{N_i} \rme^{\ii a_{u_i}}
~, \quad 
W_{F,i} =\sum_{f_i=1}^{n_i} \rme^{-\ii \big(m_{f_i}+\frac{1}{2}(\epsilon_1+\epsilon_2)\big)}
~,\quad 
V_i = \sum_{I_i=1}^{k_i} \rme^{\ii\big(\chi_{I_i}-\ft{1}{2}(\epsilon_1+\epsilon_2)\big)}\,,
\end{equation}
where the $m$'s are the masses of the fundamental hypermultiplets while $\eu$ and $\ed$ are the parameters of the
$\Omega$-background \cite{Nekrasov:2002qd, Nekrasov:2003rj}.
In addition to these, we also have the characters associated to the Lorentz group, which are given by
\begin{equation}
T_1 = \rme^{\ii\epsilon_1}~,\qquad T_2 = \rme^{\ii\epsilon_2}~.
\end{equation}
For a quiver model specified by the data $\{n_i, c_{ij}\}$, the character for a given tableau $Y$ is expressed in terms of the 
fundamental characters (\ref{fundchar}) as follows:
\begin{equation}
\label{TY}
T_{Y} = \sum_{i,j} t_{ij} \,T_{ij} -T_F\,,
\end{equation}
with
\begin{equation}
\label{TY1}
\begin{aligned}
t_{ij} &= \delta_{ij}-c_{ij}\,\rme^{\ii\big(m_{ij}-\ft{1}{2}(\epsilon_1+\epsilon_2)\big)}~,\\
T_{ij}&=-V_i V^*_{j}(1-T_1)(1-T_2)+W_iV_j^*+V_iW_j^*T_1T_2~,\\
T_F&=\sum_i V_iW_{F,i}^*
\end{aligned}
\end{equation}
where $m_{ij}$ is the mass of the bi-fundamental hypermultiplets.
Notice that the combination $m_{ij}$, $\eu$ and $\ed$ that appears in $t_{ij}$ is such that a flip in the orientation 
of an arrow, which exchanges $c_{ij}$ and $c_{ji}$, can be reabsorbed in the redefinition 
$m_{ij}\leftrightarrow -m_{ji}$ to leave $Z_Y$ invariant.
In what follows, we will often use the notation $\widehat{m} = m + \frac{1}{2}(\epsilon_1+\epsilon_2)$.

We now focus on the $\mathrm{SU}(2)\times \mathrm{SU}(2)$ quiver considered in the main body of the paper. The field
content of this model is specified by $c_{12}=1$, $c_{21}=0$, $n_1=2$ and $n_2=2$. The vacuum expectation values
for the two SU(2) factors are $a_1$ and $a_2$.
Using the notation  $T_x = \rme^{\ii\,x}$, the fundamental characters (\ref{fundchar}) are given by
\begin{equation}
\begin{aligned}
&V_1= T_{a_1}\!\!\sum_{(r,s)\in Y_{a_1}}\!\!T_1^{r-1}\,T_2^{s-1} \,+\, T_{-a_1}\!\!\sum_{(r,s)\in Y_{-a_1}}\!\!T_1^{r-1}\,
T_2^{s-1} ~,\\
&V_2= T_{a_2}\!\!\sum_{(r,s)\in Y_{a_2}}\!\!T_1^{r-1}\,T_2^{s-1} \,+\, T_{-a_2}\!\!\sum_{(r,s)\in Y_{-a_2}}\!\!T_1^{r-1}\,
T_2^{s-1} ~,\\
&W_1 = T_{a_1}+T_{-a_1}~,\qquad W_{F,1} =T_{-\widehat{m}_1}+ T_{-\widehat{m}_2}~,\\
&W_2 = T_{a_2}+T_{-a_2}~,\qquad W_{F,2} =T_{-\widehat{m}_3}+ T_{-\widehat{m}_4}~.\phantom{\Big|} 
\end{aligned}
\end{equation}
For the quiver at hand, from (\ref{TY}) and (\ref{TY1}) we find
\begin{equation}
T_Y = T_{11}-T_{\widehat{m}_{12}}T_1^{-1}T_2^{-1}T_{12}+T_{22}
-V_1 \big(T_{\widehat{m}_1}+ T_{\widehat{m}_2}\big)-V_2\big(T_{\widehat{m}_3}+ T_{\widehat{m}_4}\big)~.
\end{equation}
$T_Y$ can be explicitly calculated for a given arrangement of Young tableaux $Y=\{Y_i\}$ and, from the exponents of its various
terms, one can read off the corresponding instanton partition function $Z_{\{Y_i\}}$. 
For instance, in the one-instanton sector we find
\begin{equation}
\begin{aligned}
Z_{(\Yfund\,, \bullet| \bullet,\, \bullet)} &= \frac{(2a_1+2a_2+2m_{12}+\epsilon)(2a_1-2a_2+2m_{12}+\epsilon)}{32\,
\epsilon_1\epsilon_2\,a_1(-2a_1-\epsilon)}\prod_{f=1}^2(2a_1+2m_{f}+\epsilon)~,\\
Z_{(\bullet,\,\Yfund\,| \bullet,\, \bullet)} &=\left[Z_{(\Yfund\,, \bullet| \bullet,\, \bullet)}\right]_{a_1\rightarrow -a_1}~,\\
Z_{(\bullet,\, \bullet|\Yfund\,, \bullet)} &= \frac{(2a_2+2a_1-2m_{12}+\epsilon)(2a_2-2a_1-2m_{12}+\epsilon)}{32\,
\epsilon_1\epsilon_2\,a_2(-2a_2-\epsilon)}\prod_{f=3}^4(2a_2+2m_{f}+\epsilon)~,\\
Z_{(\bullet,\, \bullet | \bullet,\,\Yfund)} &=\left[Z_{(\bullet,\, \bullet|\Yfund\,, \bullet)}\right]_{a_2\rightarrow -a_2}~,
\end{aligned}
\end{equation}
where we have defined 
\begin{equation}
\epsilon=\epsilon_1+\epsilon_2~.
\label{epsilon}
\end{equation}
The 1-instanton partition function is then given by $Z_1=q_1\,Z_{1,0}+q_2\,Z_{0,1}$, with
\begin{equation}
Z_{1,0}=Z_{(\Yfund\,, \bullet| \bullet,\, \bullet)} +Z_{(\bullet,\,\Yfund\,| \bullet,\, \bullet)}~,\quad
Z_{0,1}=Z_{(\bullet,\, \bullet|\Yfund\,, \bullet)}+Z_{(\bullet,\, \bullet | \bullet,\,\Yfund)}~.
\end{equation}
In the same way one can calculate the  higher instanton contributions, and obtain the instanton partition function
\begin{equation}
Z_{\mathrm{inst}} = 1+\sum_{k_1,k_2}Z_{k_1,k_2}\, q_1^{k_1}\, q_2^{k_2} 
\label{Zinst1}
\end{equation}
and the non-perturbative prepotential
\begin{equation}
F_{\mathrm{inst}}= -\epsilon_1\epsilon_2\,\log Z_{\mathrm{inst}}= \sum_{k_1,k_2}F_{k_1,k_2}\, q_1^{k_1}\, q_2^{k_2}~.
\label{Finst1}
\end{equation}
Below we tabulate the first few prepotential coefficients $F_{k_1,k_2}$ computed along the lines described above. We write
the results in the NS limit where we set $\epsilon_2=0$ and each $F_{k_1,k_2}$ has a further expansion of the form
\be
F_{k_1,k_2}= \sum_{n=0}^{\infty} F_{k_1,k_2}^{(n)}\, \epsilon_1^n ~.
\ee
At order $\eu^0$ we have
\begin{subequations}
\label{Fmapp}
\begin{align}
F_{1,0}^{(0)}=&~\frac{a_1^2-a_2^2}{2}+\frac{1}{2}\big(m_1m_2+2(m_1+m_2)m_{12}+m_{12}^2\big)
+\frac{m_1m_2\big(m_{12}^2-a_2^2\big)}{2a_1^2}~,
\label{F10m}\\
F_{2,0}^{(0)}=&~\frac{13a_1^4-14a_1^2a_2^2+a_2^4}{64a_1^2}+\frac{1}{64}\big(m_1^2 + 16 m_1 m_2 + m_2^2 + 
32( m_1+ m_2) m_{12}+ 18 m_{12}^2\big)\notag\\
&+\frac{m_1^2 m_2^2 + 2\big(m_1^2 + 8 m_1 m_2 + m_2^2\big)m_{12}^2+m_{12}^4 
+2a_2^2\big(m_1^2-8m_1m_2+m_2^2-m_{12}^2\big)}{64a_1^2}\notag\\
&-\frac{3\Big[2 m_1^2 m_2^2m_{12}^2+(m_1^2+m_2^2)m_{12}^4+
2a_2^2(m_1^2m_2^2-(m_1^2+m_2^2)m_{12}^2) +a_2^4(m_1^2+m_2^2)\Big]}{64a_1^4}\notag\\
&+\frac{5m_1^2m_2^2\big(m_{12}^4-2a_2^2m_{12}^2+a_2^4\big)}{64a_1^6}~,
\label{F20m}\\
F_{1,1}^{(0)}=&~\frac{a_1^2+a_2^2}{4}+\frac{1}{4}\big(m_1m_2+m_3m_4+2(m_1+m_2)(m_3+m_4)-m_{12}^2\big)\notag\\
&+\frac{m_1m_2\big(m_3m_4-m_{12}^2+a_2^2\big)}{4a_1^2}
+\frac{m_3m_4\big(m_1m_2-m_{12}^2+a_1^2\big)}{4a_2^2}
-\frac{m_1m_2m_3m_4m_{12}^2}{4a_1^2a_2^2}~.
\label{F11m}
\end{align}
\end{subequations}
At order $\eu^1$ we simply have
\begin{subequations}
\begin{align}
F^{(1)}_{1,0} &= \frac{1}{2}(m_1+m_2+2m_{12})~,\\
F^{(1)}_{2,0}&= \frac{1}{4}(m_1+m_2+2m_{12})~,\\
F^{(1)}_{1,1}&=m_1+m_2+m_3+m_4~.
\end{align}
\end{subequations}
Finally, at order $\eu^2$ we find 
\begin{subequations}
\begin{align}
F^{(2)}_{1,0}=&\,\frac{3}{8}+\frac{m_1m_2(m_{12}^2-a_2^2)}{8a_1^4}~,\\
F^{(2)}_{2,0}=& \,\frac{23}{128}-\frac{2a_2^2+m_1^2+m_2^2+2m_{12}^2}{256a_1^2}\notag\\
&+\frac{a_2^4+2a_2^2\big((m_1-m_2)^2-m_{12}^2\big) 
+m_1^2m_2^2+2m_{12}^2(m_1+m_2)^2+m_{12}^4}{64a_1^4}\notag\\
&-\frac{15\Big[a_2^4(m_1^2+m_2^4)+2a_2^2\big(m_1^2m_2^2-m_{12}^2(m_1^2+m_2^2)\big)
+2m_1^2m_2^2m_{12}^2+(m_1^2+m_2^2)m_{12}^4 \Big]}{256a_2^6}\notag\\
&+\frac{21m_1^2m_2^2(a_2^4-m_{12}^2a_2^2+m_{12}^4)}{128a_1^8}~,\\
F^{(2)}_{1,1}=&\,\frac{7}{16}+\frac{m_1m_2m_3m_4(a_1^4+a_1^2a_2^2+a_2^4)}{16a_1^4a_2^4}
+\frac{m_1m_2(a_2^2-m_{12}^2)}{16a_1^4}+\frac{m_3m_4(a_1^2-m_{12}^2)}{16a_2^4}\notag\\
&+\frac{m_1m_2m_3m_4m_{12}^2(a_1^2+a_2^2)}{16a_1^4a_2^4}~.
\end{align}
\end{subequations}
The other prepotential terms $F_{k,\ell}$ can be obtained from $F_{\ell,k}$ by the operations
\begin{equation}
a_1 \leftrightarrow a_2~,\qquad (m_1, m_2) \leftrightarrow (m_3, m_4)~, \qquad m_{12}\leftrightarrow -m_{12} ~.
\end{equation}
An important check of these results is that $F_{k,0}$ with $a_2=0$ matches exactly the $k$-instanton prepotential of the 
conformal SU(2) gauge theory with $N_f=4$ if we choose to label the Coulomb parameter of the gauge group 
by $a_1$ and take the four masses to be given by
\begin{equation}
\big(m_1,m_2, m_{12}, m_{12}\big)
\end{equation}
(see for example \cite{Billo:2010mg}, taking into account that $m_i^{\mathrm{here}}=\sqrt{2}m_i^{\mathrm{there}}$).
These calculations can be extended to higher instanton numbers without any problem. 

We conclude by recalling the structure of the perturbative part of the prepotential for the quiver theory. The basic ingredient is 
the double-Gamma function
\begin{equation}
\label{gamma2}
\gamma_{\epsilon_1, \epsilon_2}(x) :=\log\Gamma_2(x|\epsilon_1,\epsilon_2)=\frac{d}{ds} \left[\frac{\Lambda^s}{\Gamma(s)}\int_0^{\infty}\frac{dt}{t} \frac{t^s e^{-tx}}{(1-e^{-\epsilon_1 t})(1-e^{-\epsilon_2 t})} \right]_{s=0}
\end{equation}
where $\Lambda$ is an arbitrary mass scale.
For large values of $x$, the function $\gamma_{\epsilon_1, \epsilon_2}$ has a series expansion of the form
\begin{equation}
\label{gamma2exp}
\begin{aligned}
\gamma_{\epsilon_1, \epsilon_2}(x) =&~ \frac{x^2}{4}\left(3-\log\frac{x^2}{\Lambda^2} \right)\,b_0 - 
x\left(1-\frac{1}{2}\log\frac{x^2}{\Lambda^2}\right)\,b_1-\frac{1}{4}\log\frac{x^2}{\Lambda^2}\,b_2\\
&~~+\sum_{n\ge 3} \frac{x^{2-n}}{n(n-1)(n-2)}\,b_n
\end{aligned}
\end{equation}
where the coefficients $b_n$'s are defined by
\begin{equation}
\frac{1}{(1-e^{-\epsilon_1 t})(1-e^{-\epsilon_2 t})} = \sum_{n=0}^{\infty}\frac{b_n}{n!}t^{n-2}~.
\end{equation}
For the SU(2)$\,\times\,$SU(2) quiver the perturbative part of the prepotential is 
\begin{equation}
\begin{aligned}
F_{\mathrm{pert}}= \eu\ed\Bigg[&\gamma_{\eu, \ed}(2a_1)+\gamma_{\eu, \ed}(-2a_1)
+\gamma_{\eu, \ed}(2a_2)+\gamma_{\eu, \ed}(-2a_2)\\
&-\sum_{f=1,2}\Big(\gamma_{\eu, \ed}(a_1+\widehat{m}_f)+\gamma_{\eu, \ed}(-a_1+\widehat{m}_f)\Big)\\
&-\sum_{f=3,4}\Big(\gamma_{\eu, \ed}(a_2+\widehat{m}_f)+\gamma_{\eu, \ed}(-a_2+\widehat{m}_f)\Big)\\
&-\gamma_{\eu, \ed}(a_1+a_2-\widehat{m}_{12}+\epsilon)-\gamma_{\eu, \ed}(-a_1+a_2-\widehat{m}_{12}+\epsilon)\\
&-\gamma_{\eu, \ed}(a_1-a_2-\widehat{m}_{12}+\epsilon)-\gamma_{\eu, \ed}(-a_1-a_2-\widehat{m}_{12}+\epsilon)\,\Bigg]
\end{aligned}
\label{Fpertapp}
\end{equation}
where we recall that $\widehat{m}$ stands for $m+\ft12\epsilon$, with $\epsilon$ defined in (\ref{epsilon}) .
The first line in the above formula represents the contribution of the two adjoint vector multiplets, the second and third lines
represent the contributions of the fundamental hypermultiplets of the two gauge groups, while the last two lines are the
contribution of the bi-fundamental matter.

This perturbative potential can be expanded for small $\eu$ and $\ed$ using (\ref{gamma2exp}).  Up to
order four in the masses and up to order two in the $\epsilon$'s we get
\begin{align}
F_{\mathrm{pert}} =&~ -\left(a_1^2+a_2^2+\frac{1}{12}\big(\epsilon^2+\eu\ed\big)\right)\log16\notag\\
&\,-\left(a_1^2-\frac{1}{2}\big(m_1^2+m_2^2\big)+\frac{1}{12}\big(2\epsilon^2-\eu\ed\big)\right)
\log\frac{a_1^2}{\Lambda^2}\notag\\
&\,-\left(a_2^2-\frac{1}{2}\big(m_3^2+m_4^2\big)
+\frac{1}{12}\big(2\epsilon^2-\eu\ed\big) \right)\log\frac{a_2^2}{\Lambda^2}\notag\\
&\,+\left(\frac{1}{2}\big(a_1+a_2\big)^2+\frac{1}{2}m_{12}^2-\frac{1}{24}\big(\epsilon^2-2\eu\ed\big)\right)
\log\frac{\big(a_1+a_2\big)^2}{\Lambda^2}\notag\\
&\,+\left(\frac{1}{2}\big(a_1-a_2\big)^2+\frac{1}{2}m_{12}^2-\frac{1}{24}\big(\epsilon^2-2\eu\ed\big)\right)
\log\frac{(a_1-a_2)^2}{\Lambda^2}\notag\\
&\,-\frac{2\big(m_1^4+m_2^4\big)-\big(\epsilon^2-2\eu\ed\big)\big(m_1^2+m_2^2\big)}{24a_1^2}
-\frac{2\big(m_3^4+m_4^4\big)-\big(\epsilon^2-2\eu\ed\big)\big(m_3^2+m_4^2\big)}{24a_2^2}\notag\\
&\,+\frac{m_{12}^2\big(\epsilon^2-2\eu\ed\big) -2m_{12}^4}{24(a_1+a_2)^2}
+\frac{m_{12}^2\big(\epsilon^2-2\eu\ed\big) -2m_{12}^4}{24(a_1-a_2)^2}\notag\\
&\,+\frac{\big(m_1^4+m_2^4\big)\big(\epsilon^2-2\eu\ed\big)}{48a_1^4}
+\frac{\big(m_3^4+m_4^4\big)\big(\epsilon^2-2\eu\ed\big)}{48a_2^4}\notag\\
&\,+\frac{m_{12}^4\big(\epsilon^2-2\eu\ed\big)}{48(a_1+a_2)^2}
+\frac{m_{12}^4\big(\epsilon^2-2\eu\ed\big)}{48(a_1-a_2)^2}+\dots
\label{Fpertapp1}
\end{align}
It is easy to check that in the limit $\eu$,$\ed\to0$ we recover the expected expression of the 1-loop prepotential for
the linear quiver we have considered. Notice that only in the massless undeformed theory the dependence on the
arbitrary scale $\Lambda$ drops out, in agreement with conformal invariance.

\section{Polynomials appearing in the SW curves}
\label{secn:poly}
The fourth-order polynomial $\mathcal{P}_4$ appearing in the numerator of the SW curve (\ref{su2curve}) 
for the SU(2) $N_f=4$ theory is 
\begin{equation}
\mathcal{P}_4(t)=\sum_{\ell=0}^4 C_\ell\,t^\ell
\label{P4nf4}
\end{equation}
where
\begin{subequations}
\label{CSU2}
\begin{align}
C_0&=\frac{q^2}{4}(m_1-m_2)^2~,\\
C_1&=-q u+q m_1m_2-\frac{q^2}{2}\big[(m_1+m_2)(m_3+m_4)+m_1^2+m_2^2\big]~,\\
C_2&=u+qu+\frac{q}{2}\big[(m_1+m_2)(m_3+m_4)-2m_1m_2-2m_3m_4\big]
+\frac{q^2}{4} \big(\sum_{f=1}^4m_f\big)^2~,\\
C_3&=-u+m_3 m_4-\frac{q}{2}\big[(m_1+m_2)(m_3+m_4)+m_3^2+m_4^2\big]~,\\
C_4&=\frac{1}{4}(m_3-m_4)^2~.
\end{align}
\end{subequations}
The sixth-order polynomial $\mathcal{P}_6$ appearing in the numerator of the SW curve (\ref{cmassg}) for the
SU(2)$\,\times\,$SU(2) quiver theory is
\begin{equation}
\mathcal{P}_6(t)=\sum_{\ell=0}^6 C'_\ell\,t^\ell
\label{P6}
\end{equation}
where
\begin{subequations}
\begin{align}
C'_0&=\frac{t_1^2\,t_2^2}{4}\,(m_1-m_2)^2~,\\
C'_1&=-t_1\,t_2^2\,(u_1-m_1m_2)+\frac{t_1^2\,t_2}{4}\,\Big(m_{12}^2-2m_1^2-2m_2^2+2m_{12}(m_1+m_2+m_{12})\Big)\cr
&~~-\frac{t_1^2\,t_2^2}{4}\,\Big(m_{12}^2+2(m_1+m_2+m_{12})\,\sum_{f=1}^4m_f-4m_1m_2\Big)~,\\
C'_2&=\frac{t_1\,t_2}{4}\,\Big(4(u_1+u_2)-7m_{12}^2-2m_{12}(m_1+m_2)-4m_1m_2\Big)+t_2^2\,u_1\cr
&~~+\frac{t_1\,t_2^2}{2}\Big(2u_1+(m_1+m_2+m_{12})(m_3+m_4+m_{12})+m_{12}(m_3+m_4)-2m_1m_2\Big)\cr
&~~+\frac{t_1^2}{4}\,\big(m_1+m_2-m_{12}\big)^2+\frac{t_1^2\,t_2^2}{4}\,\Big(m_{12}+m_1+m_2+m_3+m_4\Big)^2\cr
&~~-\frac{t_1^2\,t_2}{4}\,\Big(3m_{12}^2+2m_{12}(m_1+m_2+m_{12})-4(m_1+m_2)\sum_{f=1}^4m_f+4m_1m_2\Big)\\
C'_3&=-\frac{t_1}{4}\Big(4u_2+m_{12}^2-2m_{12}(m_1+m_2)\Big)-t_2\big(u_1+u_2-m_{12}^2\big)\cr
&~~-\frac{t_1\,t_2}{2}\Big(2u_1+2u_2-6m_{12}^2-m_{12}(m_1+m_2-m_3-m_4)+2(m_1+m_2)(m_3+m_4)\cr
&~~~~~\qquad-2m_1m_2-2m_3m_4\Big)-\frac{t_1^2}{2}\big(m_1+m_2-m_{12}\big)\big(m_1+m_2+m_3+m_4-m_{12}\big)\cr
&~~-\frac{t_2^2}{4}\Big(4u_1+m_{12}^2+2m_{12}(m_3+m_4)\Big)+\frac{t_1^2\,t_2}{2}\Big(m_{12}-\sum_{f=1}^4m_f\Big)
\Big(m_{12}-\sum_{f=1}^4m_f\Big)\cr
&~~-\frac{t_1\,t_2^2}{2}\big(m_{12}+m_3+m_4\big)\big(m_{12}+m_1+m_2+m_3+m_4\big)~,\\
C'_4&=u_2\!+\!\frac{t_1}{2}\Big(2u_2+m_{12}^2-m_{12}(2m_1+2m_2+m_3+m_4)\!+\!(m_1+m_2)(m_3+m_4)-2m_3m_4\Big)\cr
&~~+\frac{t_2}{4}\Big(4u_1+4u_2-7m_{12}^2+2m_{12}(m_3+m_4)-4m_3m_4\Big)\cr
&~~+\frac{t_1^2}{4}	\Big(m_{12}^2-2m_{12}\sum_{f=1}^4m_f+2\!\sum_{f<f'}m_fm_{f'}+\sum_{f=1}^4m_f^2\Big)
+\frac{t_2^2}{4}\big(m_{12}+m_3+m_4\big)^2\cr
&-~~\frac{t_1\,t_2}{4}\Big(5m_{12}^2-2m_{12}(m_3+m_4)-4(m_1+m_2)(m_3+m_4)-4m_3^3-4m_3m_4-4m_4^2\Big)~,\\
C'_5&=\,-u_2+m_3m_4-\frac{t_1}{4}\,\Big(m_{12}^2+2(m_3+m_4-m_{12})\,\sum_{f=1}^4m_f-4m_3m_4\Big)\cr
&~~+\frac{t_2}{4}\,\Big(m_{12}^2-2m_3^2-2m_4^2-2(m_3+m_4-m_{12})m_{12}\Big)~,\\
C'_6&=\,\frac{1}{4}(m_3-m_4)^2~,
\end{align}
\end{subequations}
where $t_1=q_1q_2$ and $t_2=q_2$. 

\section{Some useful integrals}
\label{secn:integral}
The calculation of the periods of the Seiberg-Witten differential $\lambda$ requires the evaluation of integrals of the following
types
\begin{equation}
\begin{aligned}
I_1=\frac{1}{\pi}\int_0^{z}\sqrt{\frac{z-t}{t}}\frac{f(t)}{q-t}\,dt~~~~\mbox{for}~|q|<1~,
\end{aligned}
\label{integral1}
\end{equation}
and
\begin{equation}
\begin{aligned}
I_2=\frac{1}{\pi}\int_0^{z}\sqrt{\frac{z-t}{t}}\frac{f(t)}{1-t}\,dt
\end{aligned}
\label{integral2}
\end{equation}
where $f(t)$ is a function admitting a Taylor expansion $\sum_nf_n\,t^n$.
Using the identities
\begin{equation}
\frac{f(t)}{q-t}=\sum_{n=0}^\infty\frac{t^n}{q^{n+1}}\left(f(q)-\sum_{\ell=n+1}^\infty \!f_\ell\, q^\ell\right)
\end{equation}
and 
\begin{equation}
\int_0^{z}\sqrt{\frac{z-t}{t}}\,t^n=(-1)^n\,\pi\,\binom{1/2}{n+1}\,z^{n+1}~,
\label{intn}
\end{equation}
we can prove that
\begin{equation}
I_1= f(q)-\sqrt{\frac{q-z}{q}}\,f(q)-\sum_{n=0}^\infty\sum_{\ell=0}^\infty(-1)^n\binom{1/2}{n+1}f_{n+\ell+1}
\,z^{n+1}\,q^{\ell}~.
\label{I1}
\end{equation}
On the other hand, from 
\begin{equation}
\frac{f(t)}{1-t}=\sum_{n=0}^\infty\sum_{\ell=0}^n f_\ell \,t^n
\end{equation}
and (\ref{intn}), we have
\begin{equation}
I_2=\sum_{n=0}^\infty\sum_{\ell=0}^n(-1)^n\binom{1/2}{n+1}f_{\ell}
\,z^{n+1}~.
\label{I2}
\end{equation}

These results can be used to compute the periods of the Seiberg-Witten differential. For example in the SU(2) $N_f=4$ theory 
considered in Section~\ref{secn:Nf4}, we can rewrite the last term of (\ref{anf4}) as
\begin{equation}
J=\frac{\sqrt{C}}{\pi(1-q)}\int_0^{e_2}\sqrt{\frac{e_2-t}{t}}\,\left(\frac{\sqrt{e_3-t{\phantom{(}}}}{q-t}
-\frac{\sqrt{e_3-t{\phantom{(}}}}{1-t}\right)dt = \frac{\sqrt{C}}{1-q}\big(I_1-I_2\big)
\label{J}
\end{equation}
where $I_1$ and $I_2$ are as in (\ref{integral1}) and (\ref{integral2}) with $z=e_2$ and $f(t)=\sqrt{e_3-t{\phantom{(}}}$. 
Then, from (\ref{I1}) and (\ref{I2}) we get
\begin{align}
J &= \frac{\sqrt{C}}{1-q}\Bigg(\sqrt{e_3-q{\phantom{(}}}-\sqrt{\frac{(e_2-q)(q-e_3)}{q}}
+\sum_{n=0}^\infty\sum_{\ell=0}^\infty
(-1)^\ell\binom{1/2}{n+1}\!\binom{1/2}{n+\ell+1}\frac{e_2^{n+1}\,q^\ell}{e_3^{n+\ell+1/2}}\notag\\
&~~~~~~~~~~~~-\sum_{n=0}^\infty\sum_{\ell=0}^n
(-1)^{(n+\ell)}\binom{1/2}{n+1}\binom{1/2}{\ell}\frac{e_2^{n+1}}{e_3^{\ell-1/2}}\Bigg)~.
\label{afinapp}
\end{align}
This is the result used to obtain (\ref{afin}) in the main text.

In the quiver theory described in Section~\ref{secn:quiver} we had to compute the integral (see (\ref{a1mass})
\begin{equation}
J'=\frac{1}{\pi}\int_0^{\zeta_1}
 \sqrt{\frac{\zeta_1-t}{t}}~\sqrt{\frac{u_2(\widehat{\zeta}-t)}{(1-t)(1-q_2t)}}\,\frac{dt}{q_1-t}
\end{equation}
which is again of the type $I_1$ with $z=\zeta_1$, $q=q_1$ and
\begin{equation}
f(t)=\sqrt{\frac{u_2(\widehat{\zeta}-t)}{(1-t)(1-q_2t)}}~.
\label{fft}
\end{equation}
Using (\ref{I1}) we then find
\begin{equation}
J'=\sqrt{\frac{u_2(\widehat{\zeta}-q_1)}{(1-q_1)(1-q_1q_2)}}-
\sqrt{\frac{u_2(q_1-\zeta_1)(\widehat{\zeta}-q_1)}{q_1(q_1-1)(q_1q_2-1)}}
-\sum_{n,\ell=0}^\infty(-1)^n \binom{1/2}{n+1}f_{n+\ell+1}\,\zeta_1^{n+1}\,q_1^\ell
\label{J'}
\end{equation}
where the $f_n$'s are the Taylor expansion coefficients of the function (\ref{fft}).
This is the result used to obtain (\ref{a1actual}) in the main text.

\section{Conformal Ward identities}
\label{appT}

The chiral blocks that are relevant for the discussion in Sections~\ref{secn:AGT} and \ref{prepotAGT} are
\begin{equation}
\begin{aligned}
\Big\langle T(z)\prod_{i=0}^{n+2}V_{\alpha_i}(z_i)\Big\rangle &=
\sum_{i=0}^{n+2}\left(\frac{\Delta_{\alpha_i}}{(z-z_i)^2}+\frac{1}{z-z_i}
\frac{\p}{\p z_i} \right)\,\Big\langle\prod_{i=0}^{n+2}V_{\alpha_i}(z_i)\Big\rangle~, \\
\Big\langle\!\! :\!T(z) \Phi_{2,1}(z)\!:\prod_{i=0}^{n+2}V_{\alpha_i}(z_i)\Big\rangle &=
\sum_{i=0}^{n+2}\left(\frac{\Delta_{\alpha_i}}{(z-z_i)^2}+\frac{1}{z-z_i}
\frac{\p}{\p z_i} \right)\,\Big\langle \Phi_{2,1}(z)\prod_{i=0}^{n+2}V_{\alpha_i}(z_i)\Big\rangle~.
\end{aligned}
\label{tpsi}
\end{equation}
We can simplify the right hand sides by imposing the constraints that follow from the 
global conformal invariance of the theory. For an $(n+3)$-point correlator these are:
\begin{equation}
\widehat{\Lambda}_k \,\Big\langle \prod_{i=0}^{n+2}V_{\alpha_{i}}(z_i)\Big\rangle=0\qquad 
\text{for}\quad k=-1,0,1~,
\label{Lambdas0}
\end{equation}
where 
\begin{equation}
\widehat{\Lambda}_{-1}=\sum_{i=0}^{n+2} \frac{\p}{\p z_i}~,~~~~
\widehat{\Lambda}_{0}=\sum_{i=0}^{n+2}\Big(z_i\frac{\p}{\p z_i}+\Delta_i\Big)~,~~~~
\widehat{\Lambda}_{1}=\sum_{i=0}^{n+2}\Big(z_i^2\frac{\p}{\p z_i}+2 z_i\Delta_i\Big)
\label{Lambdas1}
\end{equation}
are the generators of the global conformal group. The relations (\ref{Lambdas0}) allow to express the derivatives with respect
to, say, $z_0$, $z_{n+1}$ and $z_{n+2}$ in terms of the derivatives with respect to the remaining $n$ coordinates.
If we fix $z_0=0$, $z_{n+1}=1$ and $z_{n+2}=\infty$, we have
\begin{equation}
\begin{aligned}
&\frac{\p}{\p z_0}= - \sum_{i=1}^{n} \Big((z_i-1) \frac{\p}{\p z_i} + \Delta_{\alpha_i}\Big) 
+  \Delta_{\alpha_0} + \Delta_{\alpha_{n+1}} - \Delta_{\alpha_{n+2}}~,\\
&\frac{\p}{\p z_{n+1}}= - \sum_{i=1}^{n} \Big(z_i \frac{\p}{\p z_i} + \Delta_{\alpha_i}\Big) - 
\Delta_{\alpha_0} - \Delta_{\alpha_{n+1}} +  \Delta_{\alpha_{n+2}}~,\\
&\frac{\p}{\p z_{n+2}} = 0~.
 \end{aligned}
 \label{der}
 \end{equation}
Applying these relations to the first correlator in (\ref{tpsi}), we get
\begin{equation}
\begin{aligned}
\Big\langle T(z)\prod_{i=0}^{n+2}V_{\alpha_i}(z_i)\Big\rangle 
=& \left[\sum_{i=1}^{n}\left(\frac{\Delta_{\alpha_i}}{(z-z_i)^2}+\frac{z_i(z_i-1)}{z(z-1)(z-z_i)}\frac{\p}{\p z_i}\right)
 +\frac{\Delta_{\alpha_0}}{z^2}+ \frac{\Delta_{\alpha_{n+1}}}{(z-1)^2}\right.\\
&~~\left.-\frac{\sum_{i=1}^{n}\Delta_{\alpha_i}+\Delta_{\alpha_0} + \Delta_{\alpha_{n+1}} - \Delta_{\alpha_{n+2}}}{z(z-1)} \right]
\Big\langle\prod_{i=0}^{n+2}V_{\alpha_i}(z_i)\Big\rangle 
\end{aligned}
\label{tpsi1}
\end{equation}
where, both in the left and in the right hand side, it is understood that $z_0=0$, $z_{n+1}=1$ and $z_{n+2}=\infty$.

Proceeding in a similar way, we can rewrite the second correlator in (\ref{tpsi}) as
\begin{equation}
\begin{aligned}
\Big\langle&\!\! :\!T(z)\Phi_{2,1}(z)\!:\prod_{i=0}^{n+2}V_{\alpha_i}(z_i)\Big\rangle = 
\left[\sum_{i=1}^{n}\left(\frac{\Delta_{\alpha_i}}{(z-z_i)^2}+\frac{z_i(z_i-1)}{z(z-1)(z-z_i)}\frac{\p}{\p z_i}\right)
- \frac{2z-1}{z(z-1)}\frac{\p}{\p z}\right.\\
&\left. +\frac{\Delta_{\alpha_0}}{z^2}+ \frac{\Delta_{\alpha_{n+1}}}{(z-1)^2}
-\frac{\sum_{i=1}^{n}\Delta_{\alpha_i}+\Delta_z +\Delta_{\alpha_0} + \Delta_{\alpha_{n+1}} -
 \Delta_{\alpha_{n+2}} }{z(z-1)} \right]
\Big\langle\!\Phi_{2,1}(z)\prod_{i=0}^{n+2}V_{\alpha_i}(z_i)\Big\rangle~.
\end{aligned}
\label{tpsi2}
\end{equation}
To make contact with the discussion in Sections~\ref{secn:AGT} and \ref{prepotAGT}, we should notice that the punctures
$z_i$ have been denoted by $t_i$ and that these are related to the gauge couplings according to $q_i=t_i/t_{i+1}$.
Using this we can obtain from (\ref{tpsi1}) and (\ref{tpsi2}) the formul\ae~(\ref{phi25}) and (\ref{phi1}) of the main text.

\providecommand{\href}[2]{#2}\begingroup\raggedright
\endgroup

\end{document}